\RequirePackage{fix-cm} 
\documentclass[a4paper, twoside, reqno, dvips, 11pt]{amsart}
\usepackage{fixltx2e}     

\usepackage[latin1]{inputenc}
\usepackage[T1]{fontenc}

\usepackage{eucal}
\usepackage{esint}
\usepackage{dsfont}
\usepackage{xspace}
\usepackage{amsgen}
\usepackage{amsthm}
\usepackage{amssymb}
\usepackage{amsmath}
\usepackage{upgreek}
\usepackage{wasysym}
\usepackage{amsfonts}
\usepackage{stmaryrd}
\usepackage{mathtools}

\usepackage{mathrsfs}
\DeclareMathAlphabet{\mathscrbf}{OMS}{mdugm}{b}{n}

\usepackage{a4wide}

\usepackage[dvipsnames, table]{xcolor}
\definecolor{bckg}{RGB}{20.8, 20.8, 20.8}
\definecolor{oneblue}{rgb}{0.0, 0.0, 0.85}
\definecolor{Lightblue}{RGB}{214, 214, 214}
\definecolor{bluepigment}{rgb}{0.2, 0.2, 0.6}
\definecolor{charcoal}{rgb}{0.21, 0.27, 0.31}
\definecolor{denimblue}{rgb}{0.08, 0.38, 0.74}
\definecolor{Lightgray}{rgb}{0.89, 0.89, 0.89}
\definecolor{darkgrey}{rgb}{0.273, 0.281, 0.30}
\definecolor{darkelectricblue}{rgb}{0.33, 0.41, 0.47}

\usepackage[sort&compress, comma, square, numbers]{natbib}

\usepackage{psfrag}
\usepackage{graphicx}
\usepackage{subfigure}
\usepackage{morefloats}
\usepackage{indentfirst}

\usepackage{acronym}
\usepackage{microtype}
\usepackage[labelsep=period,%
            labelfont={bf,sf,color=bluepigment},%
            justification=raggedright]{caption}

\usepackage[usenames, dvipsnames, pdf]{pstricks}
\usepackage{epsfig}
\usepackage{pst-grad} 
\usepackage{pst-plot} 

\usepackage[colorlinks,
           urlcolor=oneblue,
           linkcolor=denimblue,
           citecolor=NavyBlue,
           bookmarksopen=false,
           pdfpagemode=UseNone,
           pagebackref]{hyperref}

\usepackage[explicit]{titlesec}

\titleformat{\section}
  {\color{NavyBlue}\Large\sffamily\bfseries}
  {}
  {0em}
  {\colorbox{bckg!5}{\parbox{\dimexpr\linewidth-2\fboxsep\relax}{\centering\thesection. #1}}}
  [\vspace*{0.33em}]

\titleformat{name=\section,numberless}
  {\color{NavyBlue}\Large\sffamily\bfseries}
  {}
  {0.0em}
  {\colorbox{bckg!10}{\parbox{\dimexpr\linewidth-2\fboxsep\relax}{\centering#1}}}
  [\vspace*{0.33em}]

\titleformat{\subsection}
  {\color{NavyBlue}\large\sffamily\bfseries}
  {}
  {0.0em}
  {\colorbox{bckg!5}{\parbox{\dimexpr\linewidth-2\fboxsep\relax}{\centering\thesubsection. #1}}}
  [\vspace*{0.33em}]

\titleformat{name=\subsection,numberless}
  {\color{NavyBlue}\Large\sffamily\bfseries}
  {}
  {0em}
  {\colorbox{bckg!10}{\parbox{\dimexpr\linewidth-2\fboxsep\relax}{\centering#1}}}
  [\vspace*{0.33em}]

\titleformat{\subsubsection}
  {\color{bluepigment}\sffamily\normalsize\bfseries}
  {\thesubsubsection}
  {0.5em}
  {#1}
  [\vspace*{0.33em}]

\titleformat{\paragraph}[runin]
  {\color{bluepigment}\sffamily\small\bfseries}
  {}
  {0em}
  {#1}

\titlespacing{\section}{1.0em}{1.5em plus 2pt minus 2pt}%
{1.0em plus 2pt minus 2pt}[0em]
\titlespacing{\subsection}{1.0em}{1.5em plus 2pt minus 2pt}%
{1.0em}[0em]
\titlespacing{\subsubsection}{1.0em}{1.5em plus 2pt minus 2pt}%
{1.0em plus 2pt minus 2pt}[0em]

\usepackage{titletoc}

\contentsmargin{0.5em}
\newlength{\tocsep} 
\titlecontents{section}[\tocsep]
  {\addvspace{10pt}\bfseries\sffamily}
  {\contentslabel[\thecontentslabel]{\tocsep}}
  {}
  {\ \titlerule*[0.75pc]{.}\ \thecontentspage}
  []
\titlecontents{subsection}[\tocsep]
  {\addvspace{8pt}\sffamily}
  {\contentslabel[\thecontentslabel]{\tocsep}}
  {}
  {\ \titlerule*[0.5pc]{.}\ \thecontentspage}
  []
\titlecontents*{subsubsection}[\tocsep]
  {\addvspace{2pt}\footnotesize\sffamily}
  {}
  {}
  {\ \titlerule*[0.35pc]{.}\ \thecontentspage}
  [\\*]

\makeatletter
\def\@setauthors{%
  \begingroup
  \def\thanks{\protect\thanks@warning}%
  \trivlist
  \centering\footnotesize \@topsep30\p@\relax
  \advance\@topsep by -\baselineskip
  \item\relax
  \author@andify\authors
  \def\\{\protect\linebreak}%
  \textsc{\normalsize\textcolor{darkelectricblue}{\authors}}%
  \ifx\@empty\contribs
  \else
    ,\penalty-3 \space \@setcontribs
    \@closetoccontribs
  \fi
  \endtrivlist
  \endgroup
}
\def\@settitle{\begin{center}%
  \baselineskip14\p@\relax
    \bfseries
    \textsc{\Large\textcolor{charcoal}{\@title}}
  \end{center}%
}
\makeatother

\usepackage{enumitem}
\setlist[description]{%
  topsep=30pt,               
  itemsep=5pt,               
  font={\bfseries\sffamily\color{NavyBlue}}, 
}

\usepackage{fancyhdr}
\usepackage{lastpage}

\newcommand*\Title{\textcolor{bluepigment}{Computation of capillary-gravity waves}}
\newcommand*\Authors{\textcolor{bluepigment}{D.~Dutykh, D.~Clamond \& A.~Dur\'an}}
\newcommand*{\plogo}{\textcolor{gray}{{\texttt{arXiv.org} / \textsc{hal}}}} 

\numberwithin{equation}{section}

\newcommand{\sur}[1]{{#1}_\text{s}}                    
\renewcommand{\bot}[1]{{#1}_\text{b}}                  

\newcommand{\depth}{d}

\newcommand{\ud}{\mathrm{d}}
\newcommand{\ui}{\mathrm{i}}
\newcommand{\ue}{\mathrm{e}}

\newcommand{\Bo}{\mathsf{Bo}}
\newcommand{\Fr}{\mathsf{Fr}}

\renewcommand{\Re}{\operatorname{Re}}
\renewcommand{\Im}{\operatorname{Im}}

\newcommand{\half}{{\textstyle{1\over2}}}

\newcommand{\ie}{\emph{i.e.\/} }

\acrodef{lm}[LM]{Levenberg--Marquardt}

\begin{document}

\title[\Title]{Efficient computation of capillary-gravity generalized solitary waves}

\author[D. Dutykh]{Denys Dutykh$^*$}
\address{Universit\'e Savoie Mont Blanc, LAMA, UMR 5127 CNRS, Campus Scientifique, 73376 Le Bourget-du-Lac Cedex, France}
\email{Denys.Dutykh@univ-savoie.fr}
\urladdr{http://www.denys-dutykh.com/}
\thanks{$^*$ Corresponding author}

\author[D. Clamond]{Didier Clamond}
\address{Universit\'e de Nice -- Sophia Antipolis, Laboratoire J. A. Dieudonn\'e, Parc Valrose, 06108 Nice cedex 2, France}
\email{diderc@unice.fr}
\urladdr{http://math.unice.fr/~didierc/}

\author[A.~Dur\'an]{Angel Dur\'an}
\address{Departamento de Matem\'atica Aplicada, E.T.S.I. Telecomunicaci\'on, Campus Miguel Delibes, Universidad de Valladolid, Paseo de Belen 15, 47011 Valladolid, Spain}
\email{angel@mac.uva.es}


\begin{titlepage}
\setcounter{page}{1}
\thispagestyle{empty} 
\noindent
{\Large Denys \textsc{Dutykh}}\\
{\it\textcolor{gray}{Universit\'e Savoie Mont Blanc, France}}\\[0.02\textheight]
{\Large Didier \textsc{Clamond}}\\
{\it\textcolor{gray}{Universit\'e de Nice -- Sophia Antipolis, France}}\\[0.02\textheight]
{\Large Angel \textsc{Dur\'an}}\\
{\it\textcolor{gray}{Universidad de Valladolid, Spain}}\\[0.16\textheight]

\colorbox{Lightblue}{
  \parbox[t]{1.0\textwidth}{
    \centering\huge\sc
    \vspace*{0.7cm}

    \textcolor{bluepigment}{Efficient computation of capillary-gravity generalized solitary waves}

    \vspace*{0.7cm}
  }
}

\vfill 

\raggedleft     
{\large \plogo} 
\end{titlepage}


\newpage
\thispagestyle{empty} 
\par\vspace*{\fill}   
\begin{flushright} 
{\textcolor{denimblue}{\textsc{Last modified:}} \today}
\end{flushright}


\newpage
\maketitle
\thispagestyle{empty}


\begin{abstract}

This paper is devoted to the computation of capillary-gravity solitary waves of the irrotational incompressible Euler equations with free surface. The numerical study is a continuation of a previous work in several points: an alternative formulation of the Babenko-type equation for the wave profiles, a detailed description of both the numerical resolution and the analysis of the internal flow structure under a solitary wave. The numerical code used in this study is provided in open source for those interested readers.

\bigskip
\noindent \textbf{\keywordsname:} Solitary surface waves; capillary-gravity waves; Euler equations; generalised solitary waves.


\end{abstract}


\newpage
\tableofcontents
\thispagestyle{empty}
\newpage


\section{Introduction}

The formation and dynamics of capillary-gravity waves on the surface of incompressible fluids constitute a research topic of permanent interest from experimental, theoretical and computational point of view \cite{Dias1999, Okamoto2001, Vanden-Broeck2010, Yang2010}. These solutions have been also studied using analytical asymptotic methods in various model equations (such as KdV5), see \cite{Boyd1998, Boyd1999, Boyd1998a} and also the contributions to the edited volume \cite{Segur1991}. This paper presents in details an efficient numerical algorithm for computing various steady capillary-gravity solitary waves for the irrotational Euler equations with a free surface. In particular, this algorithm can be used to compute generalized multi-hump solitary waves \cite{Clamond2015a}. In addition to the detailed algorithm, the present paper provides some physical characteristics of the generalized solitary waves that are not described in \cite{Clamond2015a}. The highlights can be assembled in three points.

The first one concerns the mathematical formulation of the equations for the wave profiles. Different from other approaches used in the literature to reformulate the Euler system on the free surface, \cite{Hunter1983a, Parau2005a, Vanden-Broeck2010}, the one employed in \cite{Clamond2015a} is based on the conformal mapping technique \cite{Ovsyannikov1974}, in order to transform the system into an integro-differential equation of Babenko type \cite{Babenko1987}. (This was previously developed for gravity solitary waves in \cite{Clamond2012, Clamond2012b, Dutykh2013b}.) As mentioned in \cite{Clamond2015a}, one of the advantages of this formulation is the preservation of the type of nonlinearity by the transformation \cite{Milewski2010}; it has implications on the convergence of the numerical method. As in \cite{Clamond2015a}, the numerical results on steady capillary-gravity waves presented here are obtained with a pseudo-spectral discretisation of the corresponding periodic problem  and the application of the \acf{lm} algorithm, \cite{Levenberg1944, Lourakis2005, Marquardt1963}, to the resulting discrete systems for the Fourier components of the approximation. 

A second highlight of the present paper is then a detailed description of the numerical procedure and its implementation to the model formulation. The performance of the method is shown through several numerical experiments computing different traveling waves.

Finally, the emergence of generalised solitary waves shown in \cite{Clamond2015a}, under gradually increasing values of the Bond number is used in a third highlighted point of the present study, where some properties of various physical fields under these waves are emphasised.

The manuscript is organised as follows. In Section~\ref{sec:model}, we summarise the application of the conformal mapping technique to the original Euler equations, explained in \cite{Clamond2015a} in more detail. Section~\ref{sec:numm} is devoted to a detailed description of the numerical method while the main numerical experiments are presented in Section~\ref{sec:numr}. Finally, the main conclusions and perspectives are outlined in Section~\ref{sec:concl}. With the aim of involving a wider audience, the numerical code used in computations below is available to download as open source \cite{Clamond2015b}. Thus, the claims made in this study can be easily verified by interested readers.


\section{Mathematical model}\label{sec:model}

Our starting point is the mathematical model for a potential flow induced by a solitary wave described in 
\cite{Clamond2015a}, which is summarised here. The physical assumptions involve an inviscid, homogeneous fluid 
in a horizontal channel of constant depth; the channel is modelled, above, by an impermeable free surface where 
the pressure is equal to the surface tension due to capillary forces, and bounded below by a fixed impermeable 
horizontal seabed. 

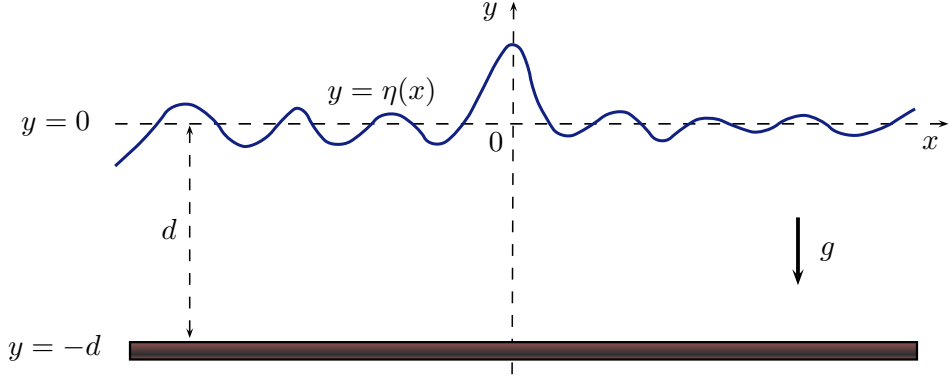
\begin{figure}
\centering
\scalebox{1} 
{
\begin{pspicture}(0,-2.4892187)(12.402813,2.5192187)
\definecolor{color22g}{rgb}{0.5490196078431373,0.3254901960784314,0.3137254901960784}
\definecolor{color22f}{rgb}{0.1803921568627451,0.1803921568627451,0.1803921568627451}
\definecolor{color105}{rgb}{0.06666666666666667,0.12941176470588237,0.5098039215686274}
\psline[linewidth=0.022cm,linestyle=dashed,dash=0.16cm 0.16cm,arrowsize=0.05291667cm 2.0,arrowlength=1.4,
arrowinset=0.4]{->}(1.3809375,0.8407813)(12.340938,0.8407813)
\psline[linewidth=0.02cm,linestyle=dashed,dash=0.16cm 0.16cm,arrowsize=0.05291667cm 2.0,arrowlength=1.4,
arrowinset=0.4]{<-}(6.6209373,2.4607813)(6.6009374,-2.4792187)
\psframe[linewidth=0.03, dimen=outer, fillstyle=gradient, 
gradlines=2000,gradbegin=color22g,gradend=color22f,gradmidpoint=0.69](11.940937,-2.0392187)(1.5609375,-2.2992187)
\usefont{T1}{ptm}{m}{n}
\rput(12.102344,0.5907813){$x$}
\usefont{T1}{ptm}{m}{n}
\rput(6.3123436,2.3307812){$y$}
\usefont{T1}{ptm}{m}{n}
\rput(6.3923435,0.61078125){$0$}
\psline[linewidth=0.02cm,linestyle=dashed,dash=0.16cm 0.16cm,arrowsize=0.05291667cm 2.0,arrowlength=1.4,
arrowinset=0.4]{<->}(2.3609376,0.82078123)(2.3609376,-1.9992187)
\psline[linewidth=0.05cm,linestyle=solid,arrowsize=0.05291667cm 2.0,arrowlength=1.4,arrowinset=0.4]{->}(10.3609376,-0.4)(10.3609376,-1.3)
\usefont{T1}{ptm}{m}{n}
\rput(10.75,-0.85){$g$}
\usefont{T1}{ptm}{m}{n}
\rput(2.0923438,-0.5492188){$d$}
\usefont{T1}{ptm}{m}{n}
\rput(0.58234376,-2.1292188){$y = -d$}
\usefont{T1}{ptm}{m}{n}
\rput(0.60234374,0.85078126){$y = 0$}
\pscustom[linewidth=0.04,linecolor=color105]
{
\newpath
\moveto(1.3809375,0.28078124)
\lineto(1.6375229,0.5188635)
\curveto(1.7658156,0.63790464)(1.9631889,0.8651648)(2.0322695,0.9733841)
\curveto(2.10135,1.0816033)(2.2937894,1.1357127)(2.4171479,1.081603)
\curveto(2.540506,1.0274936)(2.718142,0.87057585)(2.7724197,0.76776767)
\curveto(2.8266973,0.66495943)(2.979662,0.5513294)(3.0783484,0.5405072)
\curveto(3.1770349,0.5296854)(3.3645399,0.6324936)(3.4533575,0.7461237)
\curveto(3.5421755,0.8597537)(3.6852713,1.0058496)(3.739549,1.0383154)
\curveto(3.7938268,1.0707812)(3.8875794,1.0112606)(3.927054,0.91927445)
\curveto(3.9665284,0.8272882)(4.0750837,0.67578125)(4.1441646,0.6162607)
\curveto(4.2132454,0.5567401)(4.3760777,0.5513294)(4.4698305,0.6054388)
\curveto(4.563583,0.65954834)(4.726416,0.8002333)(4.795497,0.88680863)
\curveto(4.864578,0.97338396)(5.042213,1.0004388)(5.150769,0.94091827)
\curveto(5.2593246,0.8813977)(5.4073544,0.74071276)(5.446829,0.65954834)
\curveto(5.486304,0.5783838)(5.61953,0.545918)(5.7132826,0.5946167)
\curveto(5.807035,0.64331543)(5.984671,0.85434294)(6.068555,1.0166718)
\curveto(6.152438,1.1790005)(6.2856655,1.4441375)(6.3350086,1.5469457)
\curveto(6.3843517,1.6497539)(6.483039,1.8066717)(6.532382,1.8607812)
\curveto(6.5817256,1.9148908)(6.6754775,1.8932471)(6.7198863,1.8174936)
\curveto(6.7642956,1.7417401)(6.823508,1.5685894)(6.8383107,1.4711924)
\curveto(6.853114,1.373795)(6.9123254,1.1844113)(6.9567347,1.0924251)
\curveto(7.001144,1.0004388)(7.0948963,0.83811)(7.1442394,0.7677675)
\curveto(7.1935825,0.69742495)(7.32681,0.65954834)(7.410693,0.69201416)
\curveto(7.494577,0.72448)(7.6771474,0.83811)(7.7758336,0.91927445)
\curveto(7.8745203,1.0004388)(8.057091,1.0274936)(8.140975,0.9733841)
\curveto(8.224859,0.91927445)(8.367954,0.7785895)(8.427167,0.69201416)
\curveto(8.486378,0.6054388)(8.619606,0.5892059)(8.693621,0.65954834)
\curveto(8.767635,0.72989076)(8.925534,0.84352094)(9.009418,0.88680863)
\curveto(9.093302,0.93009627)(9.275871,0.92468536)(9.374558,0.87598664)
\curveto(9.473245,0.82728803)(9.660749,0.75694567)(9.749568,0.7353018)
\curveto(9.838387,0.713658)(10.011086,0.7623566)(10.094971,0.832699)
\curveto(10.178855,0.90304136)(10.361425,0.96256196)(10.460113,0.9517401)
\curveto(10.558799,0.94091827)(10.721632,0.8597537)(10.785779,0.7894113)
\curveto(10.849924,0.7190689)(11.042363,0.67037034)(11.170657,0.69201416)
\curveto(11.298949,0.713658)(11.51606,0.7948224)(11.604877,0.85434294)
\curveto(11.693696,0.91386354)(11.8121195,0.9896167)(11.900937,1.0383155)
}
\usefont{T1}{ptm}{m}{n}
\rput(4.8923435,1.2707813){$y = \eta(x)$}
\usefont{T1}{ptm}{m}{n}
\end{pspicture}
}
\caption{\small\em Definition sketch of the physical domain.}
\label{fig:sketch}
\end{figure}

The mathematical description in Cartesian coordinates moving with the wave (with $x$ as the horizontal coordinate 
and $y$ the upward vertical one) is sketched in Figure \ref{fig:sketch}: $d$ stands for the mean depth of the channel, 
the bottom is posed at $y=-d$ while $y=\eta(x)$ denotes the free surface elevation from the mean water level at $y=0$. 
The zero mean level condition of the free surface is redefined in order to cover the computations of generalised 
solitary waves (see \cite{Clamond2015a} for details).

Let $\phi$ and $\psi$ be the velocity potential and the stream function, respectively, and consider the complex 
potential $f \equiv \phi + \ui\/\psi$. They define a conformal mapping
\begin{equation}\label{conformalmap}
z \mapsto \zeta \equiv (\ui\sur{\psi} - f)/c,
\end{equation}
where $\sur{\psi}$ and $\bot{\psi}$ are the constant traces of $\psi$ at the upper and lower boundaries, respectively, 
with $-c$ standing for the mean flow velocity and satisfying
\begin{equation}
  c\ \equiv\ -\left<\,\frac{1}{\depth}\,\int_{-\depth}^\eta\,u(x,y)\ \ud\/y\,\right>\, =\ 
  \frac{\bot{\psi}-\sur{\psi}}{\depth}.
\end{equation}
The conformal mapping \eqref{conformalmap} transforms the fluid domain into the strip
\(
  -\infty\/ \leqslant\/ \alpha\/ \leqslant\/ \infty,
  -\/\depth\/ \leqslant\/ \beta\/ \leqslant\/ 0,
\)
where $\alpha \equiv \Re(\zeta)$ and $\beta \equiv \Im(\zeta)$. On this domain, the free surface $\eta$ can be 
modelled by a nonlocal Babenko-type equation of the form \cite{Clamond2015a} 
\begin{align}\label{eq:bab1}
\mathscr{C}\!\left\{B\,\eta\,-\,\frac{g\,\eta^2}{2}\,+\, \tau\,-\,\frac{\tau\,(\/1\/ + \/\mathscr{C}\{\eta\}\/)}
{\sqrt{\/(1\/+\/\mathscr{C}\{\eta\})^2\/+\/\eta_\alpha^{\,2}\/}} \right\}\ = \nonumber\\ g\,\eta\,(1\/+\/\mathscr{C}
\{\eta\})\ -\ \frac{\ud}{\ud\/\alpha}\left\{\frac{\tau\,\eta_\alpha}{\sqrt{\/(1\/+\/\mathscr{C}\{\eta\})^2\/+\/
\eta_\alpha^{\,2}\/}}\right\}\,+\ K.
\end{align}
In \eqref{eq:bab1}, $B$ is the Bernoulli constant, $\tau$ is a surface tension parameter, $K$ is an integration 
constant and $\mathscr{C}$ is a pseudo-differential operator with Fourier symbol
\begin{equation}\label{mathC}
  \widehat{\mathscr{C}}(k)\,=\,\left\{\begin{array}{lr}
  k\coth(k\depth) &\quad  (k\neq0),  \\
  1\,/\,\depth  & \quad  (k=0).
\end{array}\right.
\end{equation}
The constant $K$ is determined from the mean level condition on $\eta$. In the limit $\tau\to 0$ the Babenko 
formulation for pure gravity solitary waves derived in \cite{Clamond2012b, Dutykh2013b} is recovered.

$K=0$ for classical solitary waves because $\eta\to0$ as $x\to\pm\infty$ and the wave mass 
$\int_{-\infty}^{\infty}\eta\,\ud\/x$ is finite. For generalised solitary waves $K$ is not zero, in general, 
and it can be determined by the far field condition \cite{Clamond2015a}. For the numerical resolution of 
(\ref{eq:bab1}), we start with initial guesses such that $\eta\to0$ as $x\to\pm\infty$, so $K=0$ initially. 
When oscillatory tails emerge during the iterations, $K$ is no longer zero. In order to compute $K$ dynamically 
at each iteration, one must know the nature (and location) of the tail, that is unknown {\em a priori\/} and thus 
only an {\em a posteriori\/} computation of $K$ is practically reasonable after the algorithm has converged. 
The generalised solitary waves described in \cite{Clamond2015a} have tails with one fundamental frequency, 
and an {\em ad hoc\/} procedure is then used to compute $K$. However, the possibility of more general tails 
has not been ruled out, so a generalised algorithm for computing $K$ cannot be given until the nature of the tail 
is known. Therefore, we compute the numerical solutions of (\ref{eq:bab1}) with $K=0$, that is always 
possible without loss of generality if one releases the mean level condition (this is equivalent to redefining 
the mean water depth).    

With $K = 0$, the equation  (\ref{eq:bab1}) is reformulated as follows. Applying the operator 
$\mathscr{C}^{-1}$ to both sides of \eqref{eq:bab1} and, with $\mathscr{T}=\mathscr{C}^{-1}\,\frac{\ud}{\ud\/\alpha}$, 
we can write
\begin{align}
  \mathscr{F}\{\eta\}\ &\equiv\ \mathscr{L}\{\eta\}\ -\
  \mathscr{N}\{\eta\}\ =\ 0,\label{eq:bab}\\
  \mathscr{L}\{\eta\}\ &\equiv\ B\,\eta\ -\ g\,\mathscr{C}^{-1}\{\eta\}\ +\ \tau\,\mathscr{T}
  \{\eta_\alpha\},\label{eq:babl}\\
  \mathscr{N}\{\eta\}\ &\equiv\ g\,\mathscr{C}^{-1}\!\left\{\,\eta\,\mathscr{C}\{\eta\}\,\right\}\ 
  +\ \half\,g\,\eta^2\ +\ \frac{\tau\,(1+\mathscr{C}\{\eta\})}
  {\sqrt{\/(1\/+\/\mathscr{C}\{\eta\})^2\/+\/\eta_\alpha^{\,2}\/}}\nonumber\\
   &\quad -\ \tau\ -\ \tau\ \mathscr{T}\!\left\{\frac{\eta_\alpha}{\sqrt{\/(1\/+\/
   \mathscr{C}\{\eta\})^2\/+\/\eta_\alpha^{\,2}\/}}\,-\,\eta_\alpha\,\right\},\label{eq:babnl}
\end{align}
where we separated the linear and nonlinear (quadratic, rational) terms. The Fourier symbol of the linear operator \eqref{eq:babl} is
\begin{equation}\label{eq:Fourierl}
  \widehat{\mathscr{L}}(k)\ =\left\{\begin{matrix}\ B\ -\/\,\left(\,g\,k^{-1}\, +\,\tau\,k\,\right)
  \tanh(k\depth)&\quad (k\neq 0), \\ B\,-\,g\/d &\quad (k=0),\end{matrix}\right.
\end{equation}
In \eqref{eq:Fourierl} the dispersion relation for infinitesimal capillary-gravity waves can be recognized \cite{Dias1999, Vanden-Broeck2010}.


\section{The numerical method}\label{sec:numm}

Here, we describe a numerical procedure for discretising and solving the generalised Babenko equation \eqref{eq:bab1}.

\subsection{Pseudospectral discretisation}

For $\Lambda > 0$ sufficiently large, the periodic problem associated to \eqref{eq:bab} on $(-\Lambda, \Lambda)$ 
is discretised with Fourier collocation techniques. For $N > 1$ and on an uniform grid $\alpha_{j}\ =\ -\Lambda+jh$, 
$h\ =\ 2\Lambda/N$, $j\ =\ 1,\ldots,N$, the values $\eta(\alpha_{j})$ of the solution of \eqref{eq:bab} are approximated 
by $\eta_{h}\ =\ (\eta_{h,1},\ldots,\eta_{h,N})^\top$ satisfying
\begin{align}
  \mathscr{F}_{h}\{\eta_{h}\}\ \equiv&\ \mathscr{L}_{h}\{\eta_{h}\}\ -\ \mathscr{N}_{h}\{\eta_{h}\}\ =\ 0,\label{eq:dbab} 
  \\
  \mathscr{L}_{h}\{\eta_{h}\}\ \equiv&\  c^2\,\eta_{h}\ -\ g\, \mathscr{C}_{h}^{-1}\{\eta_{h}\}\ +\ \tau\,\mathscr{T}
  \{D_{\alpha}\eta_{h}\},\label{eq:dbabl}\\
  \mathscr{N}_{h}\{\eta\}\ \equiv&\  g\,\mathscr{C}_{h}^{-1}\!\left\{\,\eta_{h}\,\mathscr{C}_{h}\{\eta_{h}\}\,\right\}\ +\ \half\,g\,(\eta_{h})^2\ +\ \frac{\tau\,(1+\mathscr{C}_{h}\{\eta_{h}\})}{\sqrt{\/(1\/+\/\mathscr{C}_{h}\{\eta_{h}\})^2\/ + \/(D_{\alpha}\eta_{h})^{\,2}\/}}\nonumber\\ 
  &-\ \tau -\ \tau \mathscr{T}_{h}\!\left\{\frac{D_{\alpha}\eta_{h}}{\sqrt{\/(1\/+\/ \mathscr{C}_{h}\{\eta_{h}\})^2\/+\/(D_{\alpha}\eta_{h})^{\,2}\/}}\,-\,D_{\alpha}\eta_{h}\,\right\}.\label{eq:dbabnl}
\end{align}
In \eqref{eq:dbab}--\eqref{eq:dbabnl}, $D_{\alpha}$ stands for the pseudo-spectral differentiation matrix 
\cite{Boyd2000, Canuto1988, Canuto2006, Trefethen2000}, while $\mathscr{C}_{h}$ and $\mathscr{T}_{h}$ are 
discrete versions of the operators $\mathscr{C}$ and $\mathscr{T}$, respectively. They are constructed as 
$D_{\alpha}$ by using the discrete Fourier transform $F_{N}$ on $\mathds{C}^{N}$. (For example, $\mathscr{C}_{h} 
= F_{N}^{-1}W_{N}F_{N}$, where $W_{N}$ is the $N\times N$ diagonal matrix with diagonal entries given by the 
coefficients displayed in \eqref{mathC}. The operator $\mathscr{T}_{h}$ is defined in the same way.) Thus, 
the discrete system \eqref{eq:dbab}--\eqref{eq:dbabnl} is implemented in the Fourier space. Finally, the 
nonlinear terms are computed by using the Hadamard product of vectors in $\mathds{C}^{N}$. 
For these steady computations the use of the anti-aliasing rule was not necessary \cite{Boyd1997a, Boyd2000, Yang2010}. 
A major technical difficulty with nonlocal (generalised) solitary waves is that the phase of tail oscillations 
is sensitive to the position of the boundary. The possible tails are quantised by the computational box and 
since we realise the mean level condition during the computation, our method finds the corresponding solution 
for a given computational domain by making the adjustments in the mean water level.

\subsection{The Levenberg--Maquardt algorithm}

The system \eqref{eq:dbab}--\eqref{eq:dbabnl} was tentatively treated with several techniques. When $\tau = 0$ (pure gravity waves) this was successfully solved using the Petviashvili scheme \cite{Clamond2012b, Dutykh2013b}. Note that when $\tau=0$, the nonlinear term \eqref{eq:babnl} is homogeneous with degree two, while the Fourier symbol \eqref{eq:Fourierl} associated to \eqref{eq:babl} is positive for all $k$ when $B > g\depth$. These two properties are key arguments to explain this success \cite{Pelinovsky2004} (see also \cite{Alvarez2014, Alvarez2015}). However, they are not retained when $\tau\neq 0$ and the various theories about the behaviour of the Petviashvili scheme cannot be applied here.

For the general case $\tau \neq 0$, several variants of the Newton method can emerge as alternatives. The reasons for the failure of the classical implementations of this method \cite{Boyd2007} were taken into account. In our case, the translational invariance of the solitary waves implies that the corresponding Jacobian of $\mathscr{F}_h$ in \eqref{eq:dbab} is rank-deficient (in practice, it is nearly rank-deficient). This recommends to avoid not only the direct application of Newton's method but also the resolution in a least square sense with line search (the Gau\ss--Newton method). Instead of all this, the system \eqref{eq:dbab}--\eqref{eq:dbabnl} was numerically solved in a nonlinear, least square sense but with a trust region formulation. Among the standard techniques presented in the literature (see, e.g., \cite{Lawson1995} and the references therein), the most robust results were obtained employing the so-called \acf{lm} algorithm \cite{Levenberg1944, Marquardt1963}. This is one of the most widely used methods for data-fitting nonlinear problems \cite{Lourakis2005}. For the reader interest, a brief description of it in this context is given now.

Let $r:\ \mathds{R}^{N} \rightarrow \mathds{R}^{N}$ be the residual vector $r(\eta_{h})= (r_{0}(\eta_{h}), \ldots,r_{N-1}(\eta_{h}))^{\texttt{T}}\equiv\mathscr{F}_{h}\{\eta_{h}\}$, where $r_{j}(\eta_{h})$ is the residual for the $j$-th component of \eqref{eq:dbab}, $j=1,\ldots,N$ at any $\eta_{h}\in \mathds{R}^{N}$. The associated least-squares problem consists of minimising $f(\eta_{h})=(1/2)\|r(\eta_{h})\|^{2}$, where $\|\cdot\|$ stands for the Euclidean norm in $\mathds{R}^{N}$. As established in \cite{More1978}, the \acs{lm} algorithm can be formulated as a damped Gau\ss--Newton method but combined with a trust region strategy. The Gau\ss--Newton method is a variant of the Newton method with line search, where the search direction $p^{(\nu)}=p_{\mathrm{GN}}{(\nu)}$, at each Newton step $\nu$, is obtained by solving
\begin{equation}\label{eq:gn}
(J^{(\nu)})^{\texttt{T}}\,J^{(\nu)}\,p^{(\nu)}_{\textrm{GN}}\ =\ -(J^{(\nu)})^{\texttt{T}}\,r^{(\nu)},
\qquad \nu=0,1,2,\ldots,
\end{equation}
where $r^{(\nu)}, J^{(\nu)}$ stand for the residual vector and its Jacobian at the $\nu$-th iteration $\eta_{h}^{(\nu)}$, respectively. (In our case, the Jacobian was approximated with finite differences, leading in fact to a quasi-Newton method.) Instead of (\ref{eq:gn}), the \acs{lm} algorithm computes $p^{(\nu)}=p^{(\nu)}_\mathrm{LM}$ as a minimiser of the model function
\begin{equation}\label{modelf}
  m^{(\nu)}(p)\ =\ \half\,\|r(\eta_{h})\|^{2}\ +\ p^{\texttt{T}}\,(J^{(\nu)})^{\texttt{T}}\, r^{(\nu)}\ +\ \half\,p^{\texttt{T}}\,(J^{(\nu)})^{\texttt{T}}\,J^{(\nu)}\,p,
\end{equation}
subject to the condition $\|p\|\leqslant\Delta^{(\nu)}$, where $\Delta^{(\nu)}>0$ denotes the corresponding radius of the trust region where the problem is constrained to ensure convergence of the algorithm (see \cite{Madsen2004, More1978, Nocedal1999} for details). Any solution $p^{(\nu)}$ is characterised by the existence of a scalar $\lambda^{(\nu)}\geqslant 0$ (the damping parameter) satisfying
\begin{align}
  \left[\,(J^{(\nu)})^{\texttt{T}}\,J^{(k)}\ +\ \lambda^{(\nu)}\,I_{N}\,\right] p^{(\nu)}\ &=\ -\/(J^{(\nu)})^{\texttt{T}}\,r^{(\nu)},\label{cn1}\\
  \lambda^{(\nu)}\left(\Delta^{(\nu)}\, -\,\|p^{(\nu)}\|\right)\,&=\ 0,\label{cn2}
\end{align}
where $I_{N}$ is the $N\times N$ identity matrix. Condition \eqref{cn2} simply states that at least one of the nonnegative quantities, $\lambda^{(\nu)}$ or $\Delta^{(\nu)} - \|p^{(\nu)}\|$, must vanish.)

The literature contains several strategies for the computation of the damping parameter \cite{Marquardt1963, Nielsen1999, Nocedal1999}; one of those proposed in the second reference is standard and was considered here by using the MATLAB Optimization Toolbox \cite{Branch1996}: from an initial $\lambda^{(0)}$, when the step is successful, $\lambda^{(\nu+1)}$ is updated as $\lambda^{(\nu)}/10$; otherwise, $\lambda^{(\nu+1)}$ is set to $10\lambda^{(\nu)}$. Finally, the local convergence of the resulting method is linear \cite{Nocedal1999}, but close to quadratic \cite{Madsen2004}, when the damping parameter is sufficiently small.


\section{Numerical results}\label{sec:numr}

In numerical computations herein below, the parameters given in Table~\ref{tab:params} will be used. The first two concern the model and the rest is associated to the numerical procedure described above. The definition of Froude ($\Fr$) and Bond ($\Bo$) numbers is the one given in \cite{Clamond2015a}. In most of the cases, our implementation of the method will make use of the natural numerical continuation in the Bond number $\Bo$, as a way to improve the performance and to accelerate the convergence. The numerical study is focused on classical, generalised and multi-pulse solitary waves, in several physical regimes.

\begin{table}
  \begin{tabular}{l|c}
    \hline\hline
    \textit{Parameter} & \textit{Value} \\
    \hline
     Initial water depth, $\depth$ [{\sf m}] & $1.0$ \\
     Gravity acceleration, $g$ [$\mathsf{m}\,\mathsf{s}^{-2}$] & $1.0$ \\
     Number of Fourier modes, $N$ & $1024$ \\
     Initial Levenberg--Marquardt parameter, $\lambda$ & $0.05$ \\
     Half of the computational domain, $\ell$, $[-\ell, \ell]$ & $20$ \\
     Tolerance parameter on the increment norm, $\delta$ & $10^{-14}$ \\
  \hline\hline
  \end{tabular}
  \bigskip
  \caption{\small\em Physical and numerical parameters used for the computation of generalised solitary waves by continuation in the Bond number $\Bo$. The tolerance $\delta$ is the stopping criterium in the nonlinear solver.}
  \label{tab:params}
\end{table}

\subsection{Classical solitary wave of depression}

\begin{figure}
  \centering
  \subfigure{\includegraphics[width=0.485\textwidth]{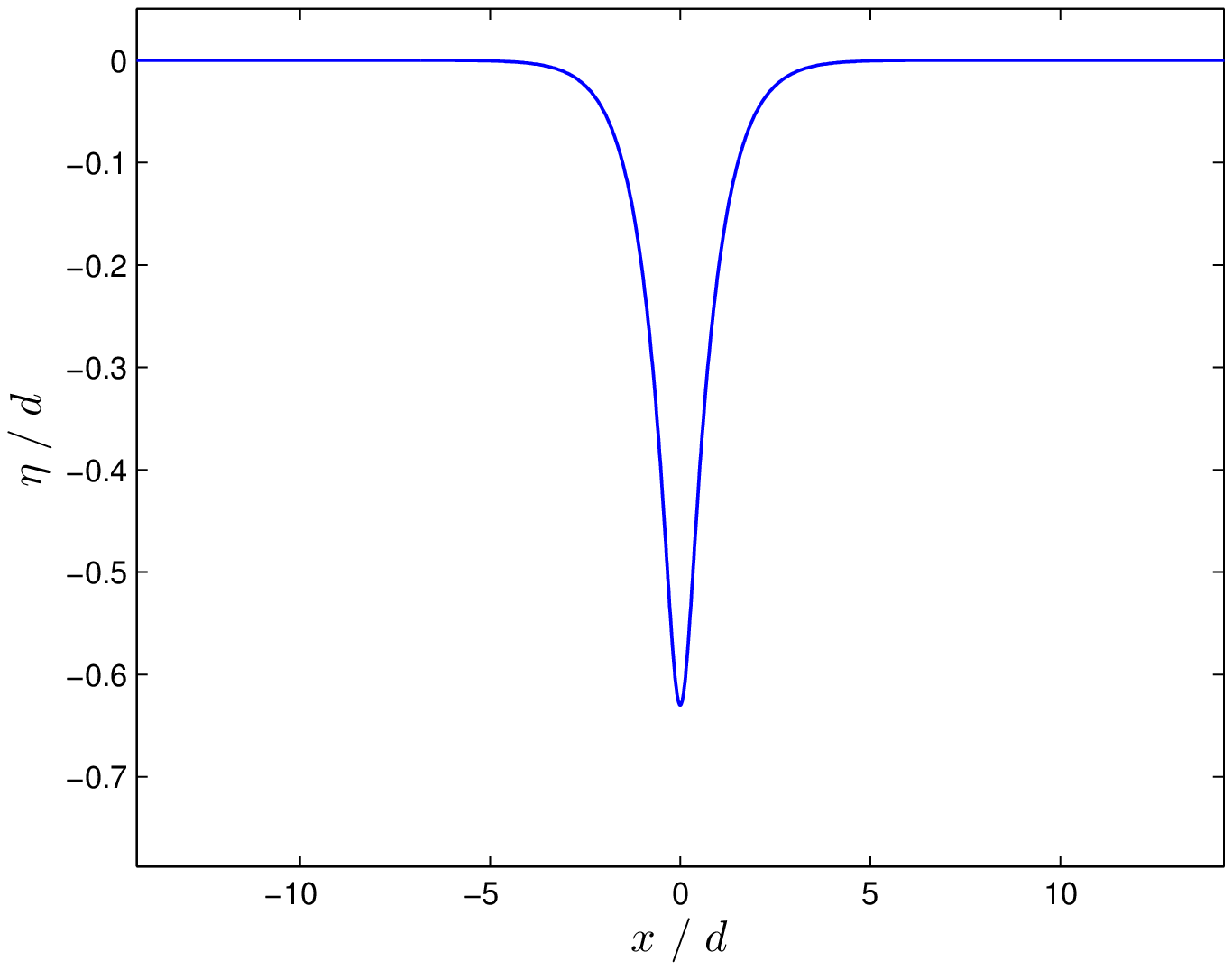}}
  \subfigure{\includegraphics[width=0.485\textwidth]{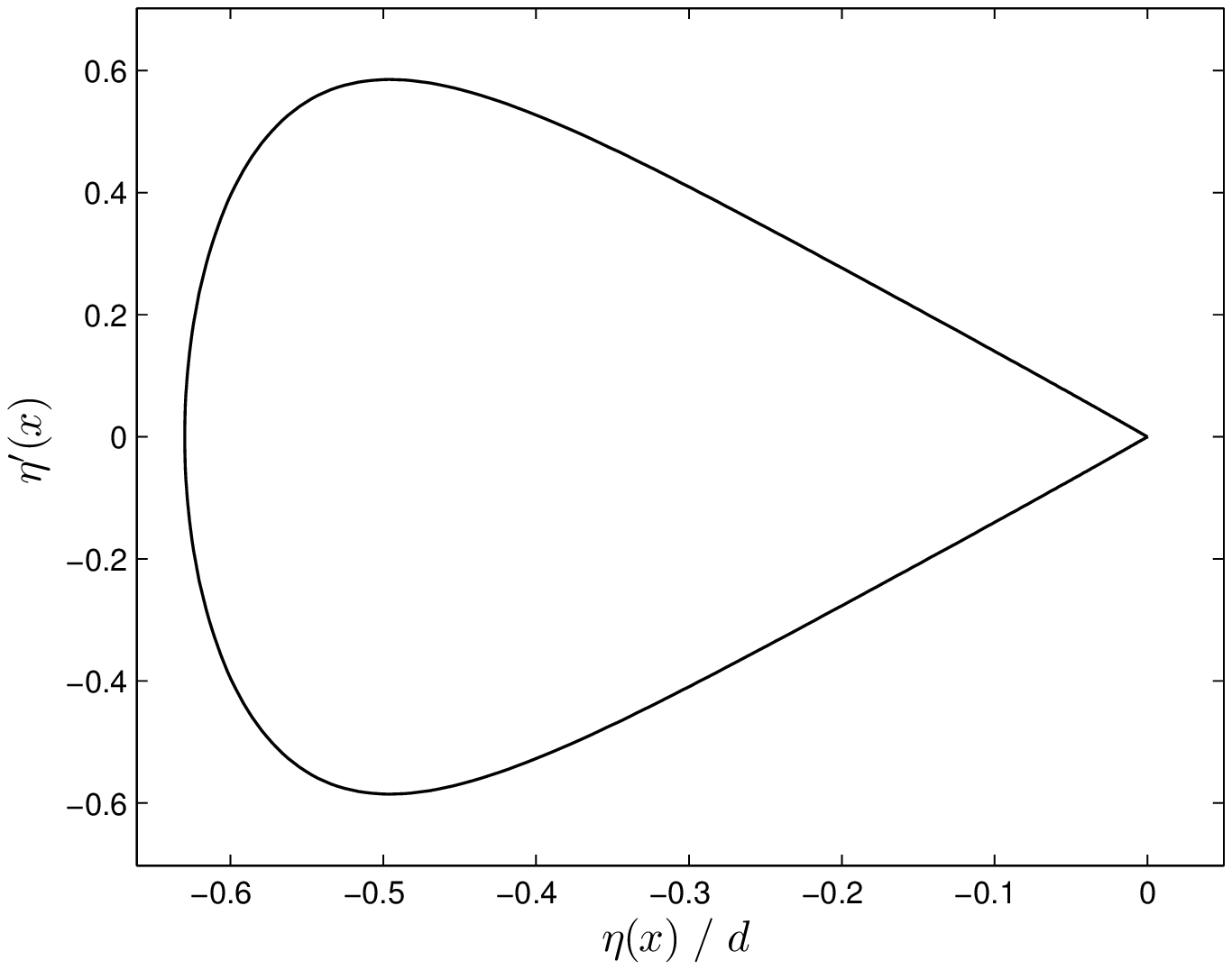}}
  \caption{\small\em Solitary wave of depression for $\Bo = 0.45$ and $\Fr = 0.6$. 
  Left: surface elevation; Right: phase portrait.}
  \label{fig:depr}
\end{figure}

The first group of numerical experiments concerns the generation of classical solitary waves of depression. With this name, we mean aperiodic solutions of \eqref{eq:bab} decaying to zero at infinity (in contrast with the generalised solitary waves, which will be considered later on) and with negative amplitude. 

For supercritical Bond numbers $\Bo>1/3$ and suitable values of the Froude number $\Fr<1$, the existence of isolated solitary waves of depression is known, analytically and numerically (see \cite{Champneys2002} and references therein). A first check of the code is the computation of one of these waves. This is shown in Figure~\ref{fig:depr}, corresponding to the values $\Bo=0.45$ and $\Fr=0.6$. The initial iteration of the procedure was chosen simply as a negative localised bump. There is no need to perform any post-processing, since the solution decays exponentially to zero, i.e., $\eta(x)\to 0$ as $|x|\to \infty$. So, the initial values of $\Fr$ and $\Bo$ parameters correspond perfectly to the fully converged solution.

\begin{figure}
  \centering
  \subfigure{\includegraphics[width=0.485\textwidth]{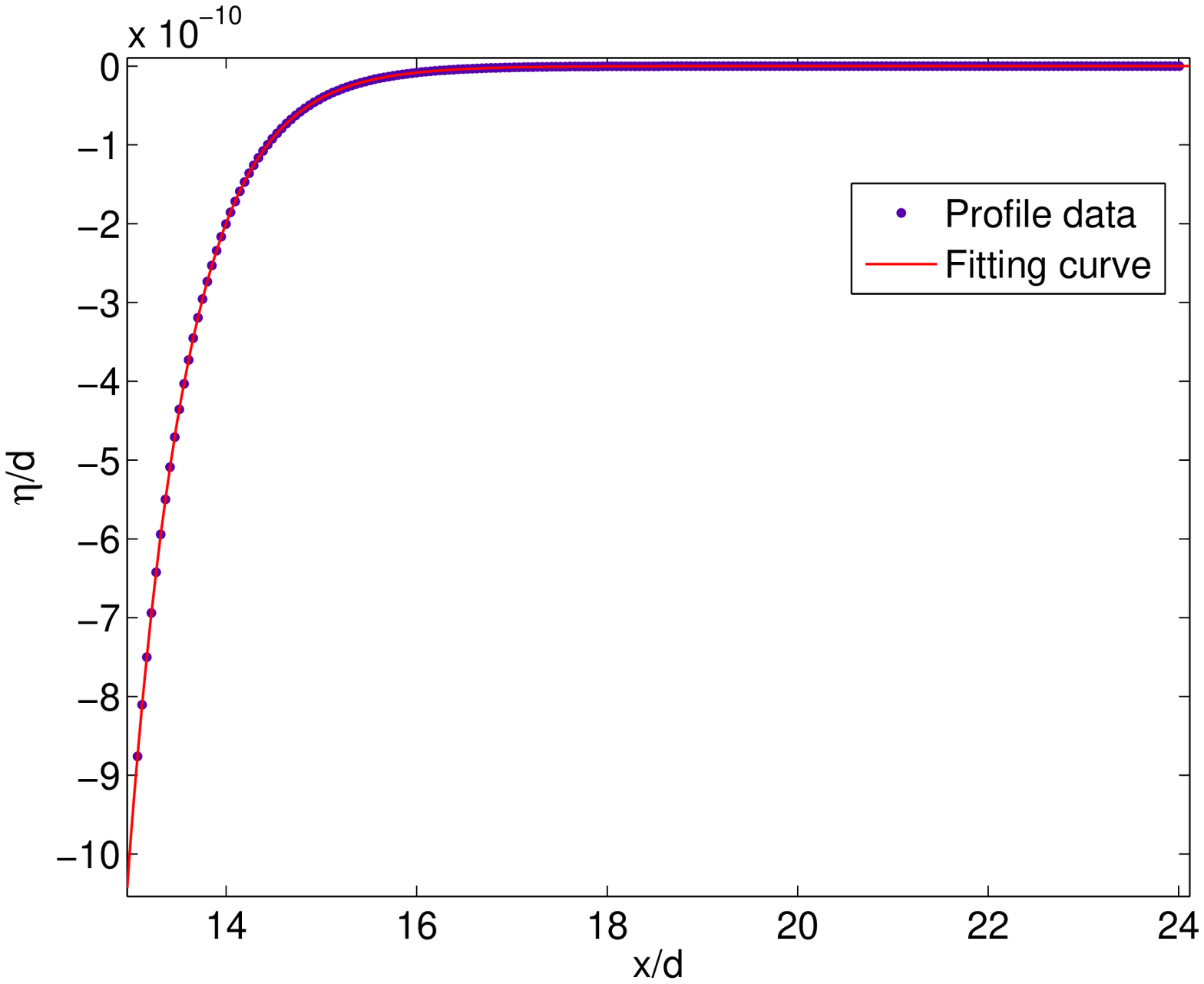}}
  \subfigure{\includegraphics[width=0.485\textwidth]{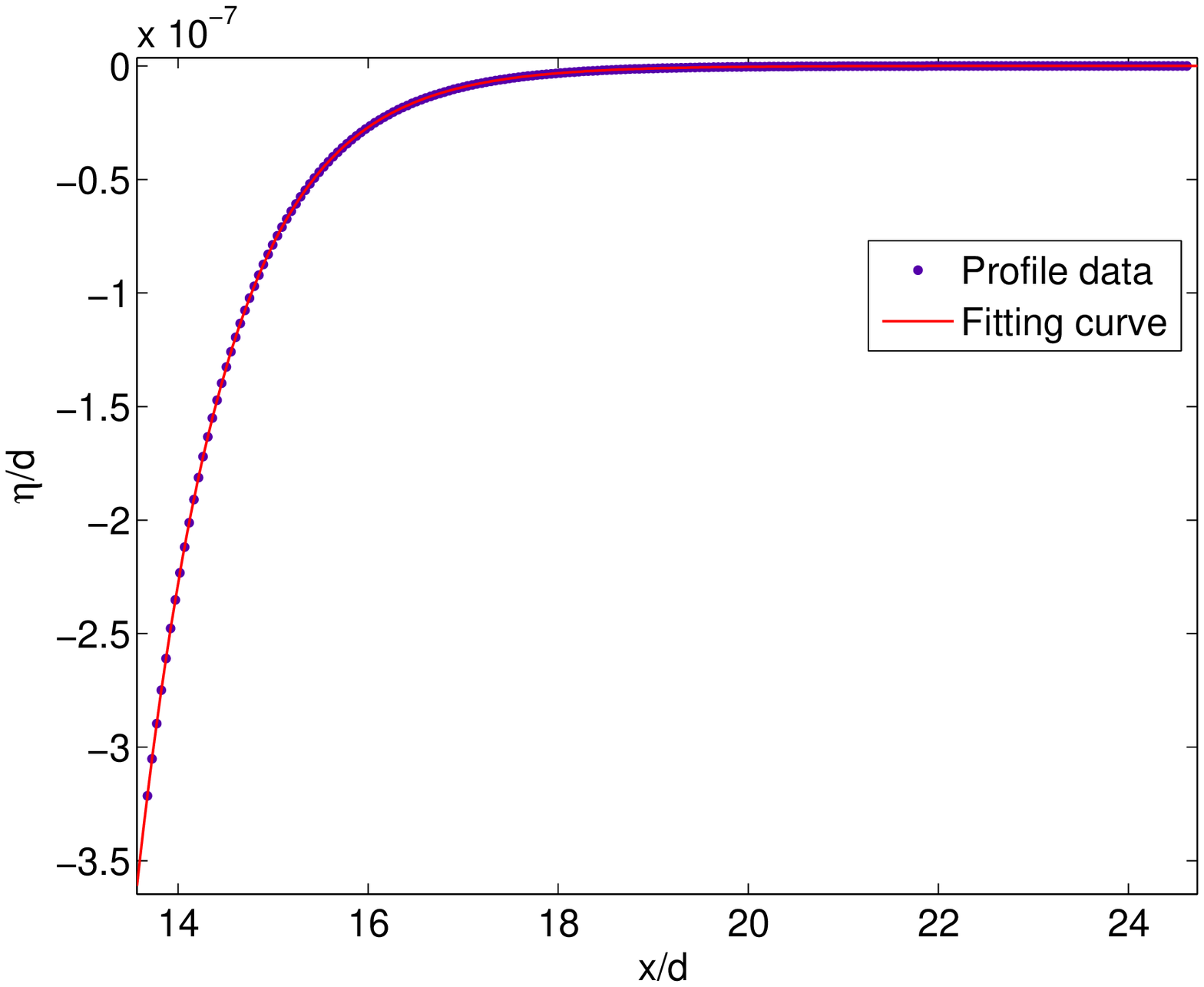}}
  \vspace{-3mm}
  \caption{\small\em Exponential fit of decay of solitary waves. Left: $\Bo = 0.46$ and $\Fr = 0.6$; Right: $\Bo = 0.46$ and $\Fr = 0.9$.}
  \label{fig:depr2}
\end{figure}

The classical character of this computed wave is confirmed by Figure~\ref{fig:depr} (right), which shows that the associated orbit in the phase portrait (computed via the spectral approximation) is homoclinic at infinity to the origin.

The exponential decay has also been numerically checked. To this end and for several profiles, the rightest part of the computed wave was fitted by an exponential function of the form $a\,\ue^{bx}$. By way of illustration, the fit corresponding to the solitary waves for $\Bo=0.46$, $\Fr=0.6$ and $\Bo=0.46$, $\Fr=0.9$ (computed on a longer interval) is shown in  Figure~\ref{fig:depr2}. For the first case, the fitting coefficients with $95\%$ confidence bound (computed with the MATLAB toolbox) are $a = -0.9186$, $b = -1.414$. The goodness of the fit was ensured by a SSE (sum of squares due to error) parameter of $\mathsf{2.024E\!-\!24}$ a R-squared parameter of $1$ and a RMSE (root mean squared error) parameter of $\mathsf{9.526E\!-\!14}$.

\begin{figure}
  \centering
  \subfigure{\includegraphics[width=0.485\textwidth]{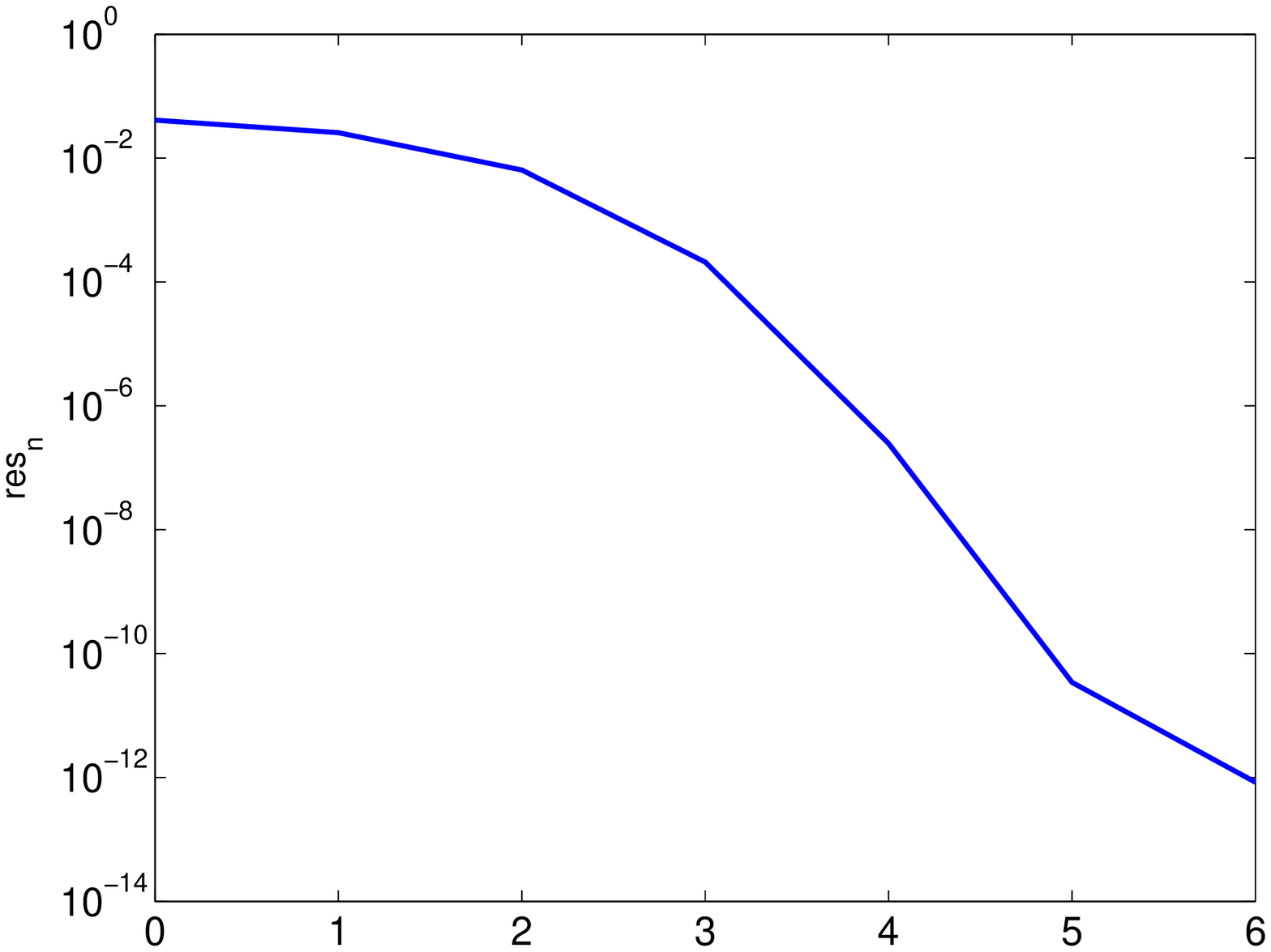}}
  \subfigure{\includegraphics[width=0.485\textwidth]{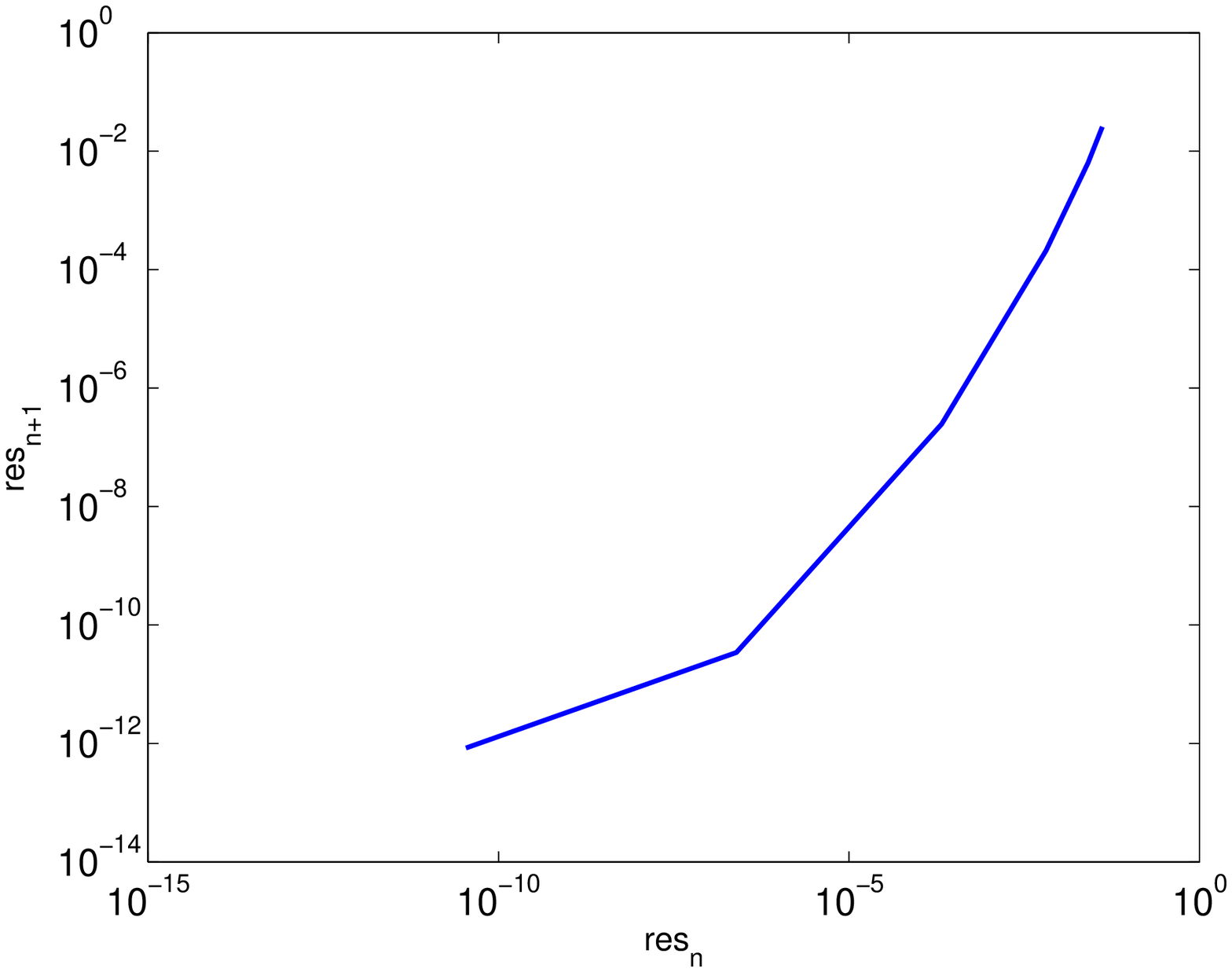}}
  \caption{\small\em Convergence of the \acs{lm} algorithm. Generation of the solitary wave of depression shown in Figure~\ref{fig:depr}. Left: Residual error \eqref{eq:RES} as function of the number of iterations; Right: Relation between two consecutive residual errors.}
  \label{fig:reserror}
\end{figure}

The convergence of the method is illustrated by the following experiments. Figure~\ref{fig:reserror} (left) displays the residual error as function of the iterations for the solitary wave of depression shown in Figure~\ref{fig:depr}. This residual error is measured, at the $n-$th iteration $\eta_{h}^{(n)}$, as
\begin{equation}\label{eq:RES}
  \mathsf{res}_{n}\ =\ \left\|\mathscr{F}_\textrm{h}\left\{\eta_{h}^{(n)}\right\}\right\|^2,
\end{equation}
where the Euclidean norm is considered and $\mathscr{F}_\textrm{h}\{\cdot\}$ is given by \eqref{eq:dbab}. The vertical scale is logarithmic and the results confirm the convergence. On the other hand, Figure~\ref{fig:reserror} (right) shows, in log-log scale, the relation between two consecutive residual errors. This serves as an estimate of the order of convergence. By fitting the logarithmic data to a straight line, the corresponding slope (with $95\%$ confidence bound) is $1.179$. When only errors up to $10^{-7}$ are considered, the corresponding fitting line has a slope of approximately $2.156$. Then, up to this level of errors, Figure~\ref{fig:reserror} (right) suggests a quadratic order of convergence, becoming linear below this error tolerance. Moreover, the residual magnitude is known to be a rather pessimistic estimation of the accuracy in pseudo-spectral methods \cite{Boyd2000}. Consequently, the convergence of the residual \eqref{eq:RES} is a robust diagnostics of the algorithm convergence.

As the Froude and Bond numbers approach the limiting values, the maximum negative excursion of the profiles of depression decreases and the code becomes less accurate. This was observed, for example, in the goodness of fit with an exponential when estimating the decay.

\begin{figure}
  \centering
  \subfigure{\includegraphics[width=0.485\textwidth]{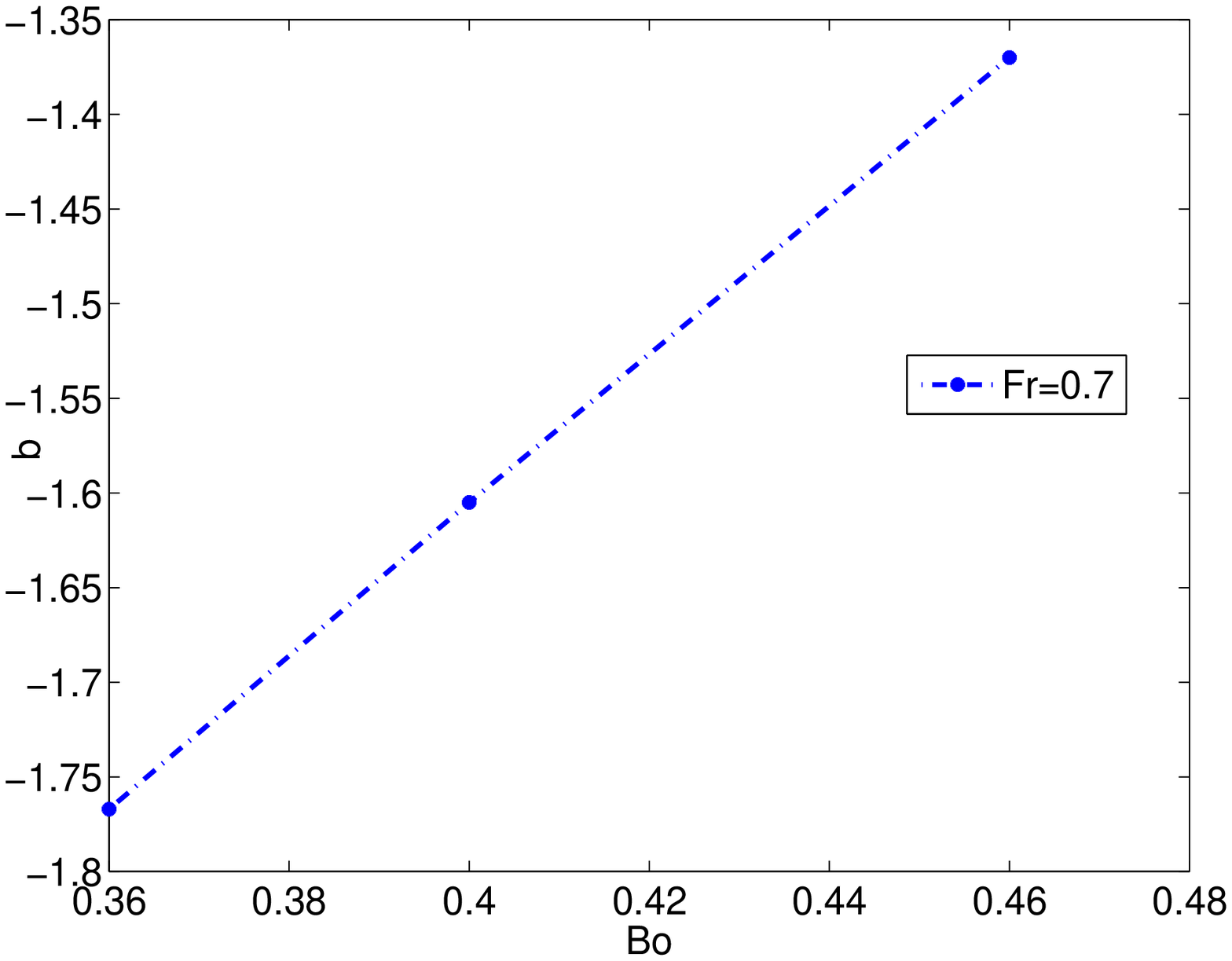}}
  \subfigure{\includegraphics[width=0.485\textwidth]{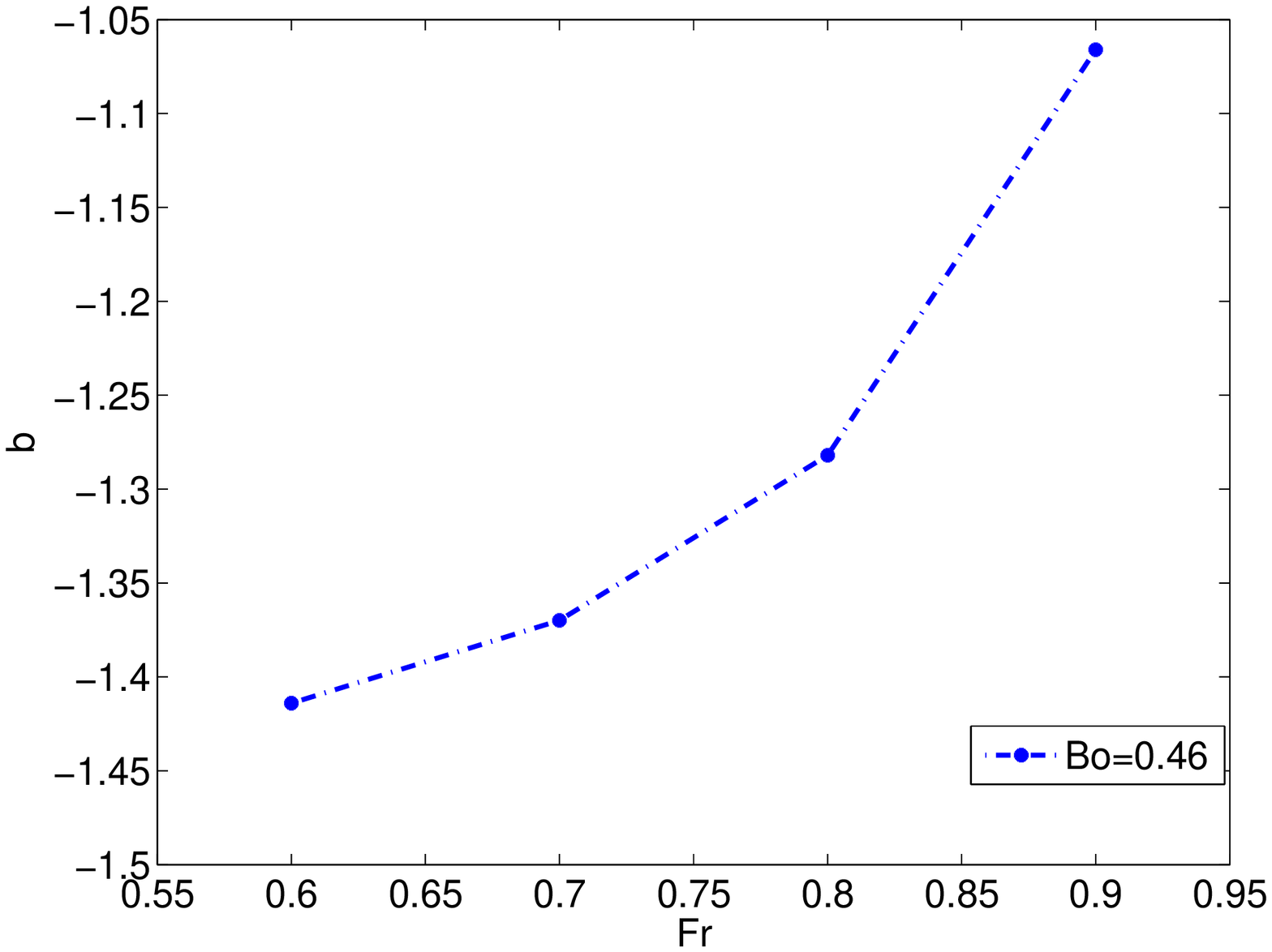}}
  \caption{\small\em Behaviour of the decay as function of $\Bo$ (left) and $\Fr$ (right).}
  \label{fig:decay}
\end{figure}

It should be noted that one may also study, by numerical means, the dependence of the asymptotic decay on the Froude and Bond numbers. This is illustrated in Figure~\ref{fig:decay}. On the left, $\Fr=0.7$ is fixed, the exponent $b$ of the fitting curve for $\Bo=0.36$, $0.4$, $0.46$ is computed and shown as function of $\Bo$. With profiles for more values of $\Bo$, this might give informations about the behaviour of the decay with respect to $\Bo$ for $\Fr$ fixed (and not very close to one). Figure~\ref{fig:decay}, right, displays the behaviour of $b$ as function of $\Fr$ by computing the corresponding values for $\Fr=0.6$, $0.7$, $0.8$, $0.9$ and $\Bo = 0.46$ fixed (and not too close to $1/3$).

\begin{figure}
  \centering
  \subfigure[$\phi_s + \Fr\cdot x$]{
  \includegraphics[width=0.485\textwidth]{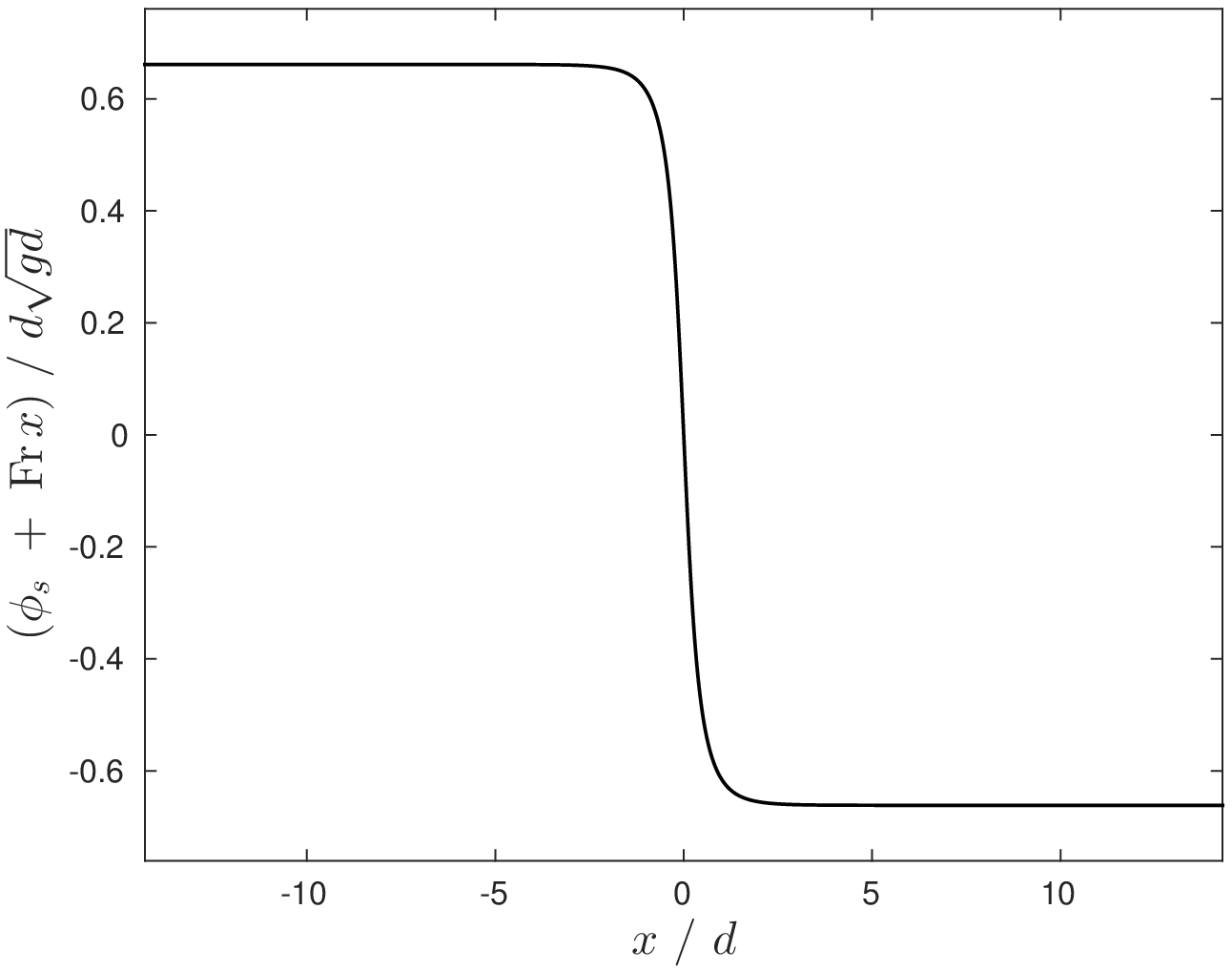}}
  \subfigure[$\psi_s - \psi_b + \Fr\cdot(\eta+d)$]{
  \includegraphics[width=0.485\textwidth]{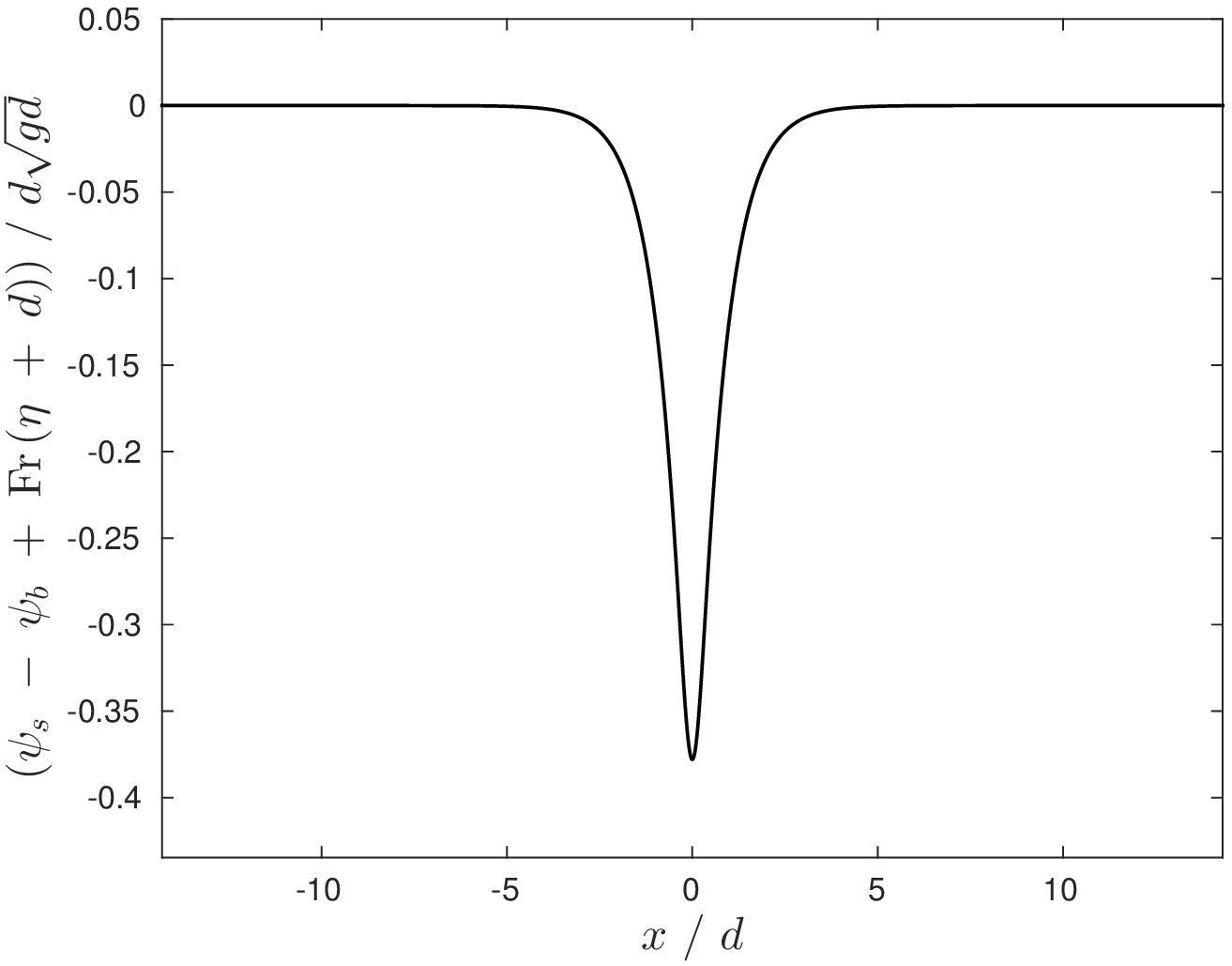}}
  \subfigure[$u_s + \Fr$]{
  \includegraphics[width=0.485\textwidth]{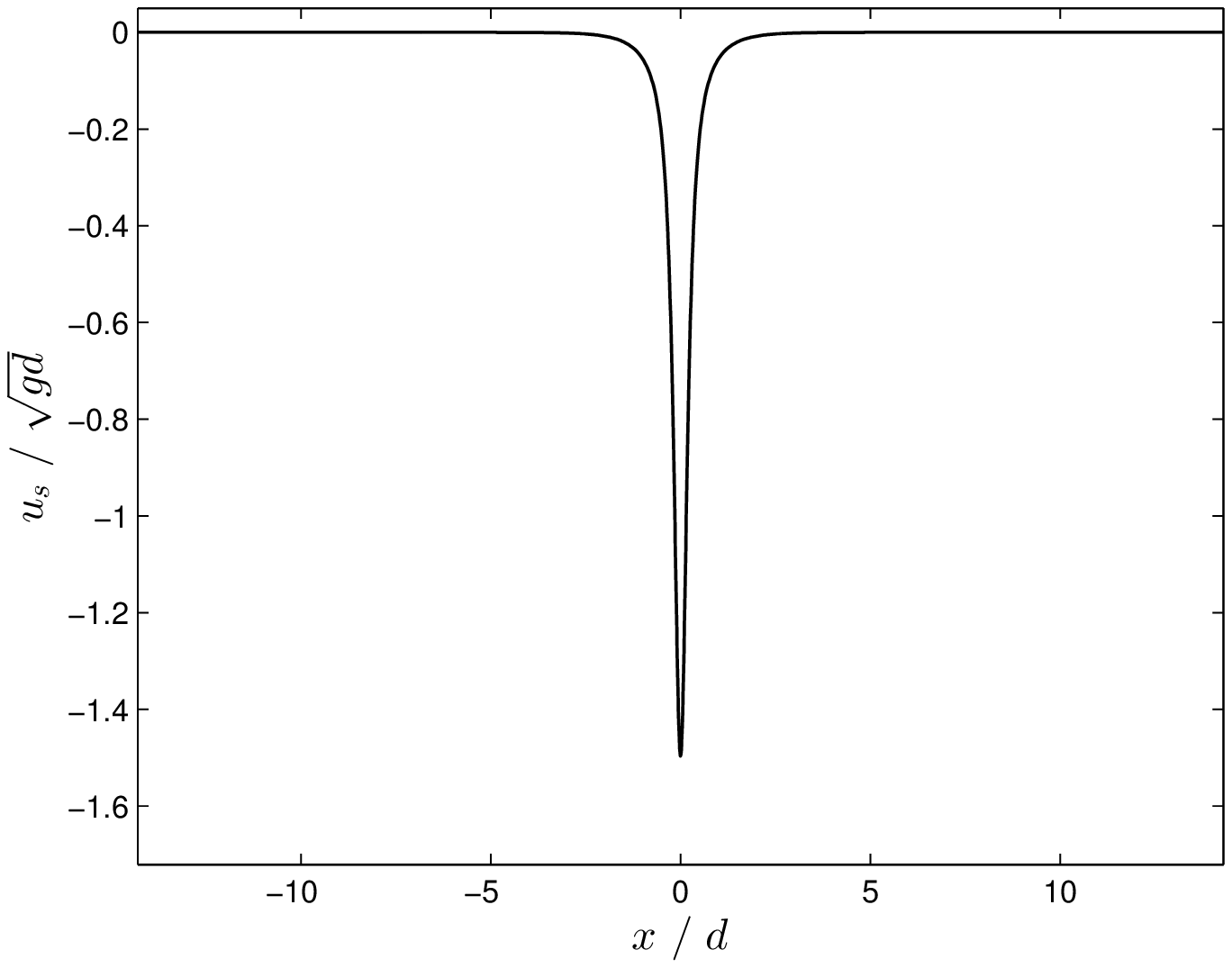}}
  \subfigure[$v_s(x)$]{
  \includegraphics[width=0.485\textwidth]{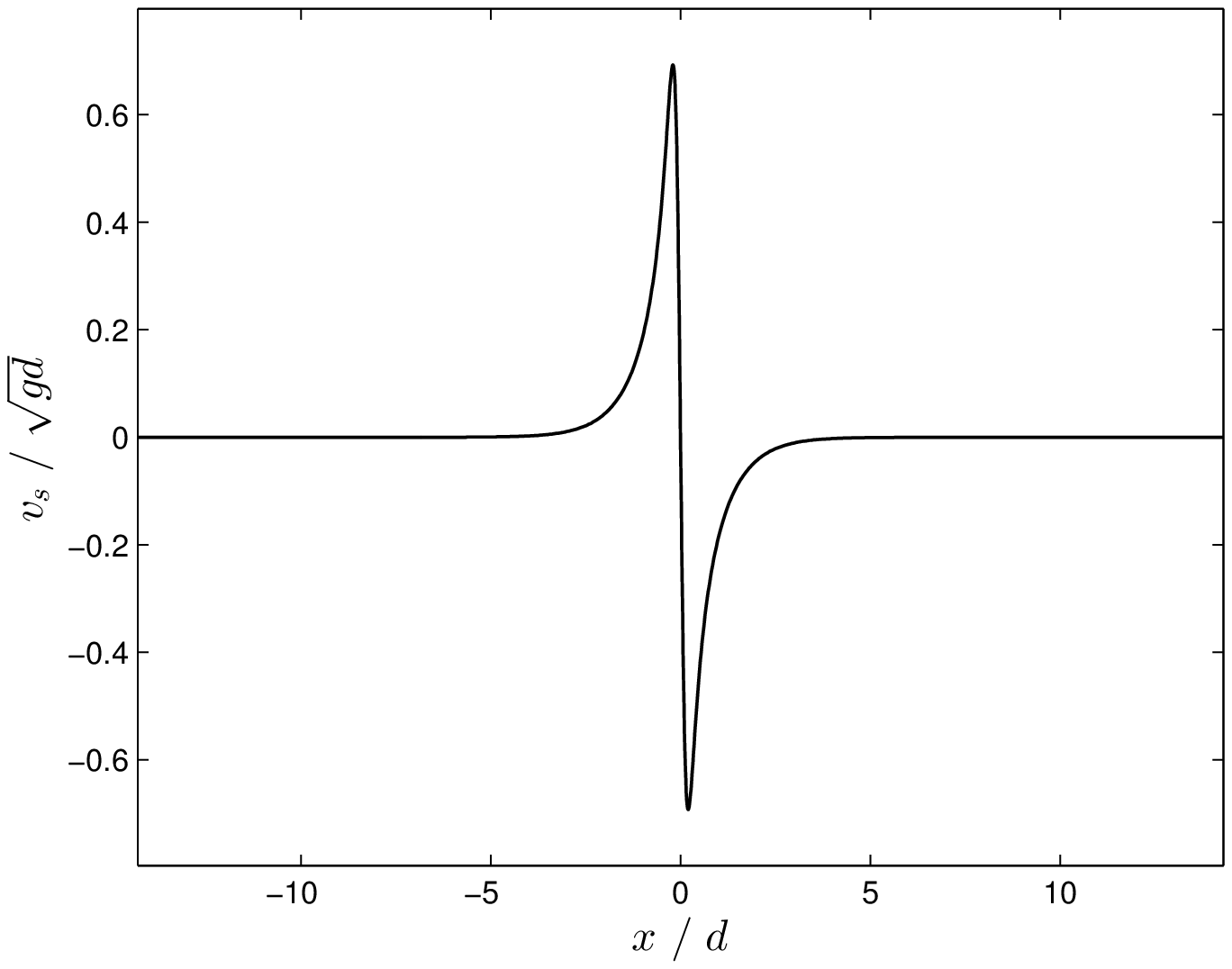}}
  \caption{\small\em Velocity potential (a), stream function (b), horizontal (c) and vertical (d) velocities at the free surface for the solitary wave of depression represented on Figure~\ref{fig:depr}.}
  \label{fig:deprsurf}
\end{figure}

\begin{figure}
  \centering
  \subfigure[$\phi + \Fr\cdot x$]{
  \includegraphics[width=0.485\textwidth]{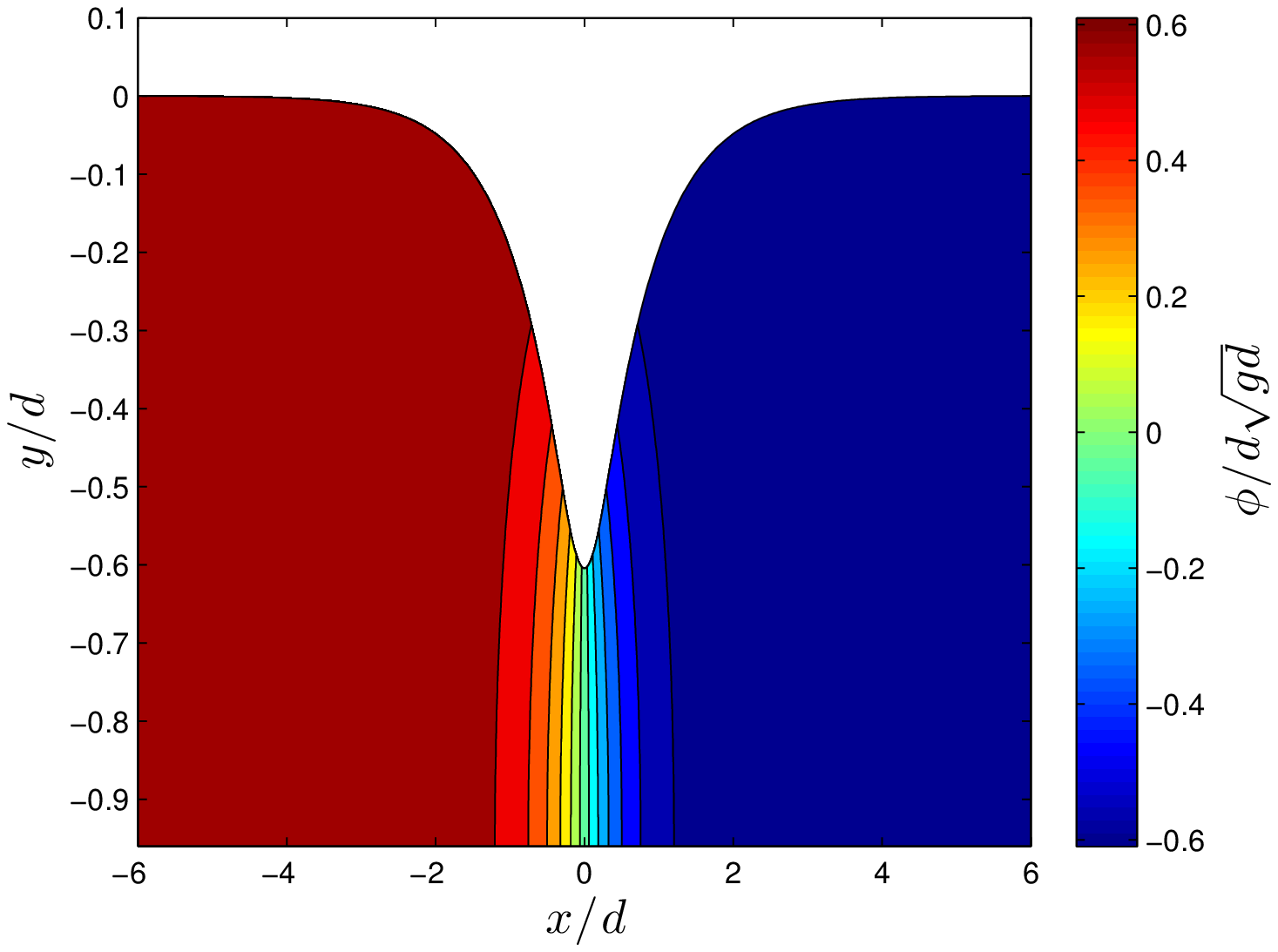}}
  \subfigure[$\psi - \bot{\psi} + \Fr\cdot(y + d)$]{
  \includegraphics[width=0.485\textwidth]{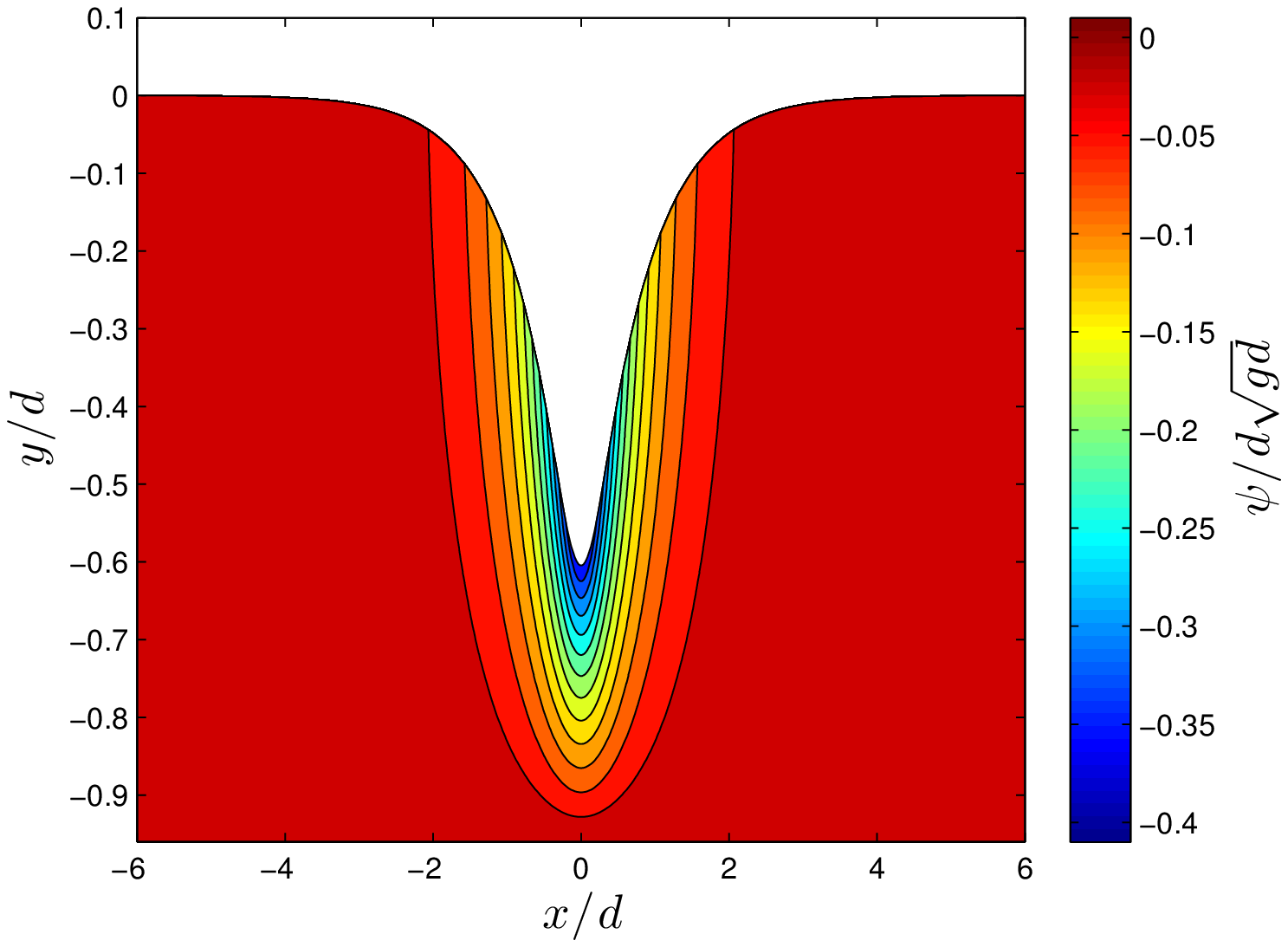}}
  \caption{\small\em The velocity potential (left) and the stream function (right) in the fluid domain under a solitary wave of depression which is represented on 
  Figure~\ref{fig:depr}.} \label{fig:depr_pot}
\end{figure}

\begin{figure}
  \centering
  \subfigure[$P$]{
  \includegraphics[width=0.485\textwidth]{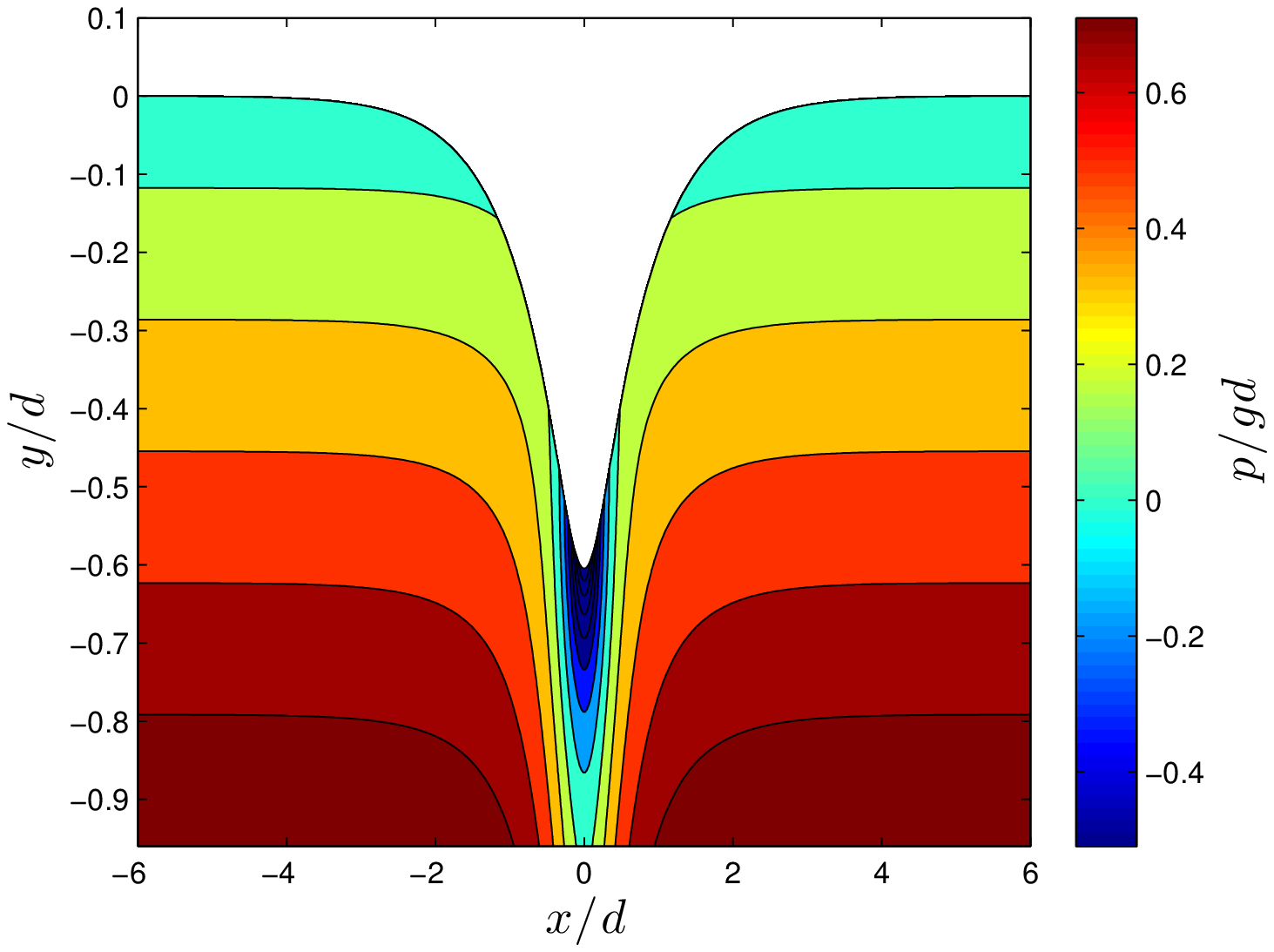}}
  \subfigure[$P+gy$]{
  \includegraphics[width=0.485\textwidth]{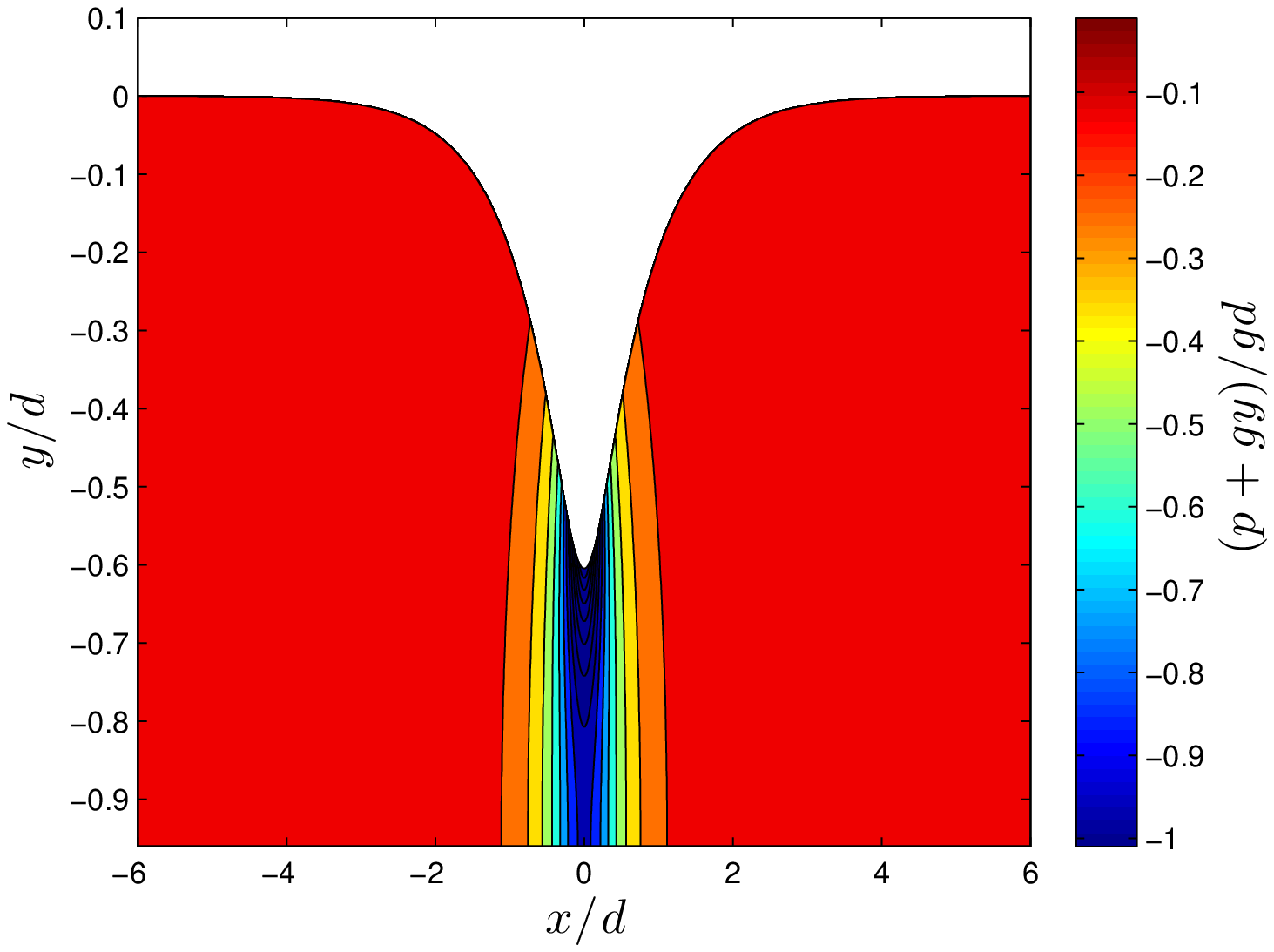}}
  \caption{\small\em Total (left) and dynamic (right) pressures in the fluid domain under a solitary wave of depression which is represented on Figure~\ref{fig:depr}.}
  \label{fig:depr_press}
\end{figure}

\begin{figure}
  \centering
  \subfigure[$u + \Fr$]{
  \includegraphics[width=0.485\textwidth]{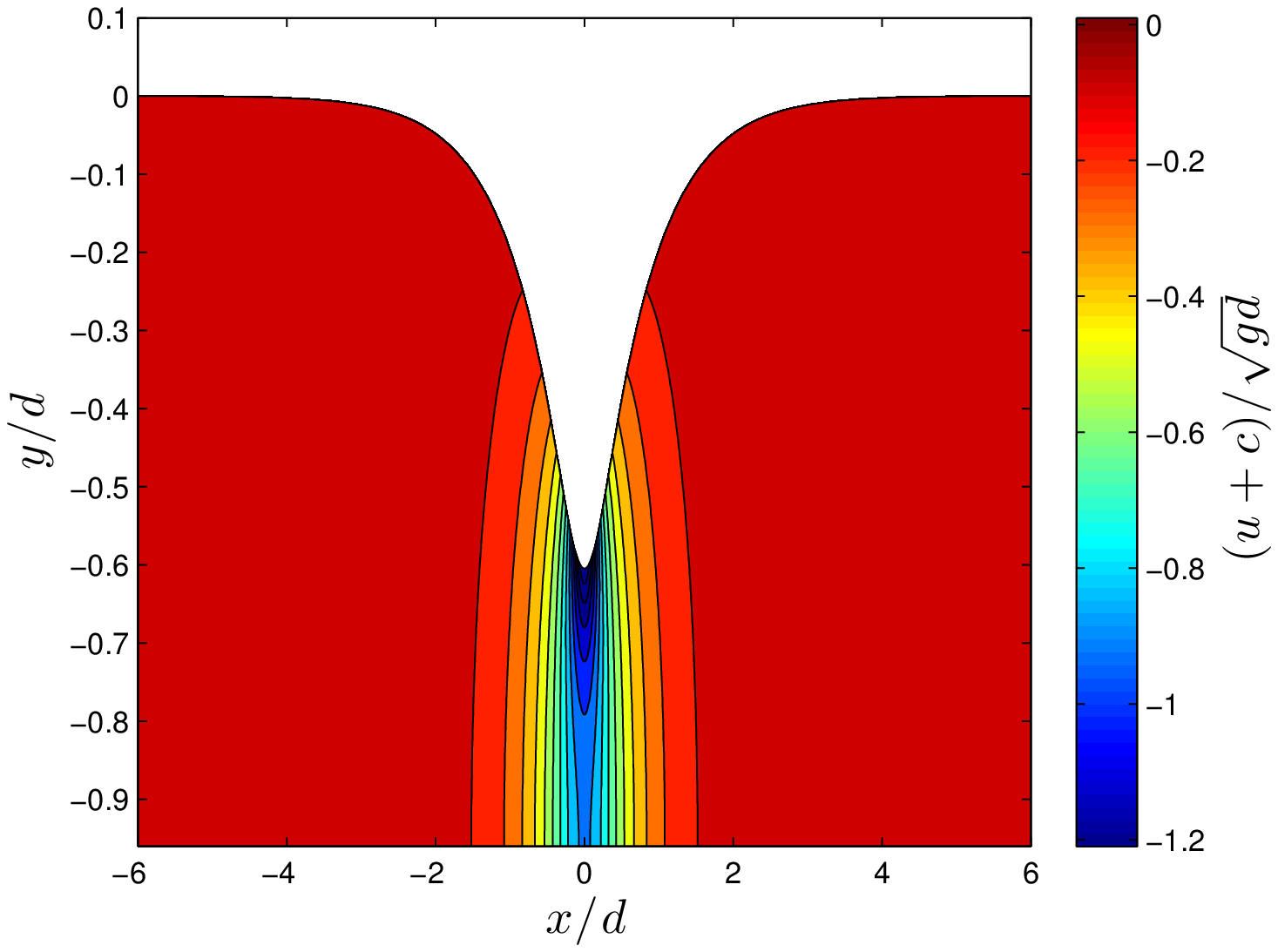}}
  \subfigure[$v$]{
  \includegraphics[width=0.485\textwidth]{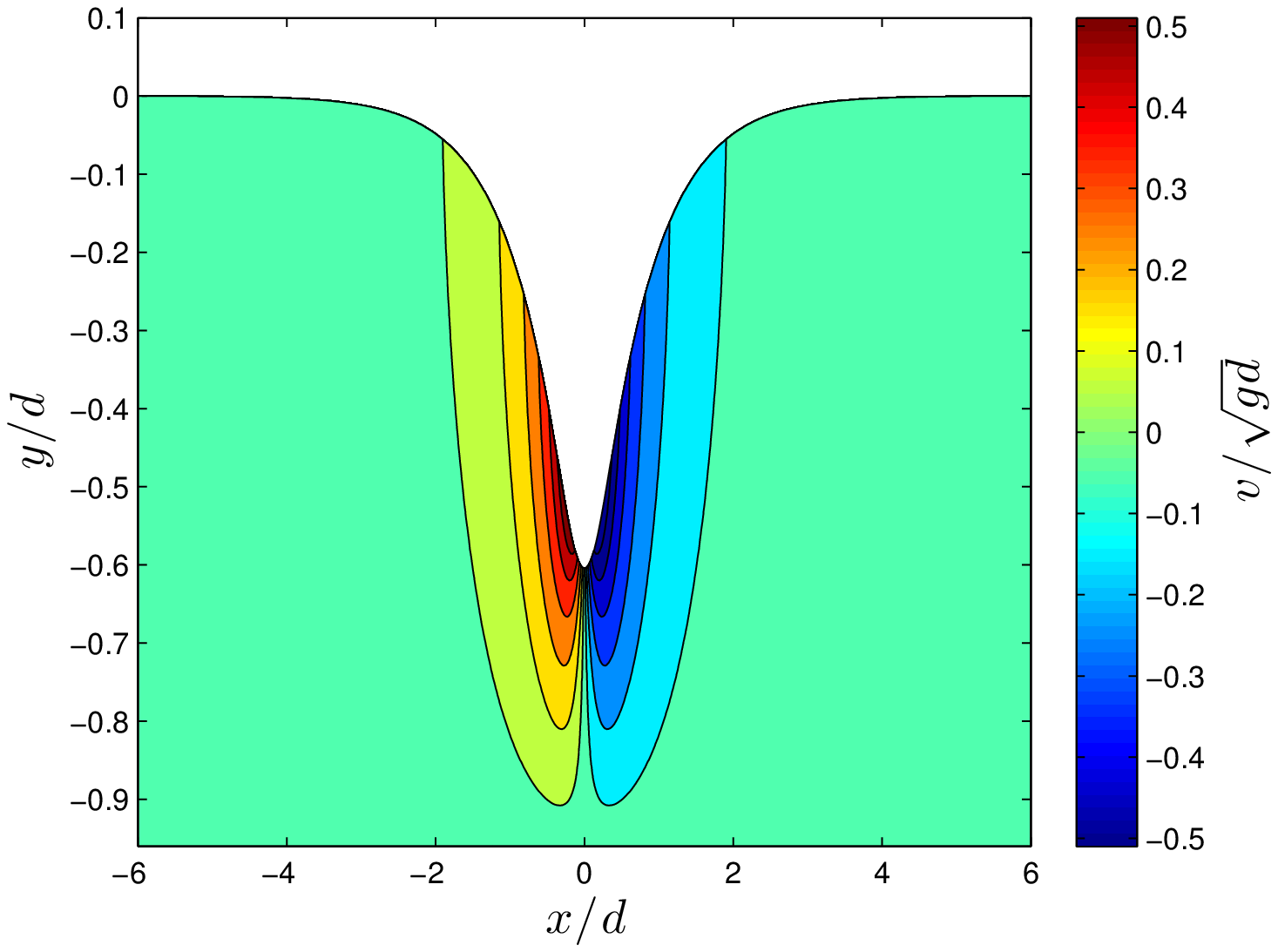}}
  \caption{\small\em Horizontal (left) and vertical (right) velocities distribution in the fluid domain under a solitary wave of depression which is represented on Figure~\ref{fig:depr}.}
  \label{fig:depr_speed}
\end{figure}

\begin{figure}
  \centering
  \subfigure[$a_x$]{
  \includegraphics[width=0.485\textwidth]{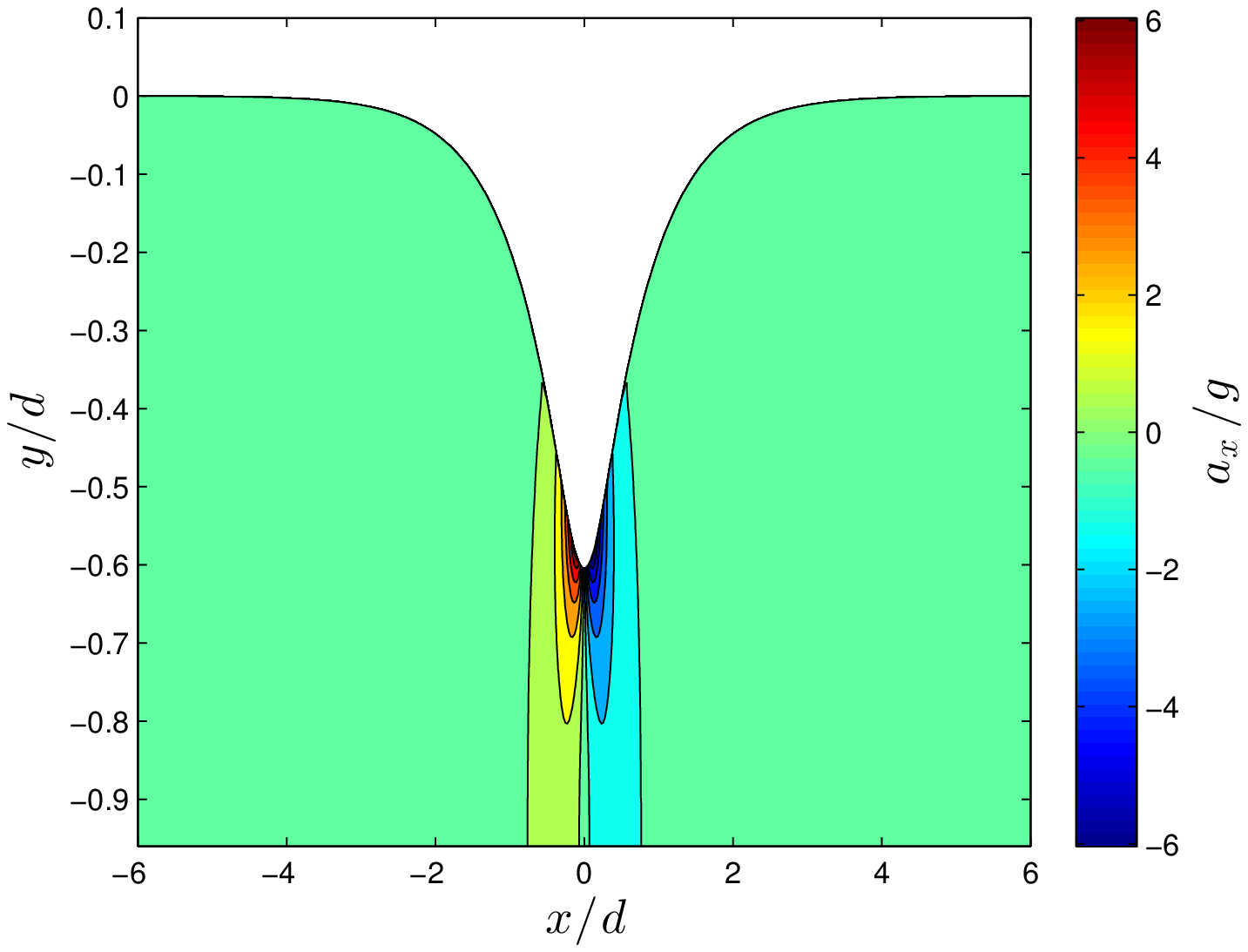}}
  \subfigure[$a_y$]{
  \includegraphics[width=0.485\textwidth]{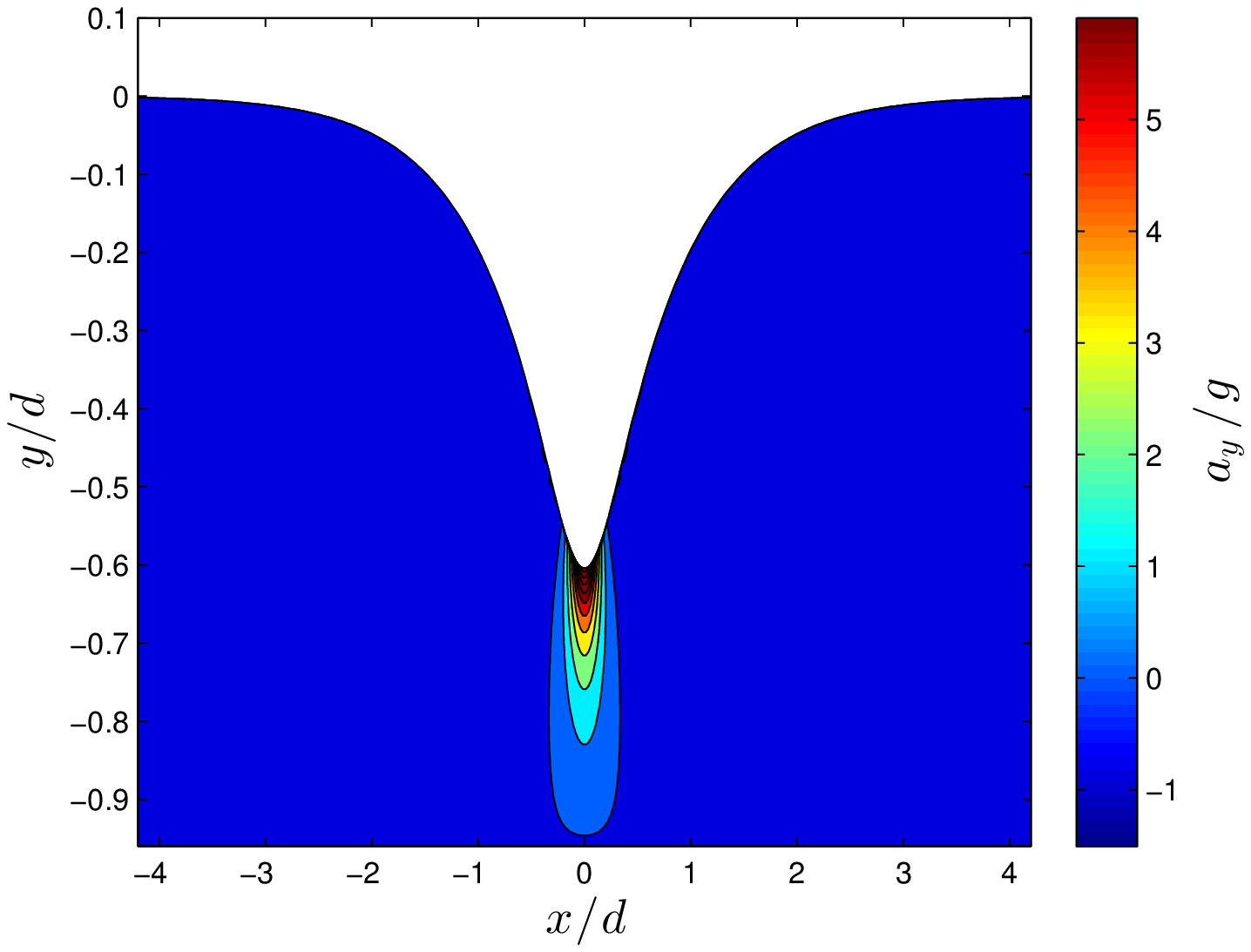}}
  \caption{\small\em Horizontal (left) and vertical (right) accelerations distribution in the fluid domain under a solitary wave of depression which is represented on Figure~\ref{fig:depr}.}
  \label{fig:depr_accel}
\end{figure}

This subsection ends by visualising the internal hydrodynamics under classical solitary waves. This is illustrated by considering the computed wave of Figure~\ref{fig:depr}. Thus, the velocity potential and the stream function are depicted in Figure~\ref{fig:depr_pot}, the total and dynamic pressures are represented in Figures~\ref{fig:depr_press}, while horizontal and vertical speed and acceleration distribution are displayed in Figures~\ref{fig:depr_speed} and \ref{fig:depr_accel}, respectively. All are implemented by using the Cauchy integral formula, which allows to compute all physical fields of interest in the bulk of the fluid, \cite{Dutykh2013b}. The traces of the velocity potential, stream function, horizontal and vertical velocities are depicted in Figure~\ref{fig:deprsurf}.

\subsubsection*{Multi-modal classical solitary waves}

Another type of solutions, which can be computed by our code, consists of multi-modal solitary waves. The existence of these solutions in the KdV5 equation was shown analytically and numerically in \cite{Champneys1997}. We refer to \cite{Clamond2015a} for more details on these solutions in case of the full Euler equations.


\subsection{Generalised solitary waves}

\begin{table}
  \begin{tabular}{|c|c||c|c|}
    \hline\hline
    {\textit{Coefficients}} &{\textit{g.o.f.}}&{\textit{Coefficients}} &{\textit{g.o.f.}}  \\
    \hline
      $\omega=7.938$&SSE$=1.075\times 10^{-6}$& $\alpha_{1}=0.02517$ &SSE$=4.885\times 10^{-6}$\\
      $a_{0}=0.006348$&R-squared$=1$& $\beta_{1}=7.938$ &R-squared$=0.9999$\\
      $a_{1}=-0.01396$&RMSE$=7.005\times 10^{-5}$& $\gamma_{1}=-0.5795$ &RMSE$=0.0001536$\\  
      $b_{1}=0.02097$&& $\alpha_{2}=0.01307$ &\\    
      $a_{2}=0.0005767$&& $\beta_{2}=0.2713$ &\\   
      $b_{2}=0.001378$&& $\gamma_{2}=3.506$ &\\ 
      && $\alpha_{3}=0.0067$ &\\     
      && $\beta_{3}=0.3922$ &\\     
      && $\gamma_{3}=4.712$ &\\ 
      && $\alpha_{4}=4.093\times 10^{-5}$ &\\     
      && $\beta_{4}=8.565$ &\\     
      && $\gamma_{4}=1.909$ &\\  
      && $\alpha_{5}=2.585\times 10^{-7}$ &\\     
      && $\beta_{5}=7.117$ &\\     
      && $\gamma_{5}=3.197$ &\\  
      && $\alpha_{6}=0.001485$ &\\     
      && $\beta_{6}=15.88$ &\\     
      && $\gamma_{6}=0.3004$ &\\   
    \hline\hline
  \end{tabular}
  \bigskip
  \caption{\small\em Coefficients and goodness of fit for the fitting curves of Figure~\ref{fig:osc_fit1}: $f(x) = a_{0} + a_{1}\cos(\omega x) + b_{1}\sin(\omega x) + a_{2}\cos(2\omega x) + b_{2}\sin(2\omega x)$; $f(x) = \sum_{j=1}^{6}\alpha_{j}\sin(\beta_{j}x + \gamma_{j})$.}
  \label{tab:fit}
\end{table}

The question of existence of classical solitary waves of elevation is not solved, to our knowledge, although some references suggest a negative answer \cite{Champneys2002}. For $\Bo<1/3$ and $\Fr>1$, what is known is the existence of generalised solitary waves, \ie solitary waves that are homoclinic to exponentially small amplitude oscillatory waves. Furthermore, the amplitude of the corresponding asymptotic oscillations are of order less than
\begin{equation}\label{eq:order}
  \exp\left(-\/\omega\,s\,\sqrt{\,\frac{1/3\,-\,\Bo}{|\,\Fr^{-2}\,-\,1\,|}\,}\,\right),
\end{equation}
for some $s\in (0,\pi)$ and where $\omega\neq 0$ satisfies the equation $\omega\cosh(\omega) = (1 + \omega^{2}\Bo)\sinh(\omega)$ \cite{Lombardi2000}.

\begin{figure}
  \centering
  \subfigure[$\Bo = 0.08$, $\Fr_\infty = 1.14993$, $\Bo_\infty = 0.079991$]{
  \includegraphics[width=0.485\textwidth]{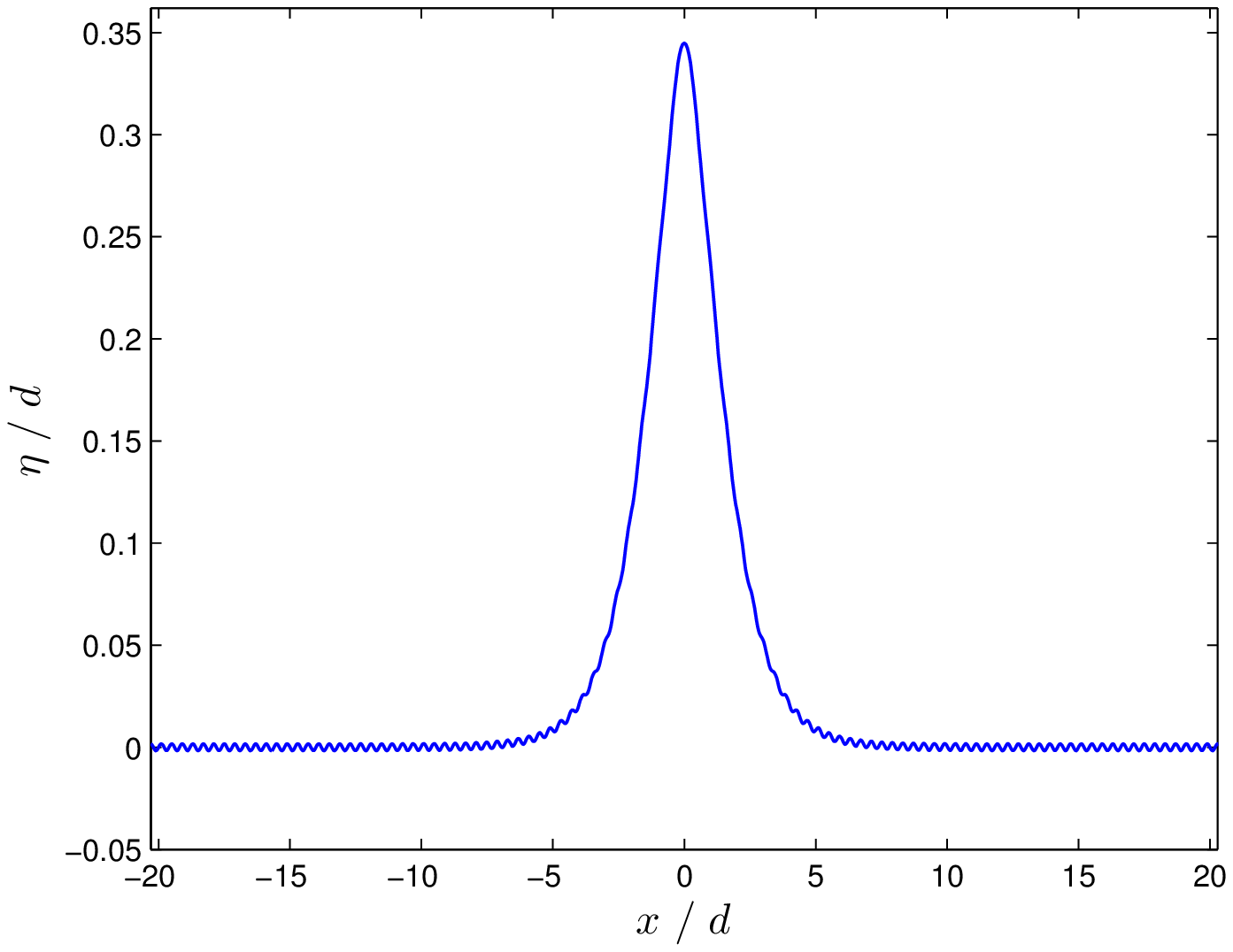}}
  \subfigure[$\Bo = 0.1$, $\Fr_\infty = 1.15894$, $\Bo_\infty = 0.099797$]{
  \includegraphics[width=0.485\textwidth]{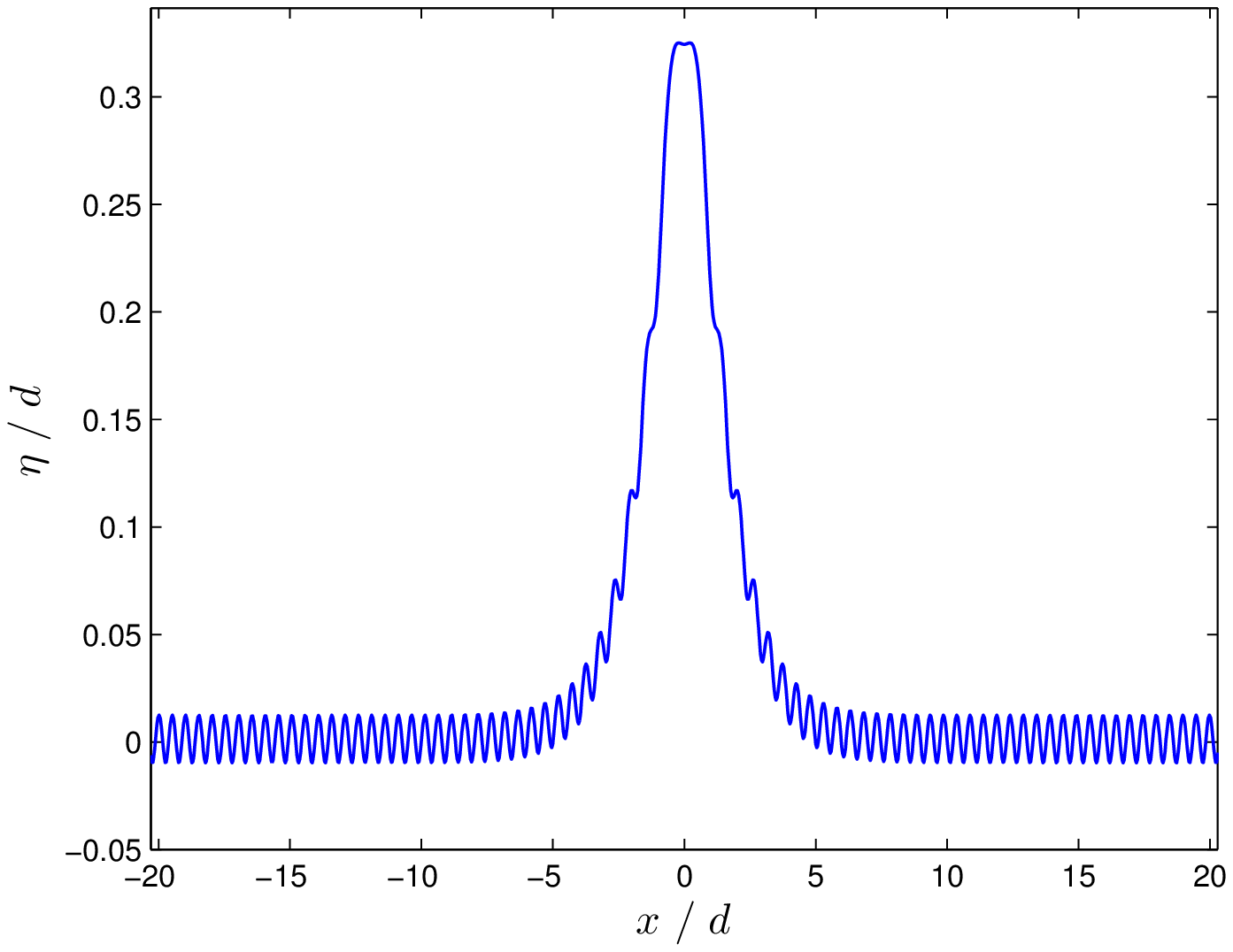}}
  \subfigure[$\Bo = 0.12$, $\Fr_\infty = 1.16376$, $\Bo_\infty = 0.11926$]{
  \includegraphics[width=0.485\textwidth]{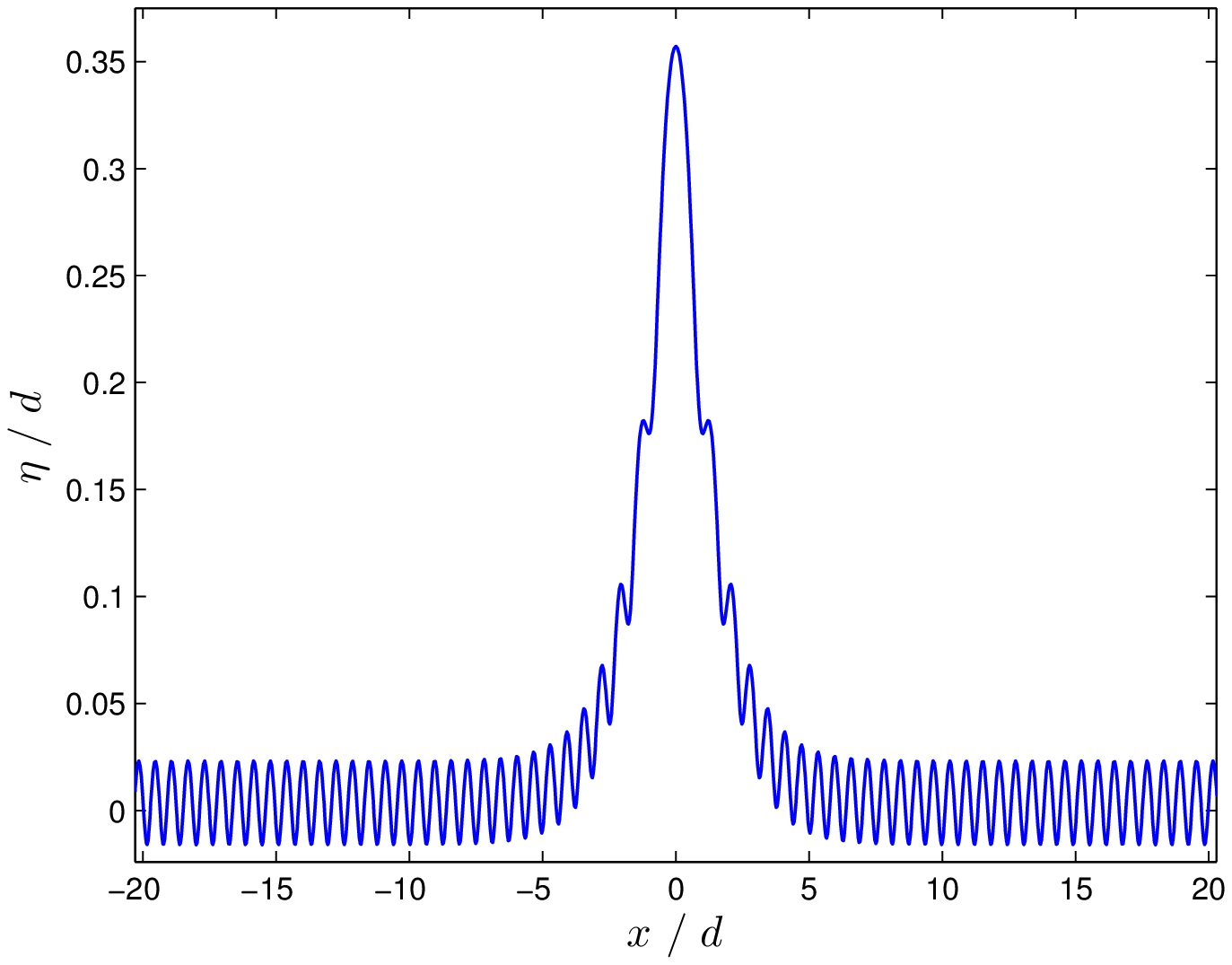}}
  \subfigure[$\Bo = 0.15$, $\Fr_\infty = 1.161479$, $\Bo_\infty = 0.14876$]{
  \includegraphics[width=0.485\textwidth]{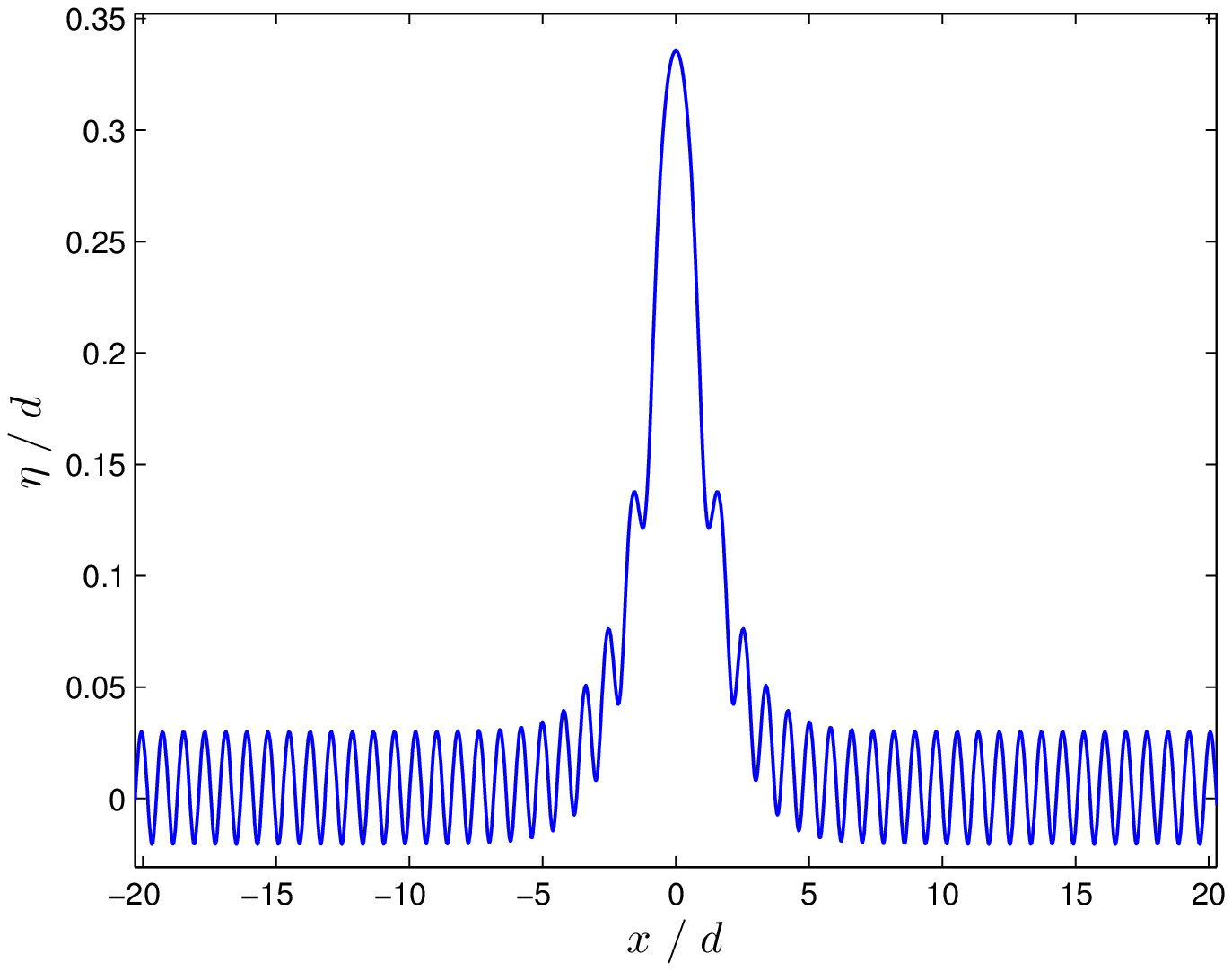}}
  \subfigure[$\Bo = 0.19$, $\Fr_\infty = 1.15994$, $\Bo_\infty = 0.18768$]{
  \includegraphics[width=0.485\textwidth]{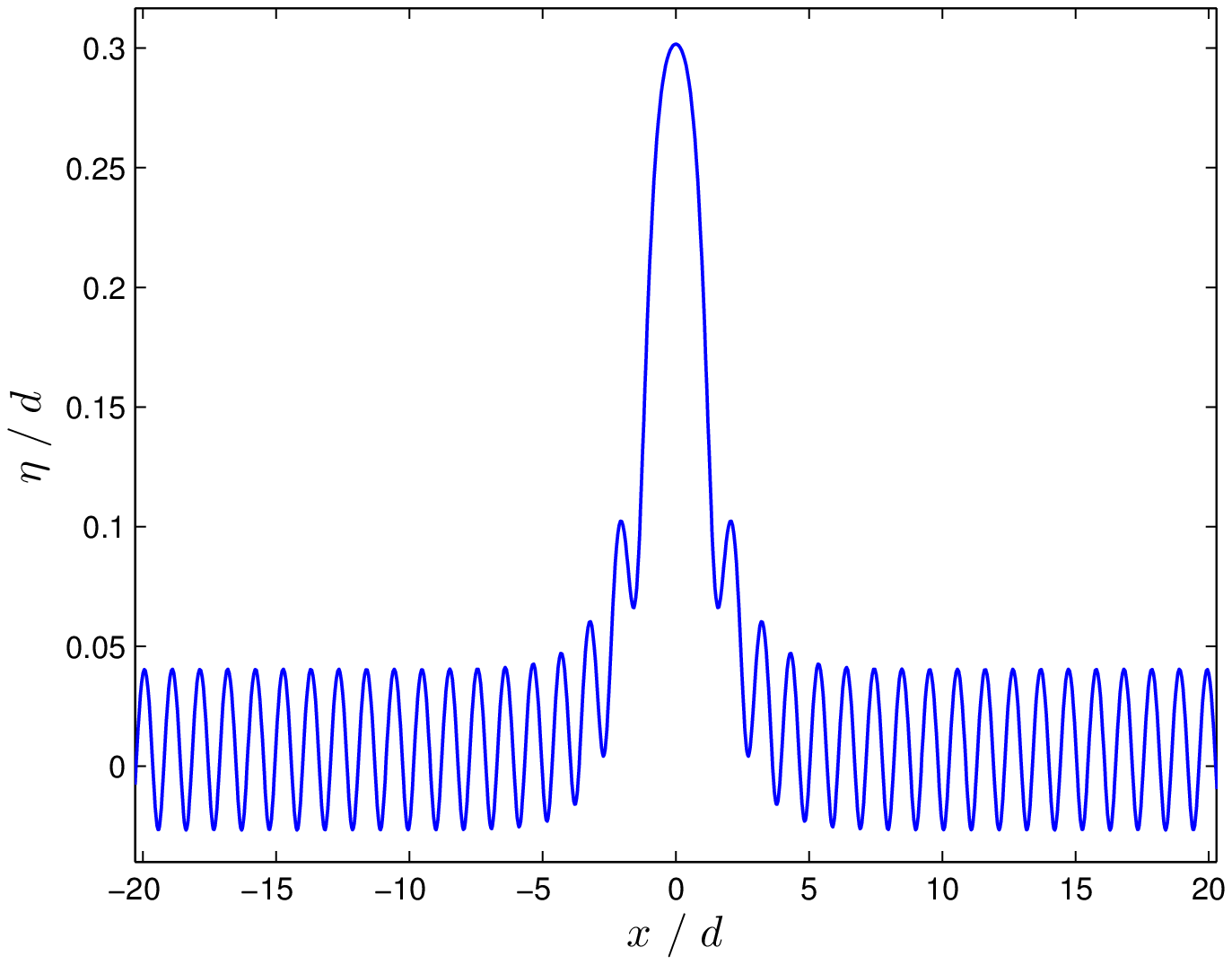}}
  \subfigure[$\Bo = 0.22$, $\Fr_\infty = 1.164997$, $\Bo_\infty = 0.216534$]{
  \includegraphics[width=0.485\textwidth]{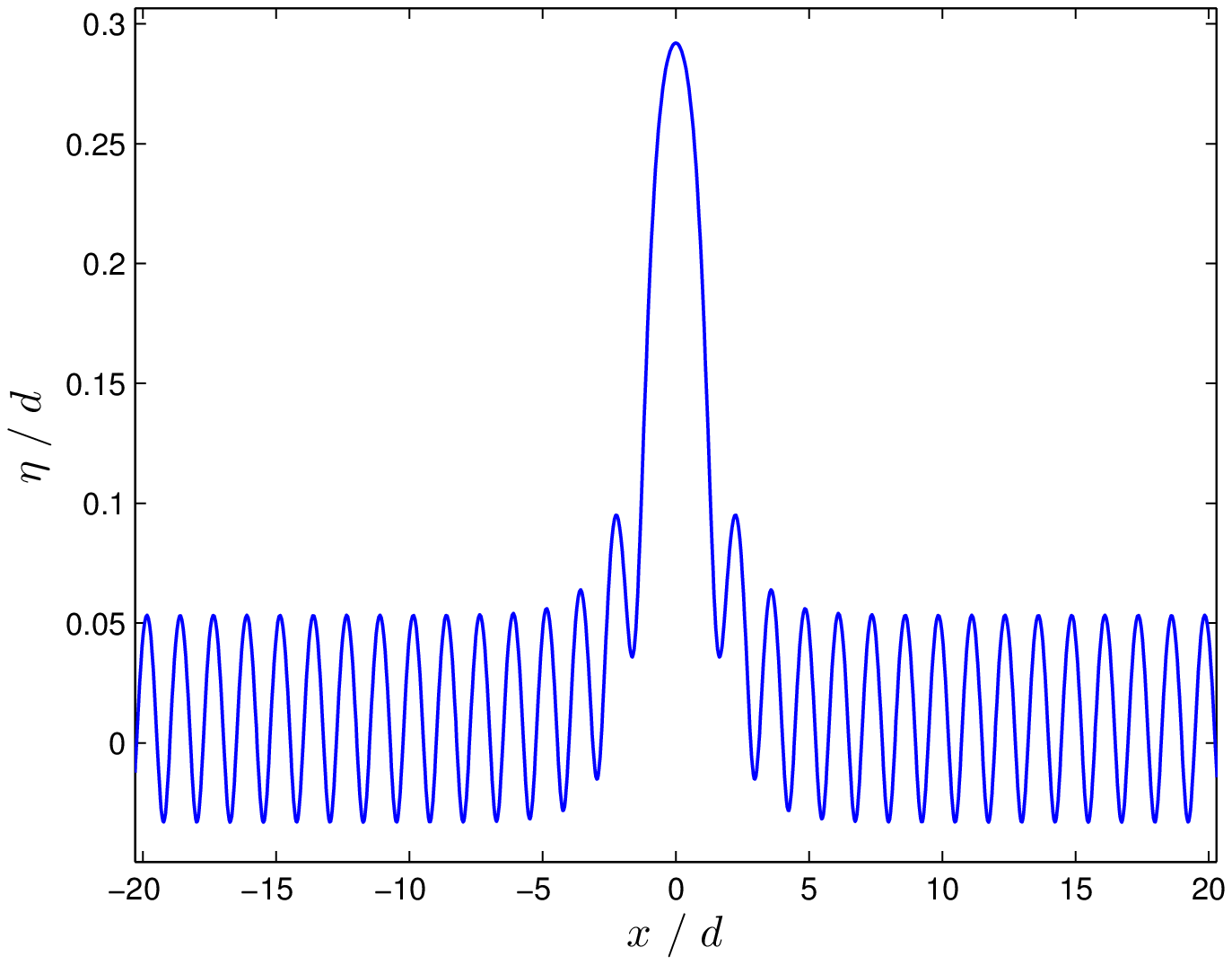}}
  \caption{\small\em Emergence of a generalised solitary wave for $\Fr = 1.15$ by natural continuation in the Bond number $\Bo$ from $0$ to $0.22$.}
  \label{fig:fr1_15cont}
\end{figure}

\begin{figure}
  \centering
  \subfigure[$\Bo = 0.08$, $\Fr_\infty = 1.14993$, $\Bo_\infty = 0.079991$]{
  \includegraphics[width=0.485\textwidth]{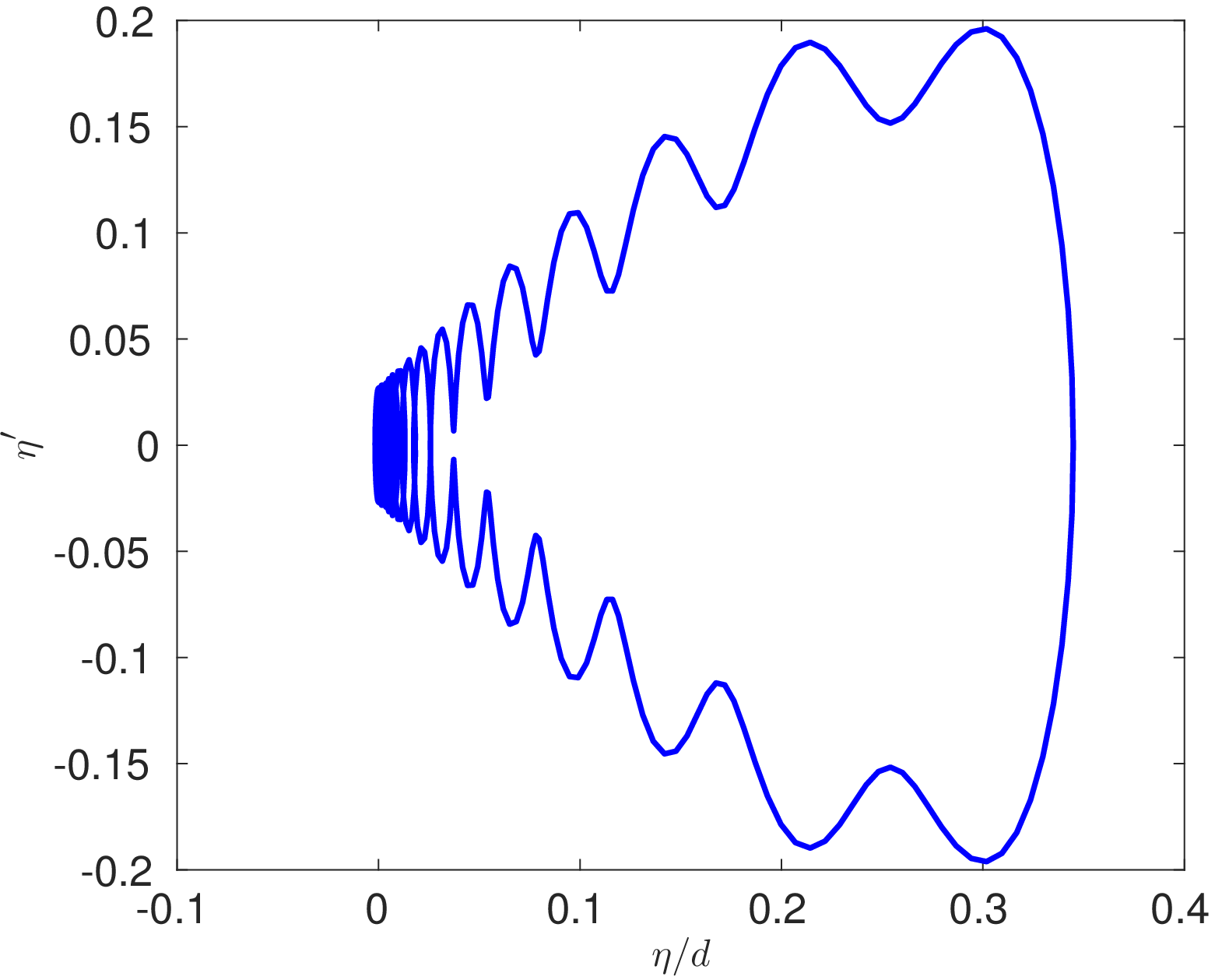}}
  \subfigure[$\Bo = 0.1$, $\Fr_\infty = 1.15894$, $\Bo_\infty = 0.099797$]{
  \includegraphics[width=0.485\textwidth]{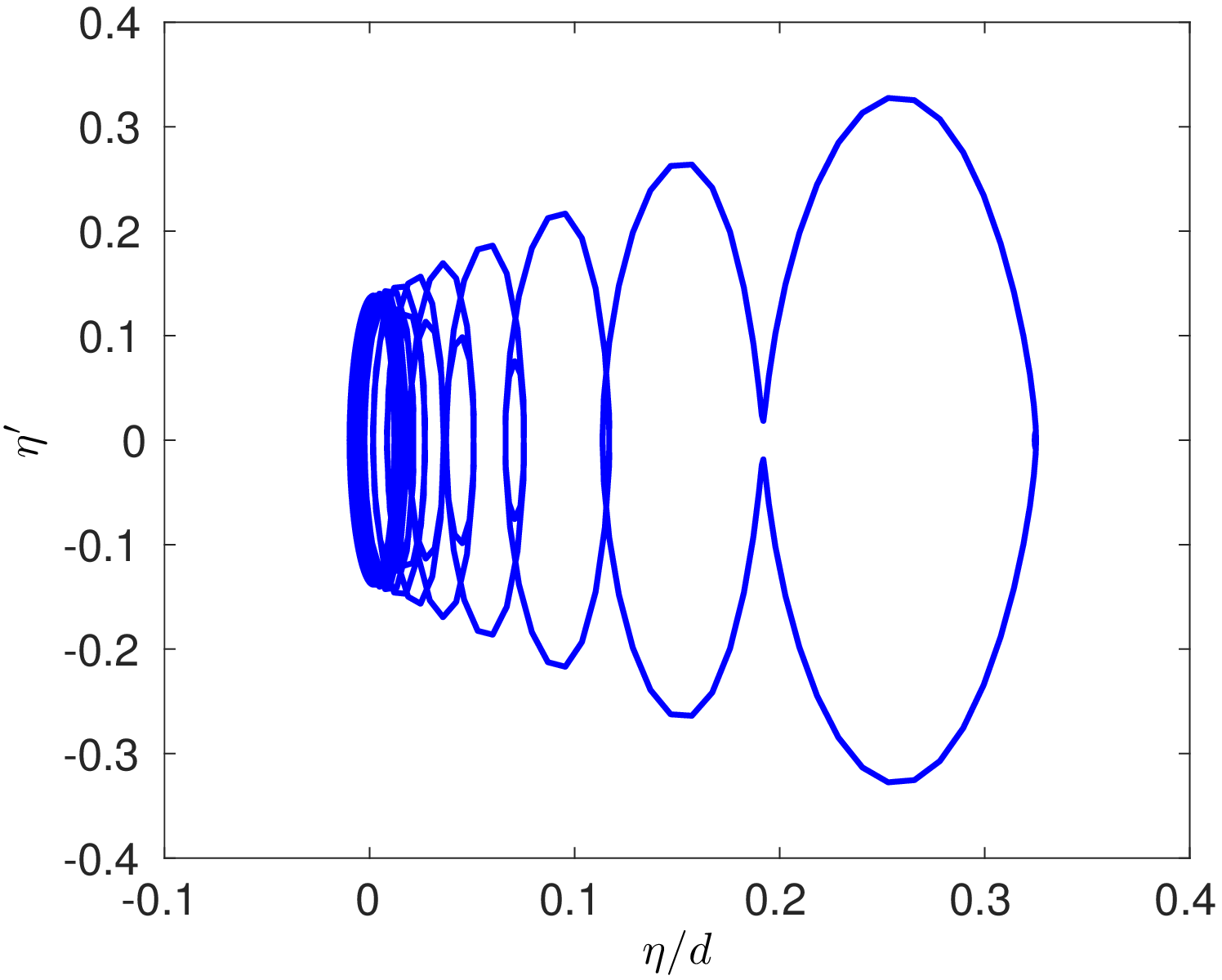}}
  \subfigure[$\Bo = 0.12$, $\Fr_\infty = 1.16376$, $\Bo_\infty = 0.11926$]{
  \includegraphics[width=0.485\textwidth]{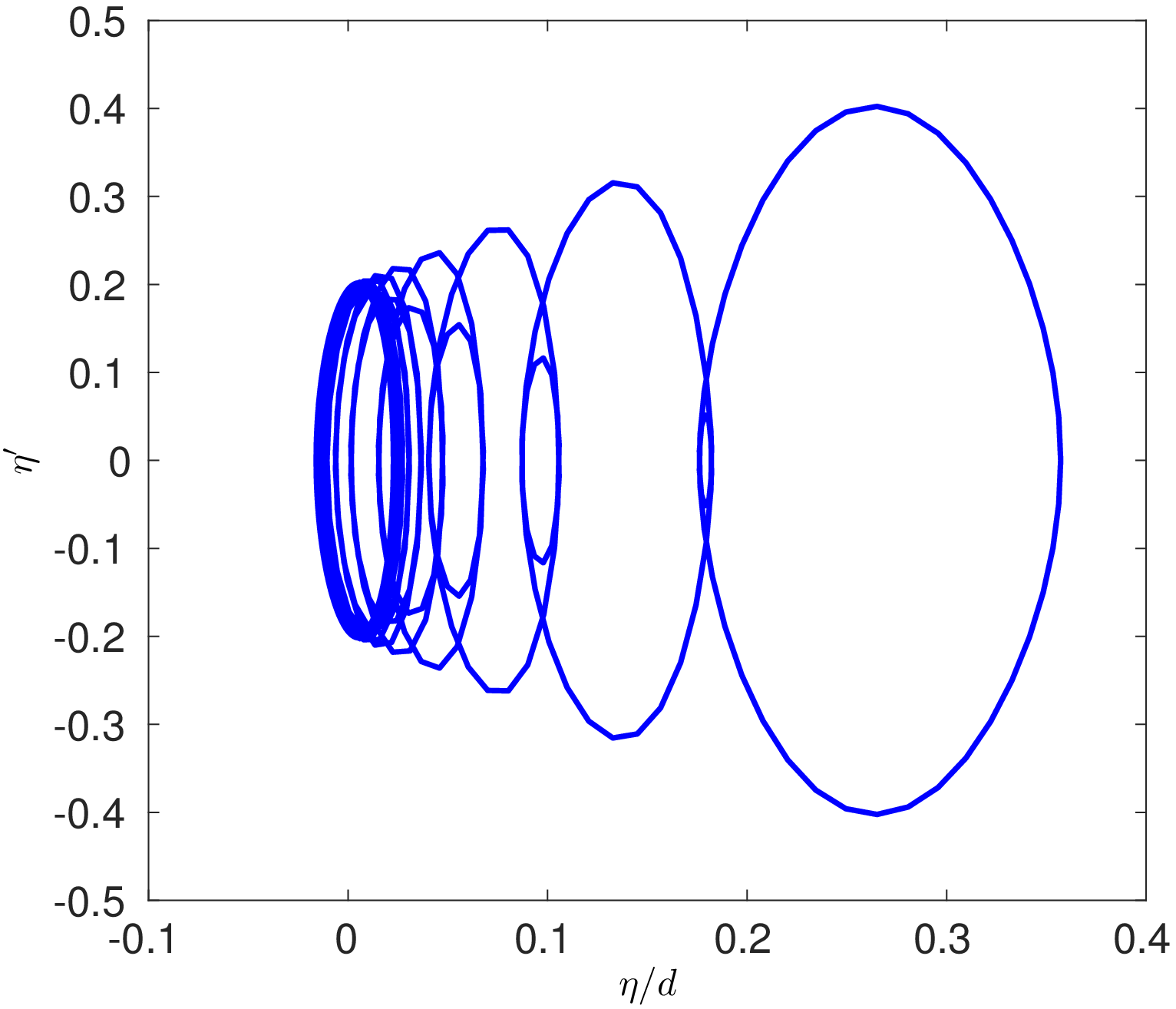}}
  \subfigure[$\Bo = 0.15$, $\Fr_\infty = 1.161479$, $\Bo_\infty = 0.14876$]{
  \includegraphics[width=0.485\textwidth]{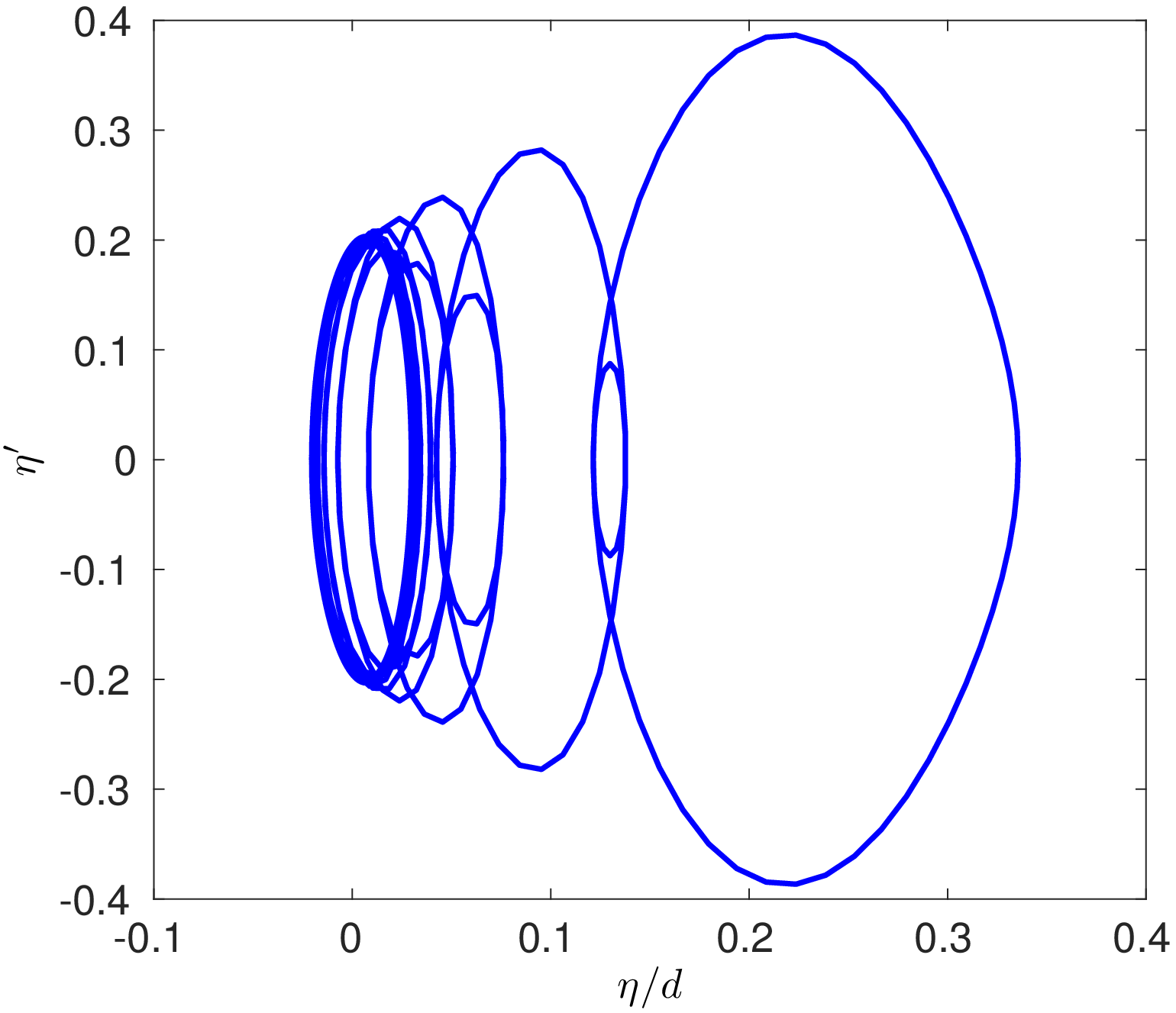}}
  \subfigure[$\Bo = 0.19$, $\Fr_\infty = 1.15994$, $\Bo_\infty = 0.18768$]{
  \includegraphics[width=0.485\textwidth]{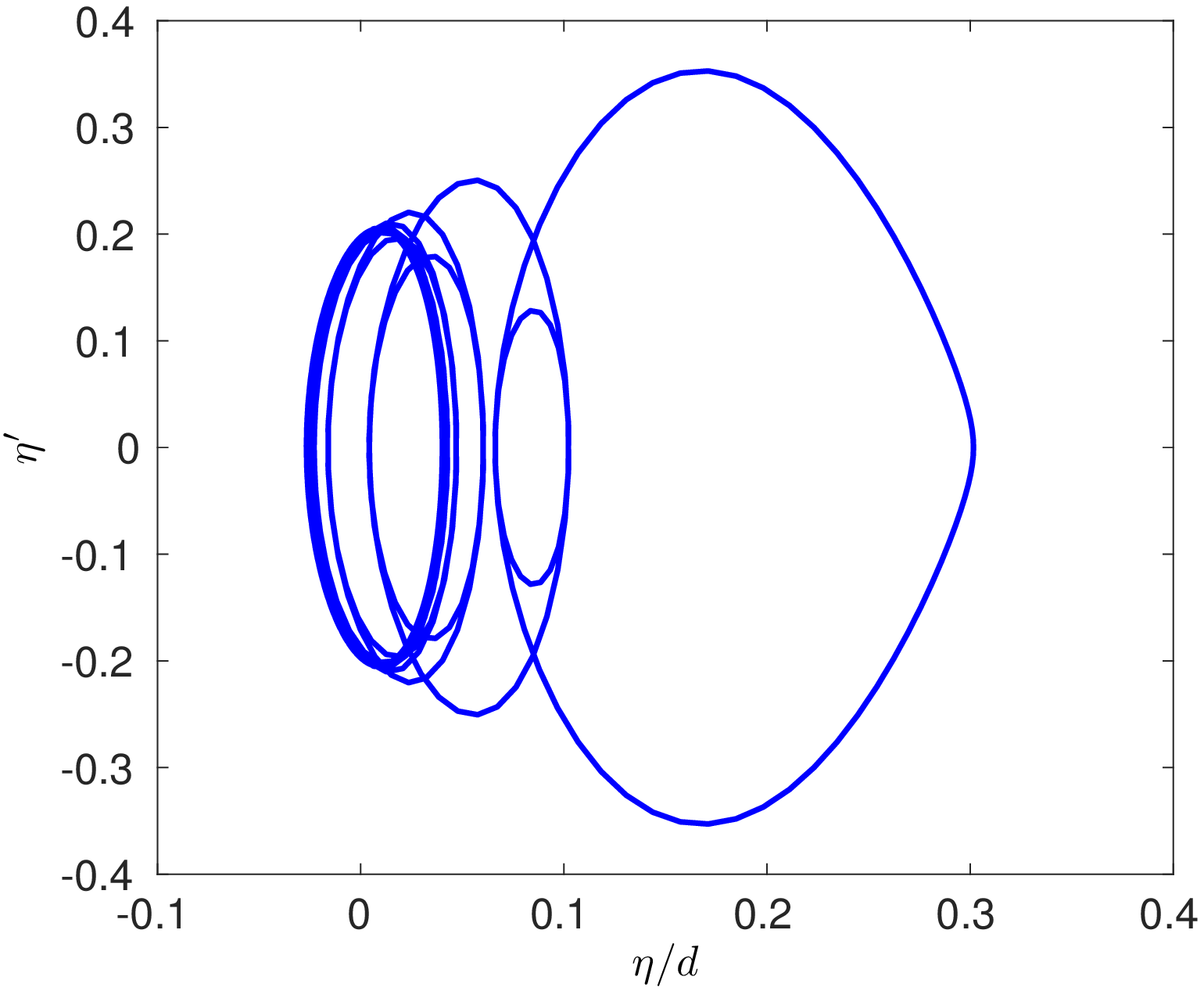}}
  \subfigure[$\Bo = 0.22$, $\Fr_\infty = 1.164997$, $\Bo_\infty = 0.216534$]{
  \includegraphics[width=0.485\textwidth]{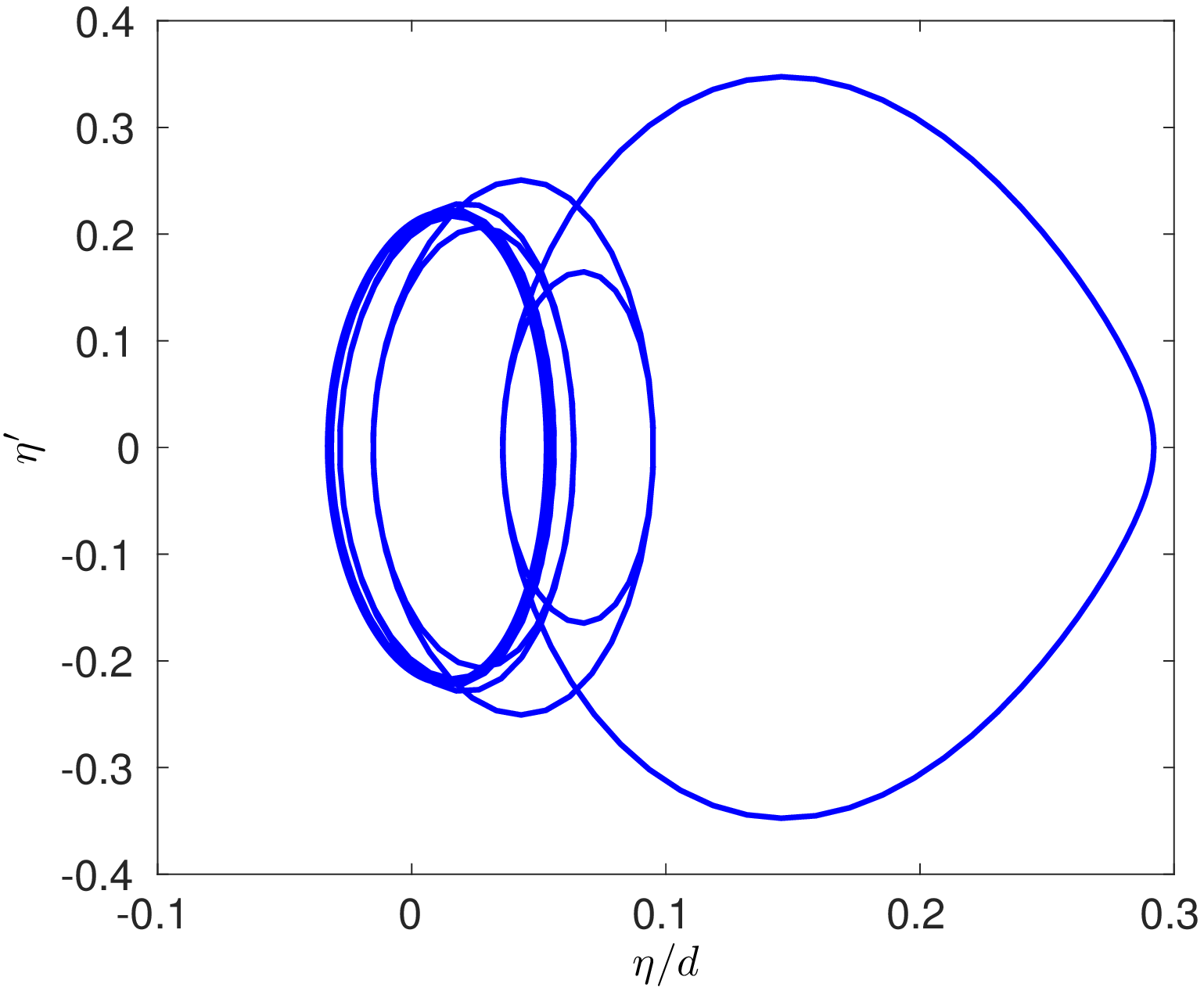}}
  \caption{\small\em Phase plots of generalised solitary waves of Figure~\ref{fig:fr1_15cont}.}
  \label{fig:fr1_15cont_pp}
\end{figure}

Here, we focus on the numerical generation of this kind of waves. Our first experiments concern the range $\Bo<1/3$ and $\Fr>1$, illustrating thus the influence of the capillarity (smaller or larger values of $\Bo$) on the computations. Taking as initial guess a third-order asymptotic solution for solitary-gravity waves \cite{Grimshaw1971} (for $\Fr > 1$ and $0 < \Bo \ll 1/3$), the method has been run in the case of strong capillarity ($\Fr = 1.15$ and $\Bo = 0.22$). The resulting numerical profiles are shown in Figure~\ref{fig:fr1_15cont}. Once the solution is computed, the Froude and Bond numbers have to be actualised to take eventually into account for the periodic tail. The methodology of post-processing is explained in \cite{Clamond2015a} and the effective values of the Froude $\Fr_\infty$ and Bond $\Bo_\infty$ numbers are provided in the corresponding captions in Figure~\ref{fig:fr1_15cont}. The corresponding phase plots are displayed in Figure~\ref{fig:fr1_15cont_pp}. They show the characteristic behaviour of the orbit of a generalised solitary wave profile, homoclinic at infinity to a periodic wave.

\begin{figure}
  \centering
  \subfigure[]{
  \includegraphics[width=0.485\textwidth]{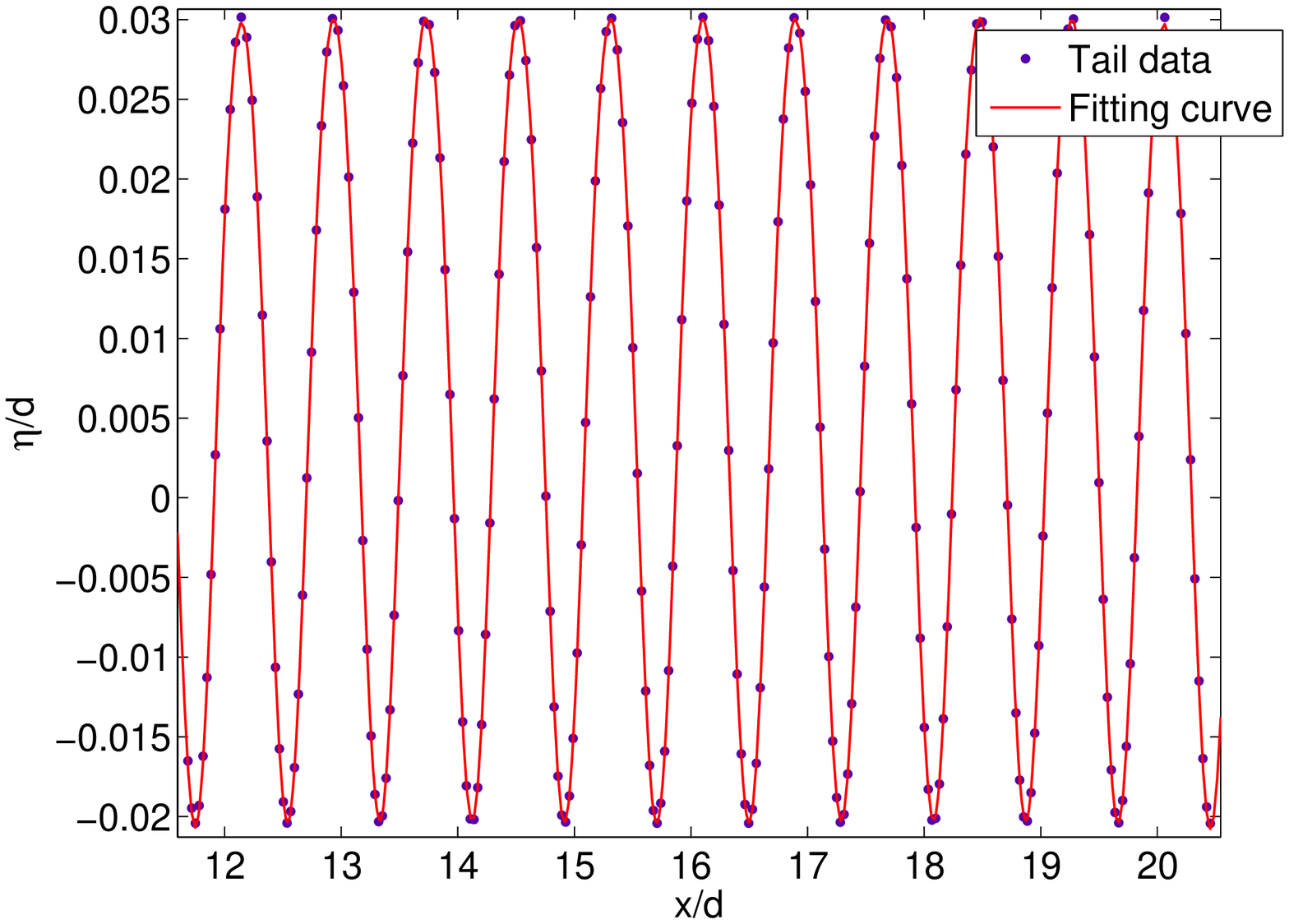}}
  \subfigure[]{\includegraphics[width=0.485\textwidth]{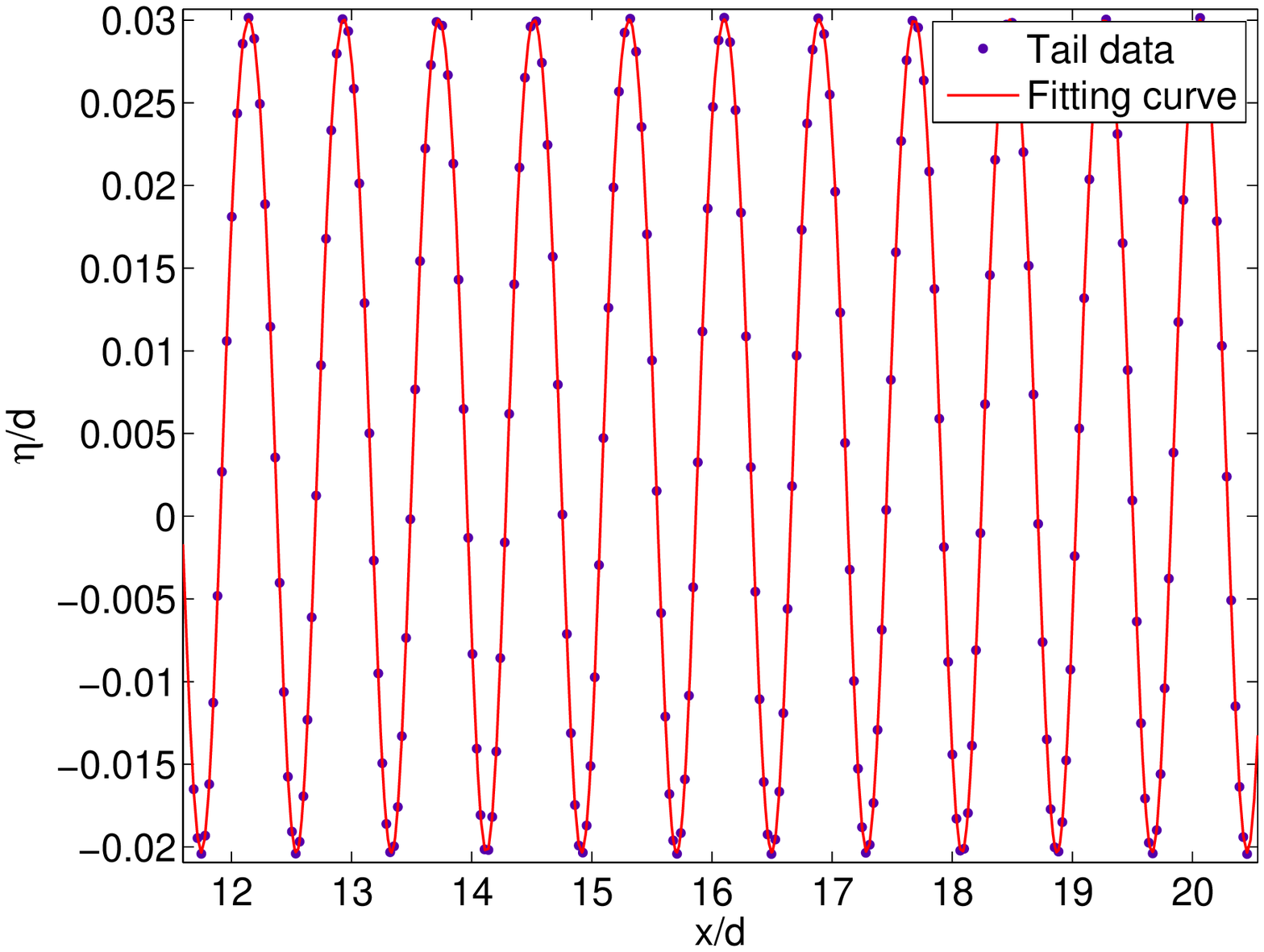}}
  \caption{\small\em Fit of the oscillatory tail for the generalised wolitary wave with $\Fr = 1.15$, $\Bo = 0.15$. Left: Fourier sum with two terms. Right: Sum of six sinusoidal functions. The coefficients are in Table~\ref{tab:fit}.}
  \label{fig:osc_fit1}
\end{figure}

From the computed profiles, some additional information concerning the oscillatory tails can be obtained. For instance, part of the profile can be taken and fitted to a sinusoidal wave in order to estimate the corresponding amplitude and wavelength. This fit is illustrated in Figure~\ref{fig:osc_fit1} which shows the data of part of the tail of Figure~\ref{fig:fr1_15cont} ({\it d}) and two different sinusoidal curves. In Figure~\ref{fig:osc_fit1}({\it a}) the data are fitted by a Fourier sum (with two terms) while in Figure~\ref{fig:osc_fit1}({\it b}) this is made by a sum of sinusoidal functions with six different wavelengths. The corresponding coefficients and goodness of fit parameters are displayed in Table~\ref{tab:fit}. In the first case, the best choice among different Fourier sums corresponds to taking two Fourier modes and leads to an approximate fundamental frequency of $f = w/(2\pi) = 1.2634$. On the other hand, in the second fitting curve we have $\beta_1 = \omega$, $\beta_6 = 2\omega$ and the influence of the rest of the frequencies is limited by the amplitudes of the corresponding sinusoidal terms. Thus the two fits give the same essential information about the oscillatory tail: an approximate amplitude of $0.02517$ and wavelength of $\lambda = 1/f = 0.7915$.

\begin{figure}
  \centering
  \subfigure[]{\includegraphics[width=0.485\textwidth]{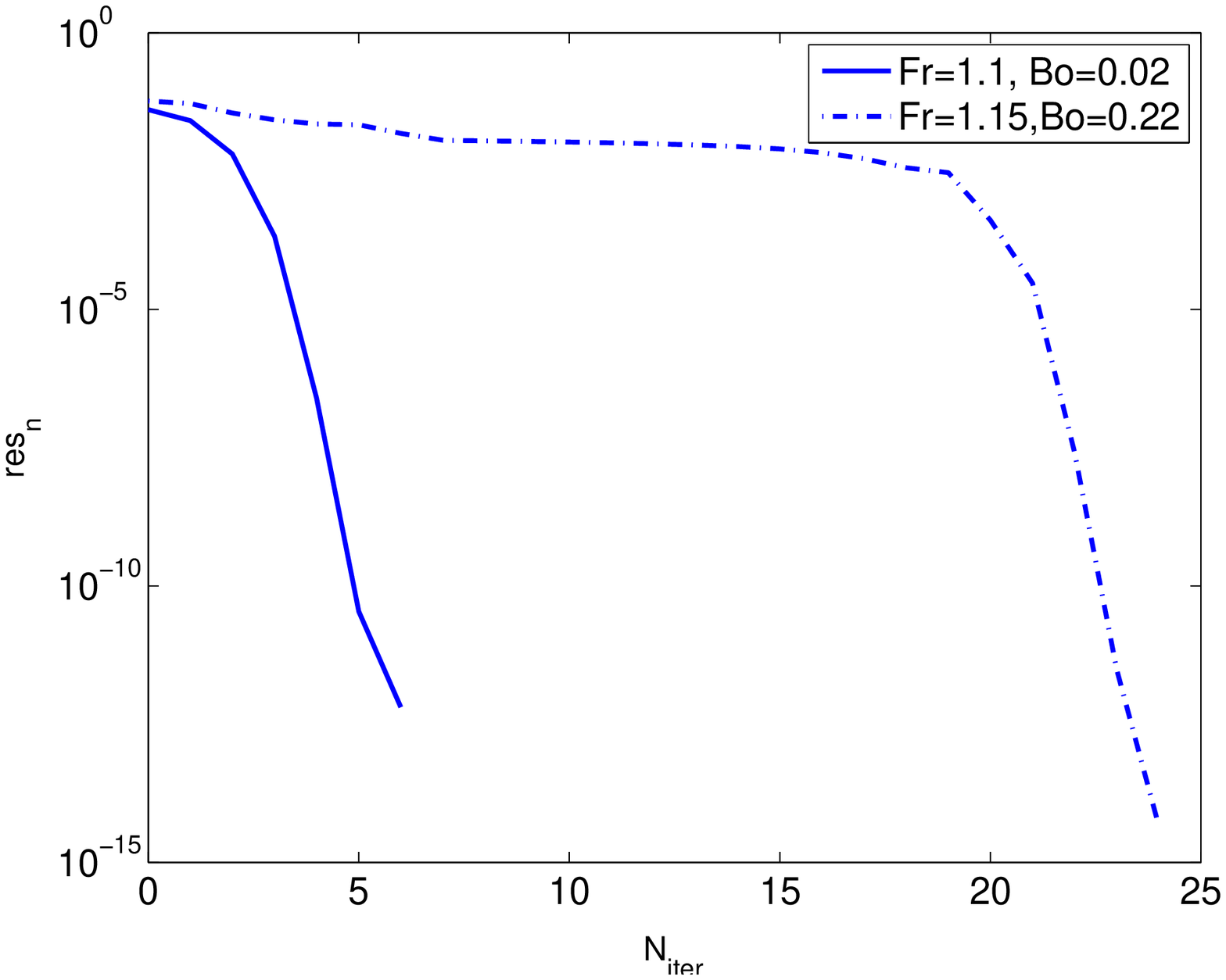}}
  \subfigure[]{\includegraphics[width=0.485\textwidth]{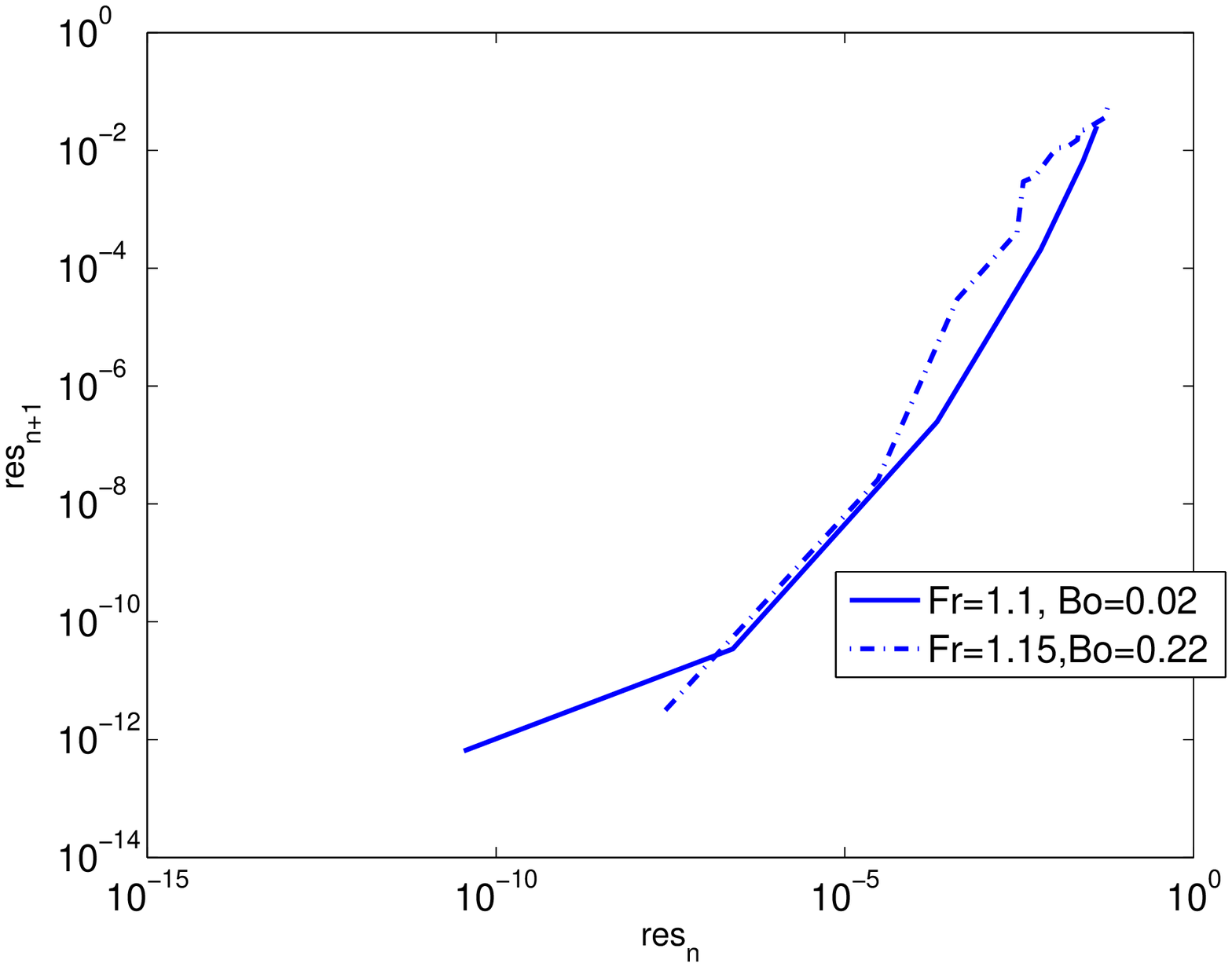}}
  \caption{\small\em Convergence of the \acs{lm} algorithm. Generation of generalised solitary waves. Solid line: $\Fr = 1.1$, $\Bo = 0.02$; dashed-dotted line: $\Fr = 1.15$, $\Bo = 0.22$. Left: Residual error \eqref{eq:RES} as function of the number of iterations (semi-logarithmic scale). Right: Relation between two consecutive residual errors \eqref{eq:RES} (log-log scale).}
  \label{fig:gres1}
\end{figure}

The convergence process is illustrated as in the previous Section. Figure~\ref{fig:gres1} displays the logarithm of the residual error \eqref{eq:RES} as function of the number of iterations. Note that the method performs better in the case of the nonlinear wave with weaker capillary effects, when the oscillations at infinity of the wave is of smaller amplitude, see \eqref{eq:order}. The order of convergence is suggested by the Figure~\ref{fig:gres1} (right) (where the ratio between two consecutive errors is shown in log-log scale) in the same sense as in Figure~\ref{fig:reserror} (right): for errors above $10^{-7}$, the slopes of the corresponding fitting lines suggest quadratic convergence, being linear from this error tolerance.

\begin{figure}
  \centering
  \subfigure[$\phi\,+\,\Fr\cdot x$]{
  \includegraphics[width=0.485\textwidth]{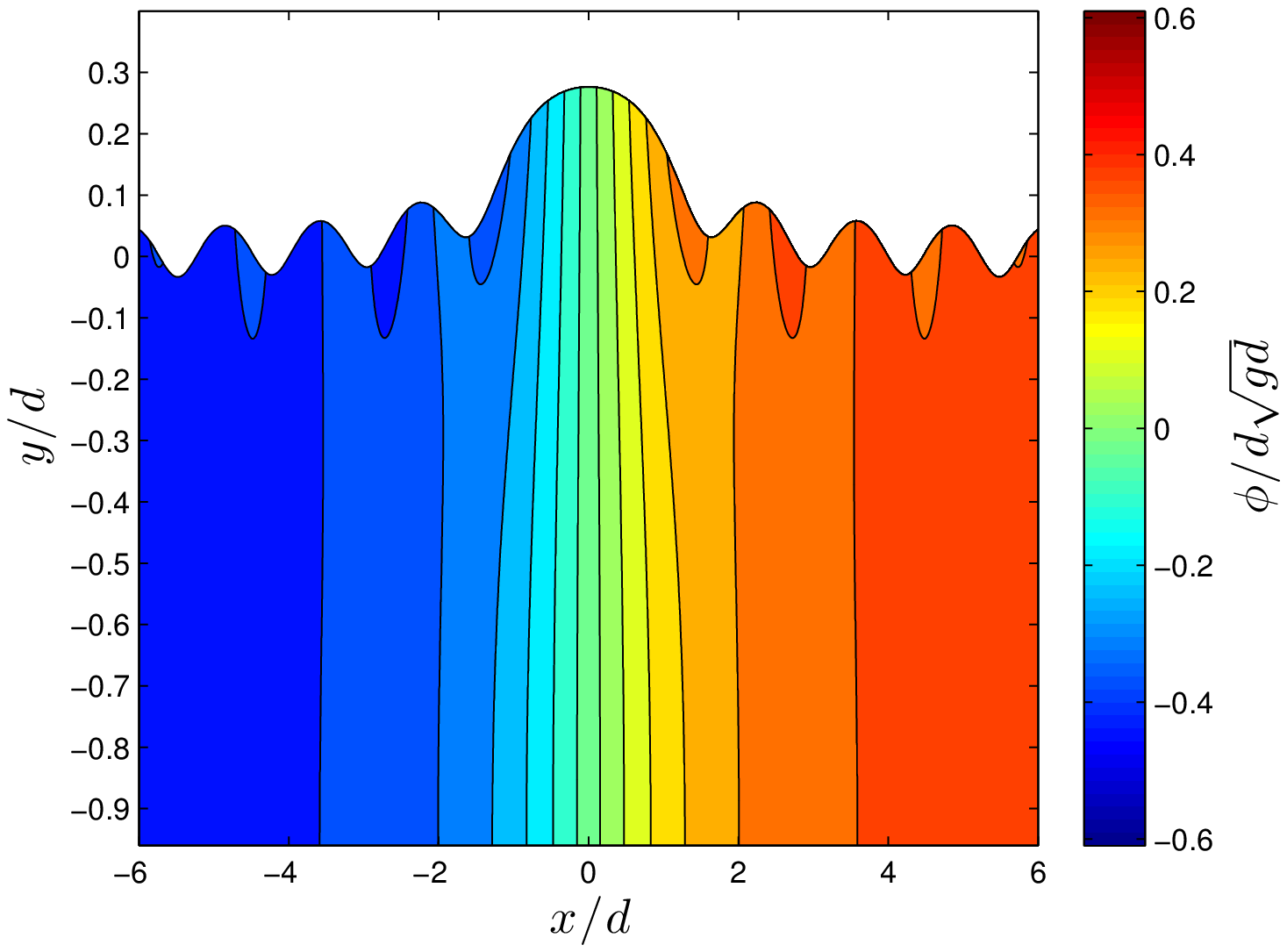}}
  \subfigure[$\psi\,-\,\bot{\psi}\,+\,\Fr\cdot (y+d)$]{
  \includegraphics[width=0.485\textwidth]{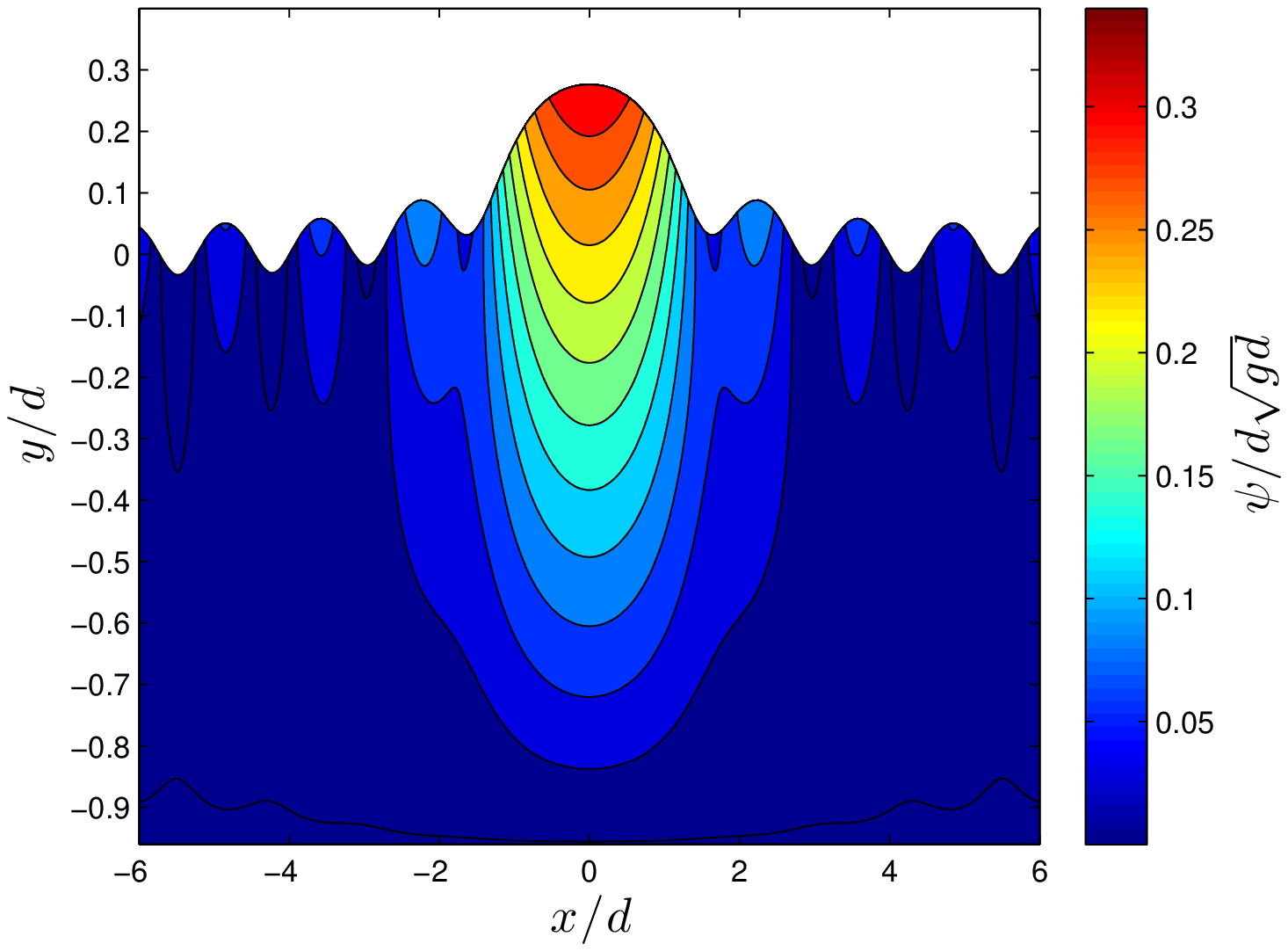}}
  \caption{\small\em The velocity potential (left) and the stream function (right) iso-pressures in the fluid domain under the generalised solitary wave displayed on Figure~\ref{fig:fr1_15cont}{\it f}.}
  \label{fig:fieldsPot}
\end{figure}

\begin{figure}
  \centering
  \subfigure[$P$]{
  \includegraphics[width=0.485\textwidth]{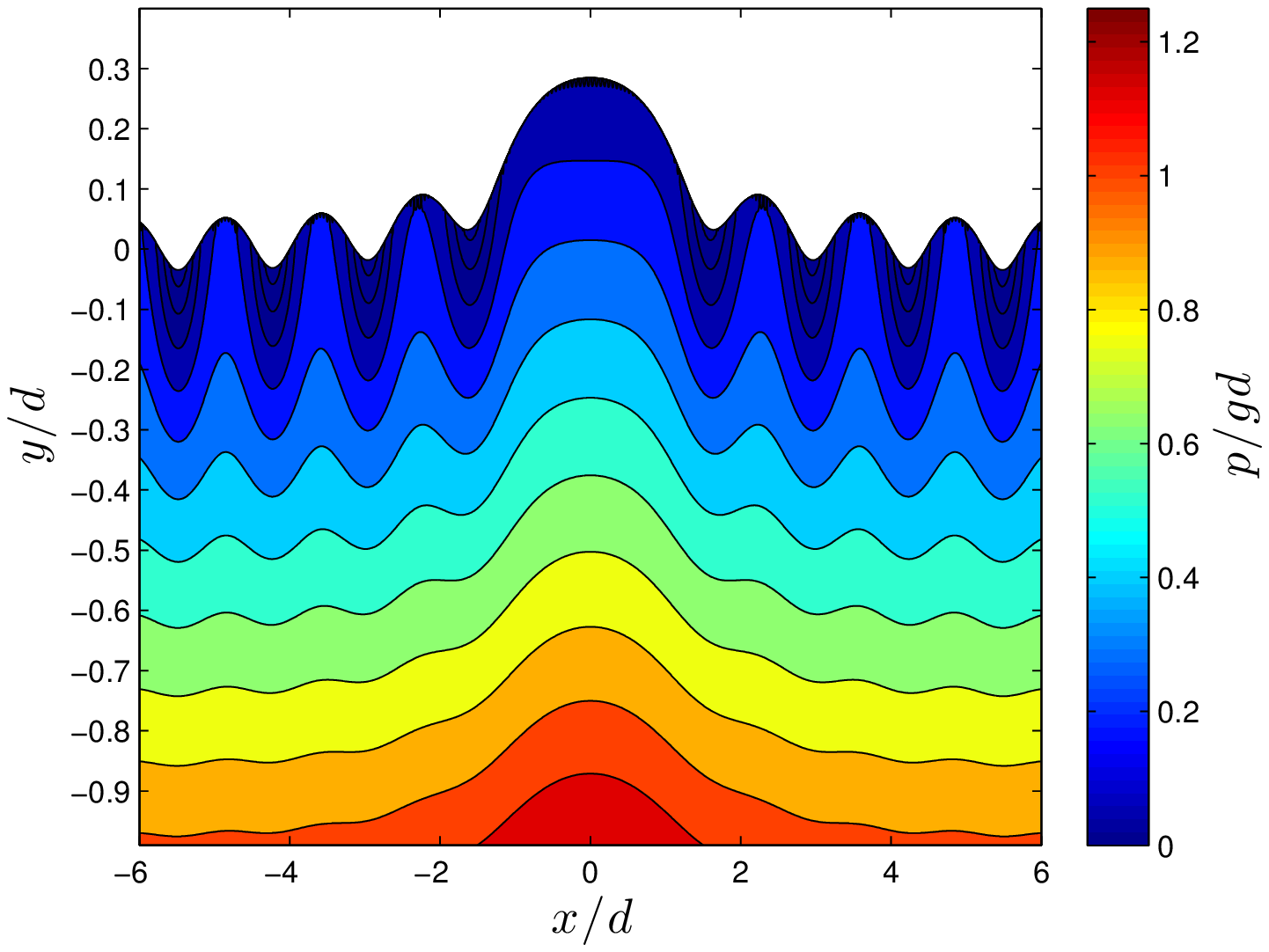}}
  \subfigure[$P+gy$]{
  \includegraphics[width=0.485\textwidth]{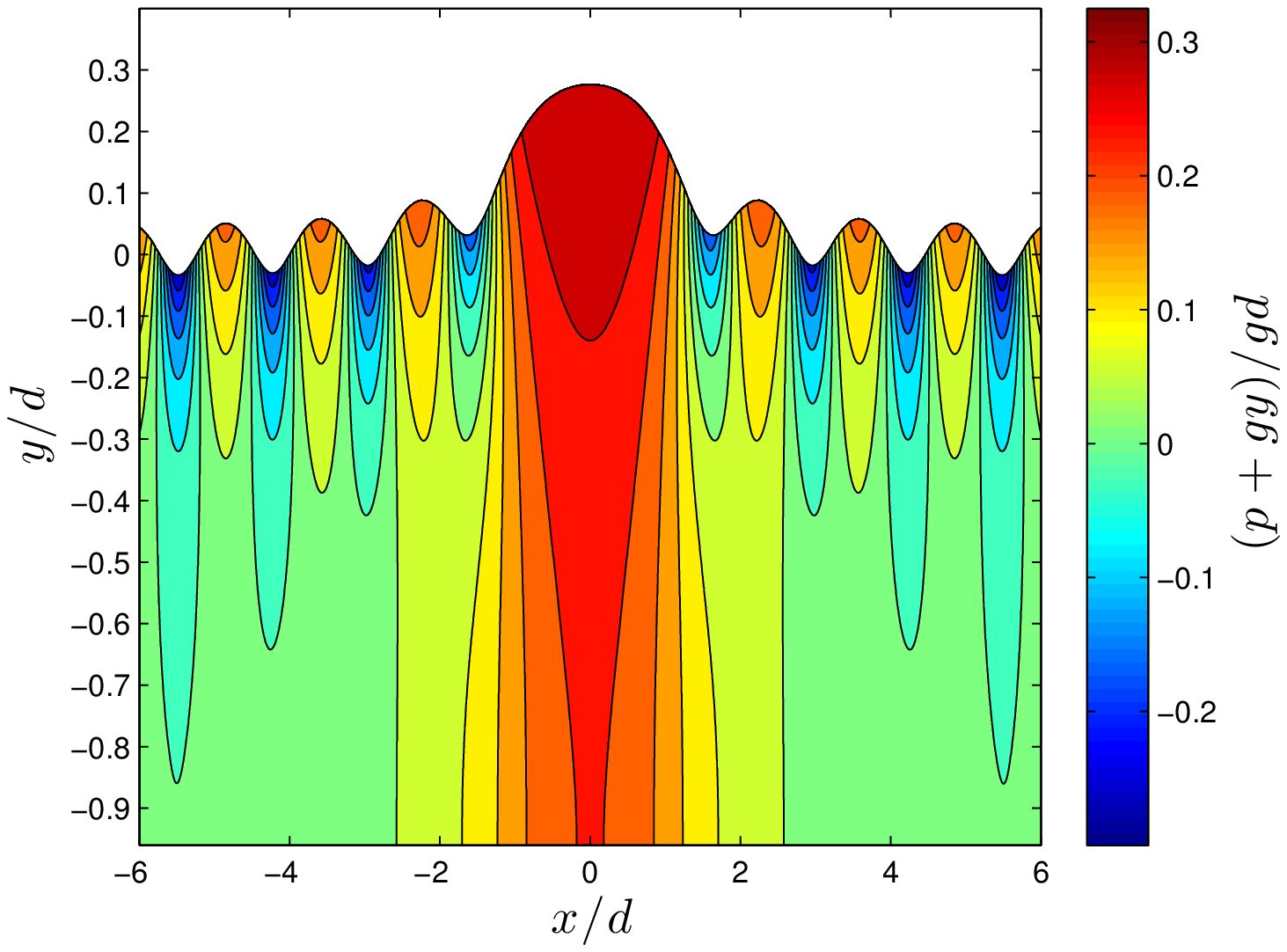}}
  \caption{\small\em Total (left) and dynamic (right) iso-pressures in the fluid domain under the generalised solitary wave displayed on Figure~\ref{fig:fr1_15cont}{\it f}.}
  \label{fig:fieldsPress}
\end{figure}

\begin{figure}
  \centering
  \subfigure[$u\,+\,\Fr$]{
  \includegraphics[width=0.485\textwidth]{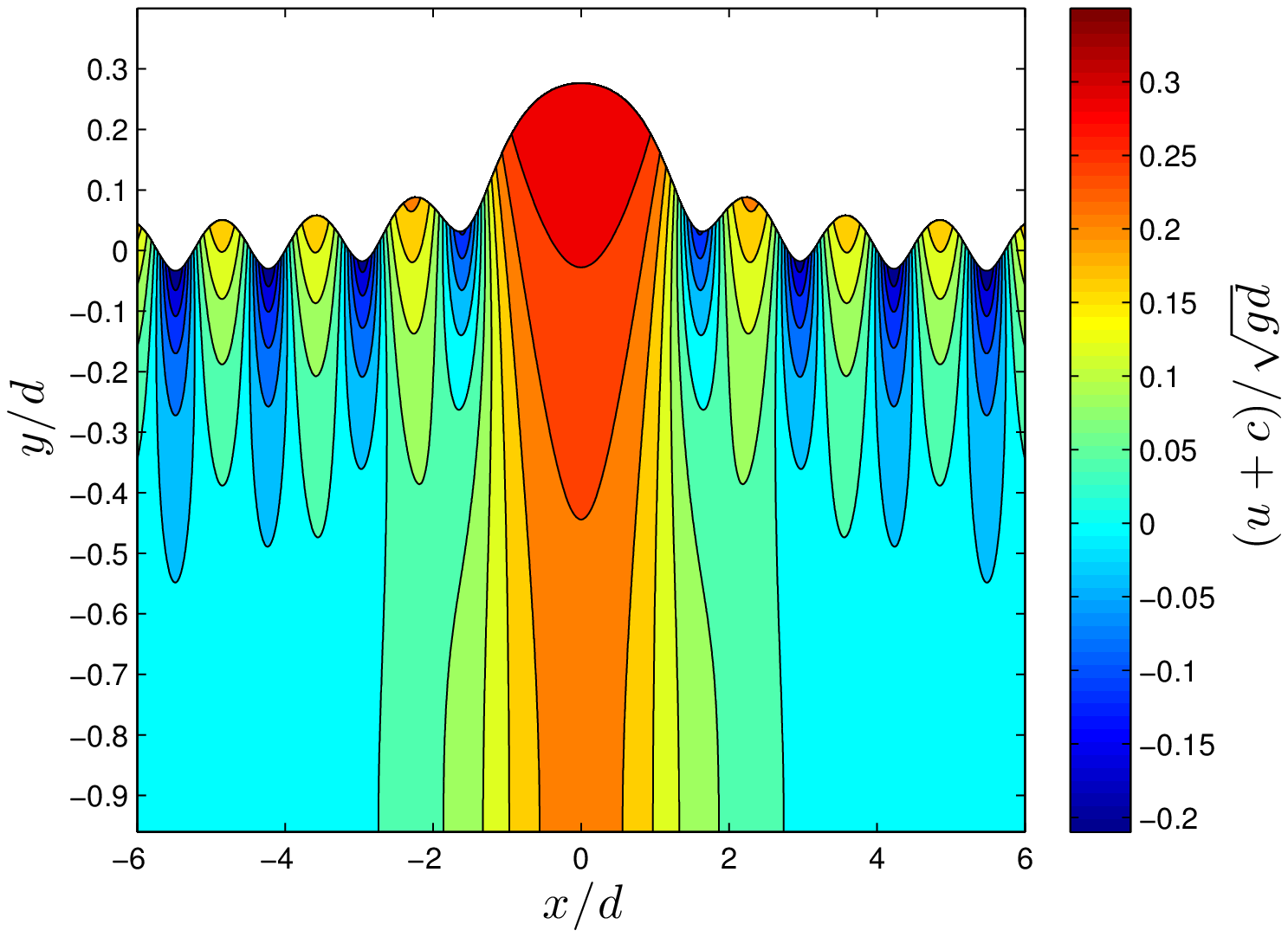}}
  \subfigure[$v$]{
  \includegraphics[width=0.485\textwidth]{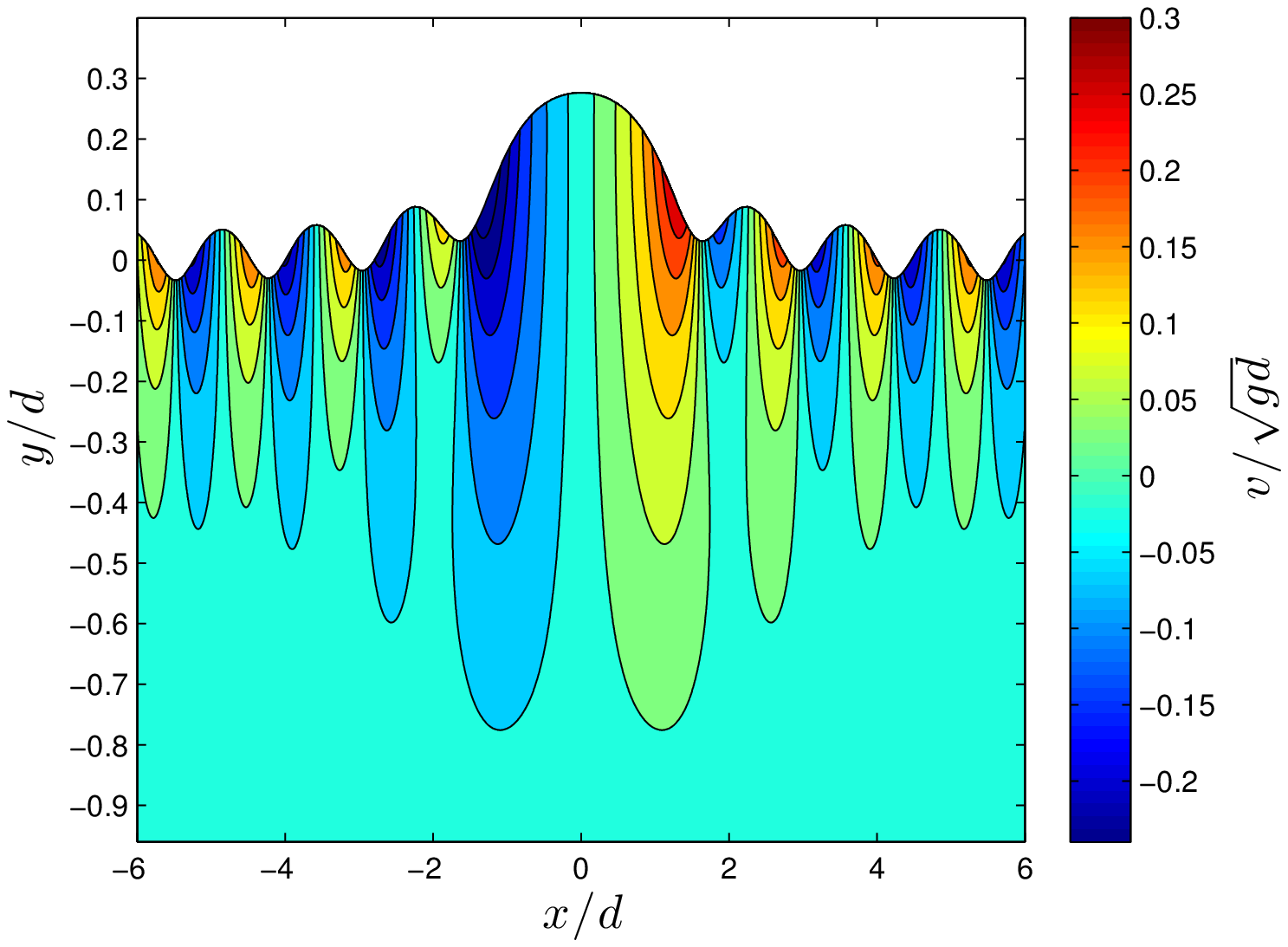}}
  \caption{\small\em Horizontal (left) and vertical (right) velocities in the fluid domain under the generalised solitary wave which displayed on Figure~\ref{fig:fr1_15cont}{\it f}.}
  \label{fig:fieldsSpeeds}
\end{figure}

\begin{figure}
  \centering
  \subfigure[$a_x$]{
  \includegraphics[width=0.485\textwidth]{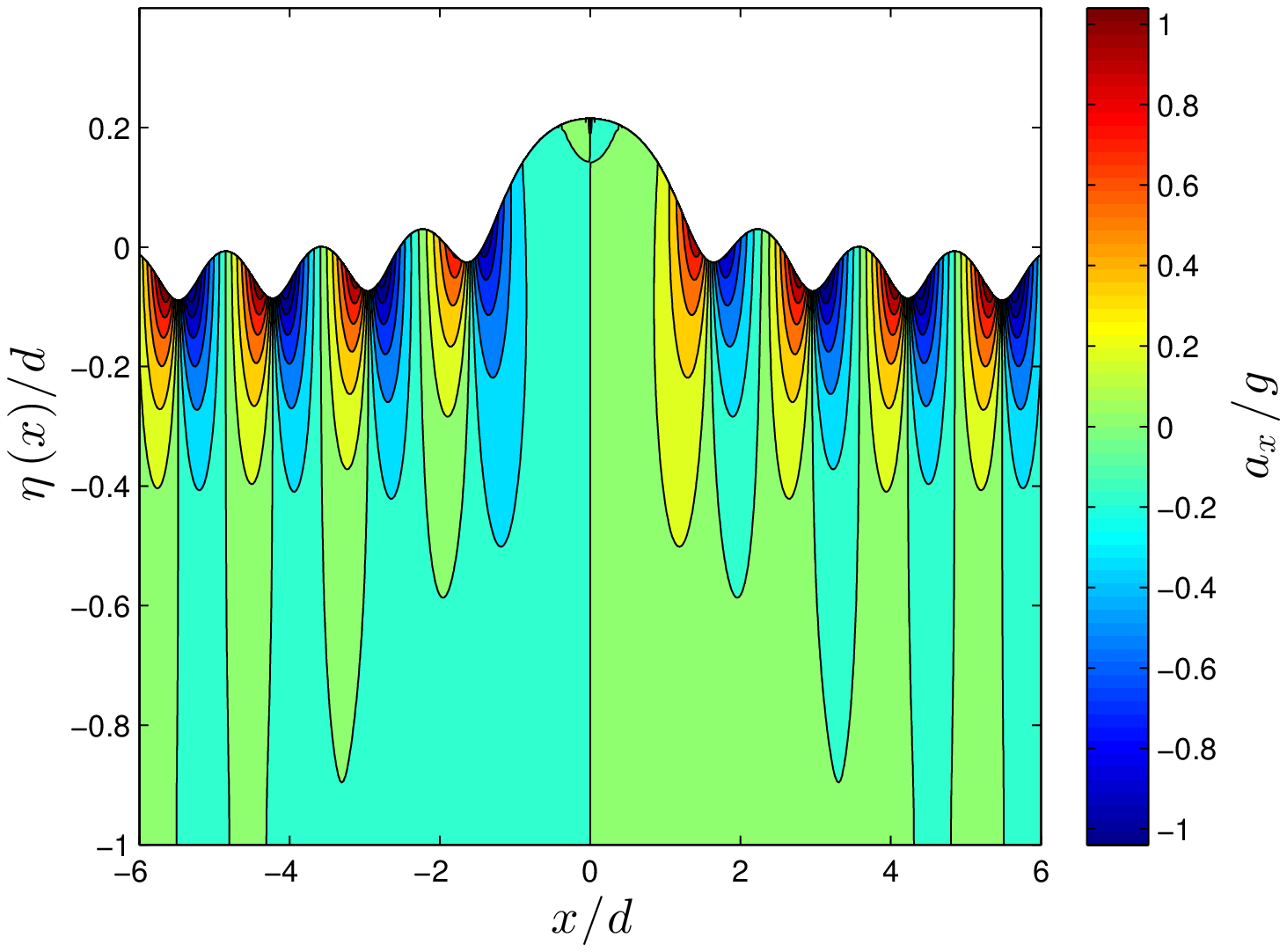}}
  \subfigure[$a_y$]{
  \includegraphics[width=0.485\textwidth]{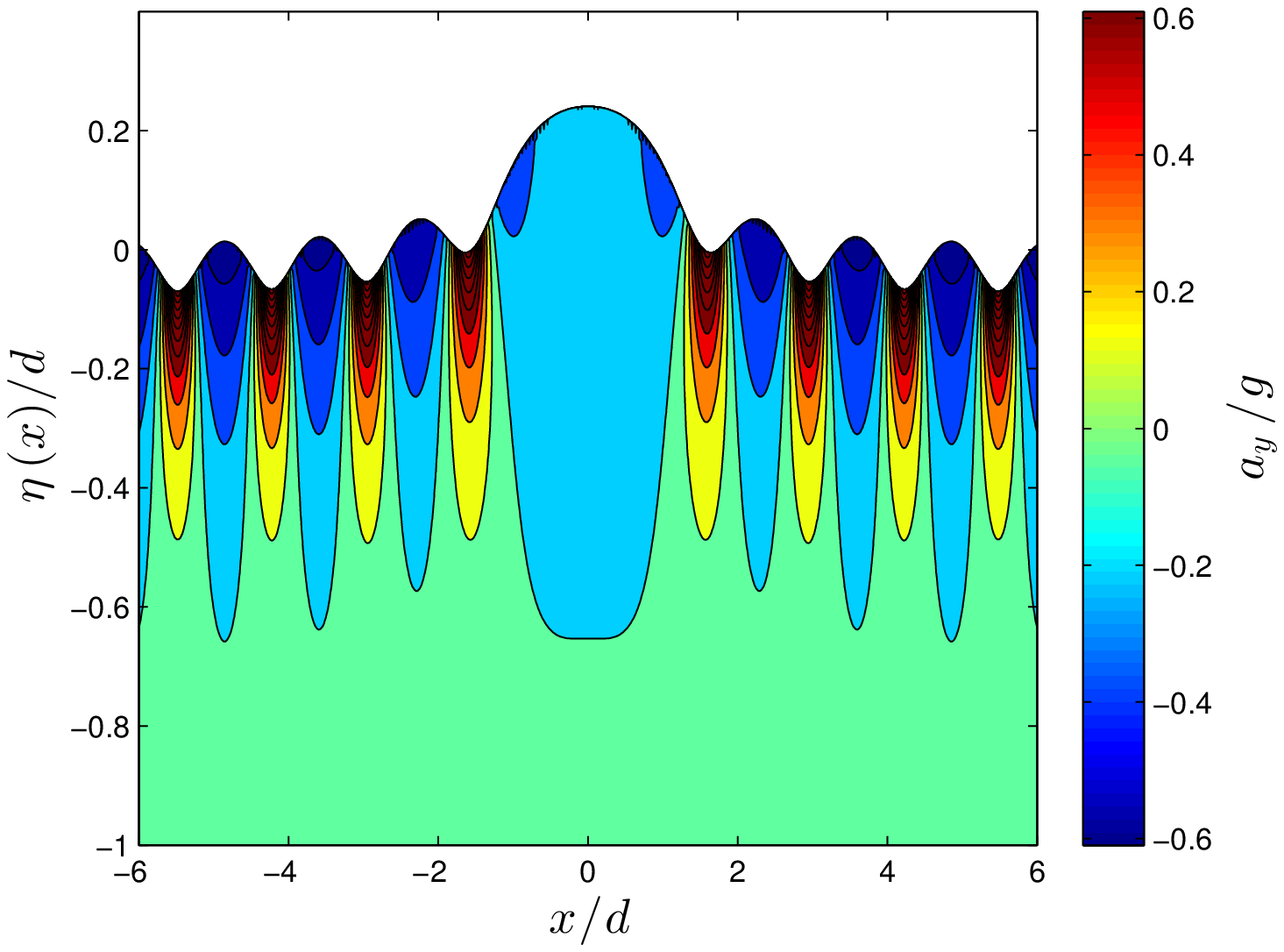}}
  \caption{\small\em Horizontal (left) and vertical (right) accelerations in the fluid domain under the generalised solitary wave displayed on Figure~\ref{fig:fr1_15cont}{\it f}.}
  \label{fig:fieldsAccel}
\end{figure}

\begin{figure}
  \centering
  \subfigure[$\phi_s(x)$]{%
  \includegraphics[width=0.485\textwidth]{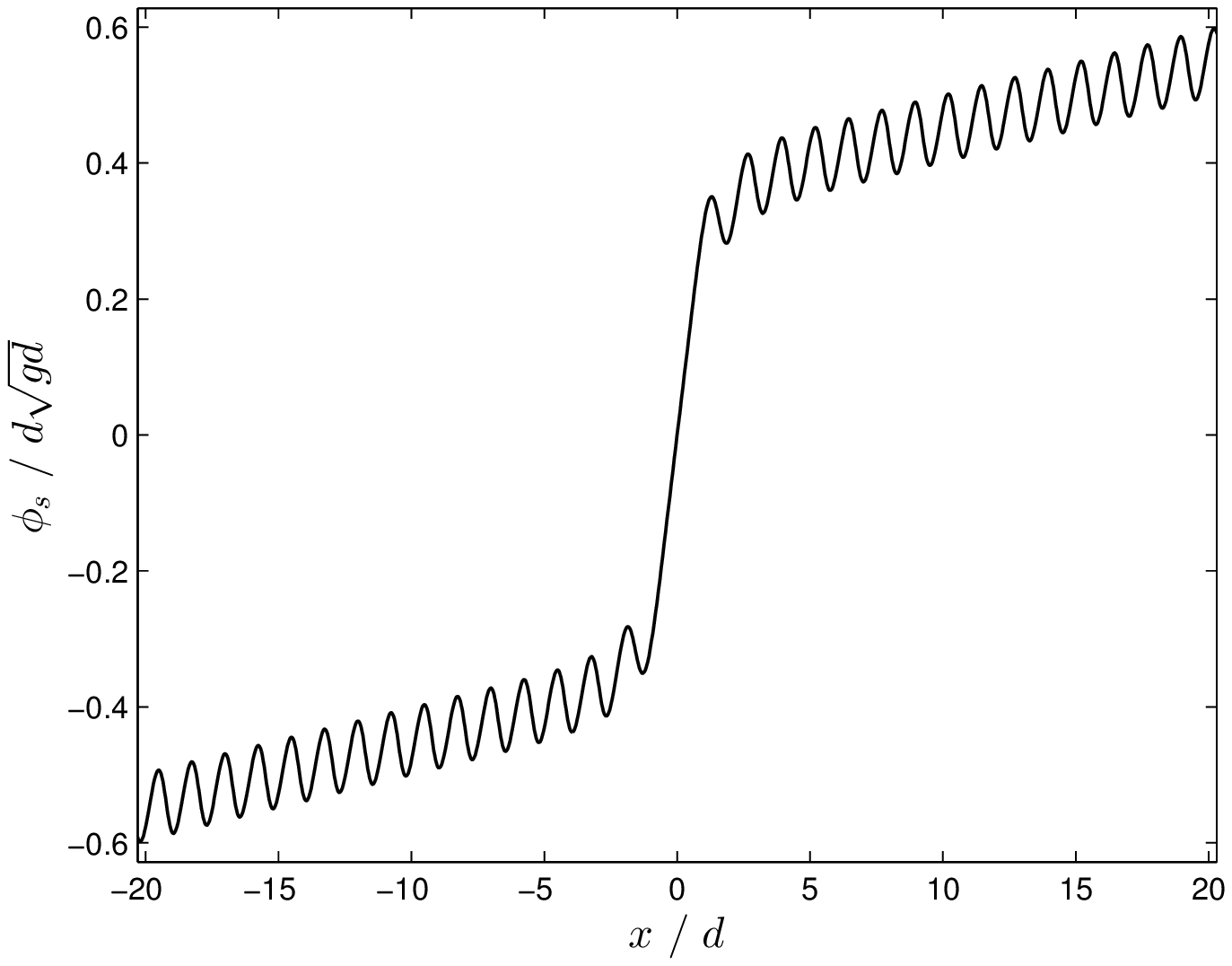}}
  \subfigure[$\psi_s(x)$]{%
  \includegraphics[width=0.485\textwidth]{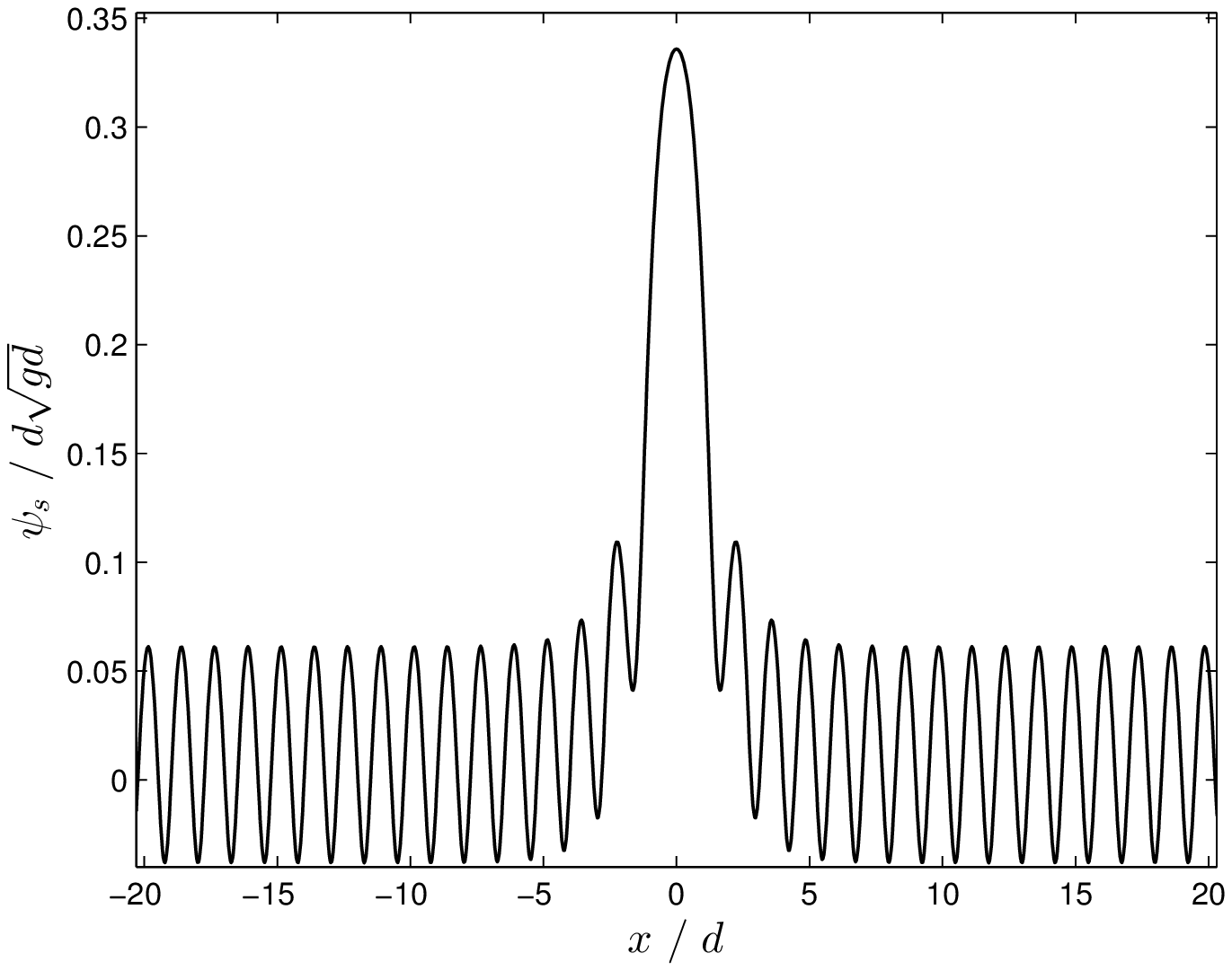}}
  \subfigure[$u_s(x)$]{%
  \includegraphics[width=0.485\textwidth]{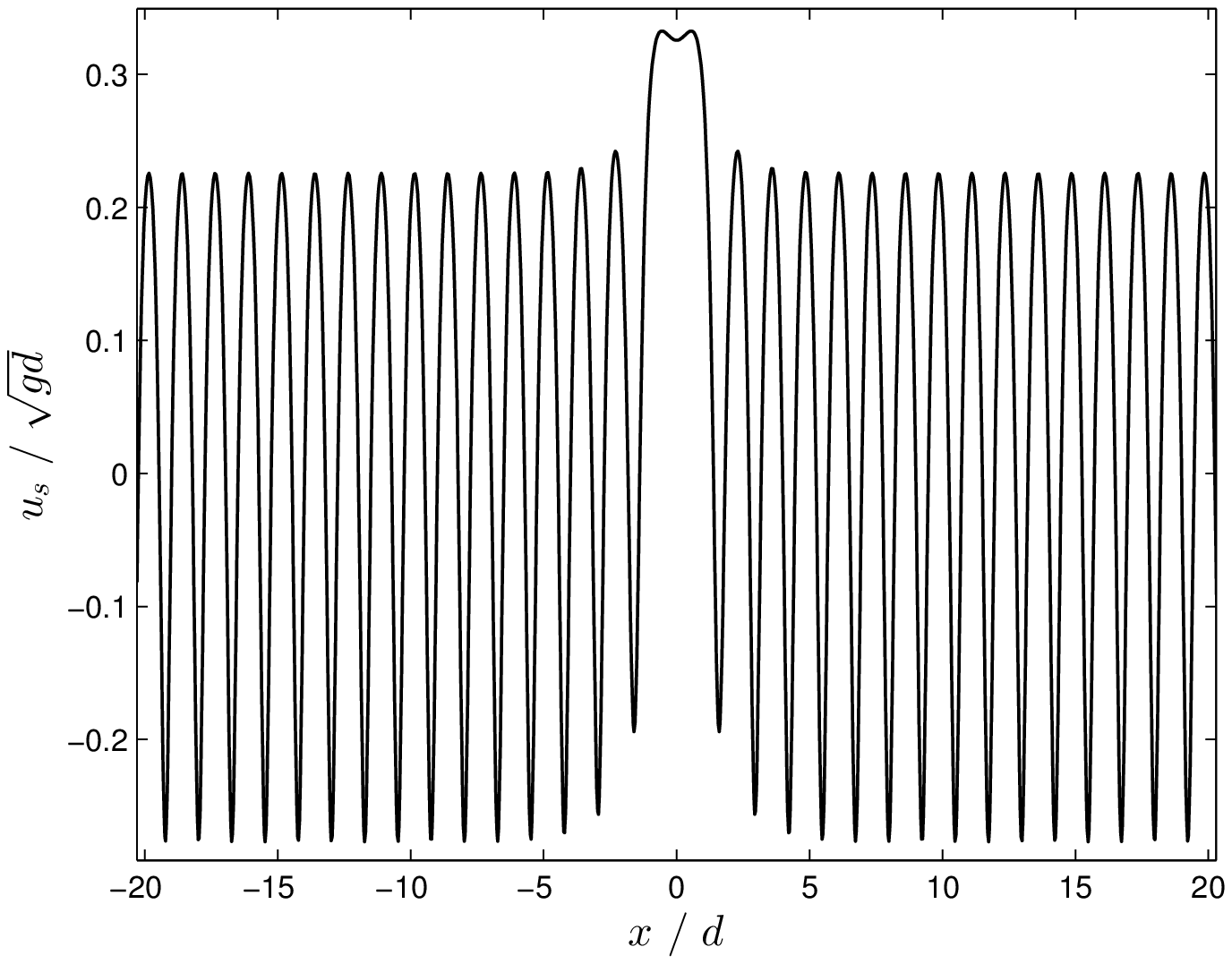}}
  \subfigure[$v_s(x)$]{%
  \includegraphics[width=0.485\textwidth]{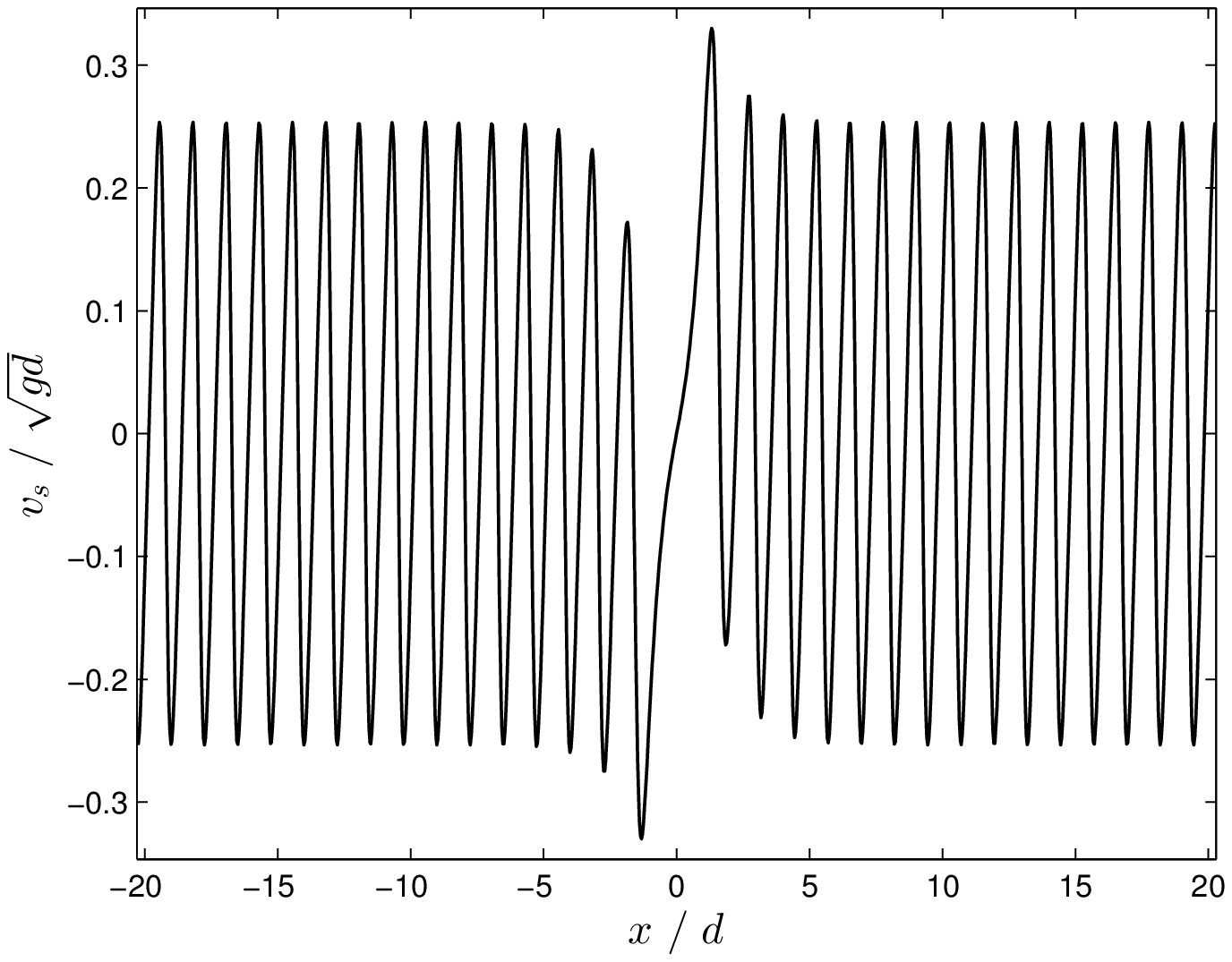}}
  \caption{\small\em The traces of the velocity potential (a), stream function (b), horizontal (c) and vertical (d) velocities at the free surface for a generalised solitary wave ($\Fr = 1.15$, $\Bo = 0.22$). The free surface profile is represented on Figure~\ref{fig:fr1_15cont}{\it f}.}
  \label{fig:surf}
\end{figure}

For the case $\Fr = 1.15$ and $\Bo = 0.22$ corresponding to the Figure~\ref{fig:fr1_15cont} ({\it f}), the internal hydrodynamics under a generalised solitary wave is illustrated in Figures~\ref{fig:fieldsPot}--\ref{fig:fieldsAccel} for the same profile, with the velocity potential (Fig.~\ref{fig:fieldsPot} left) and stream function (Fig.~\ref{fig:fieldsPot} right), total (Fig.~\ref{fig:fieldsPress} left) and dynamic (Fig.~\ref{fig:fieldsPress} right) pressures, horizontal (Fig.~\ref{fig:fieldsSpeeds} left) and vertical (Fig.~\ref{fig:fieldsSpeeds} right) velocities and accelerations (Fig.~\ref{fig:fieldsAccel}). The traces of the velocity potential, stream function, horizontal and vertical velocities on the free surface are depicted in Figure~\ref{fig:surf}.

\begin{figure}
  \centering
  \includegraphics[width=0.49\textwidth]{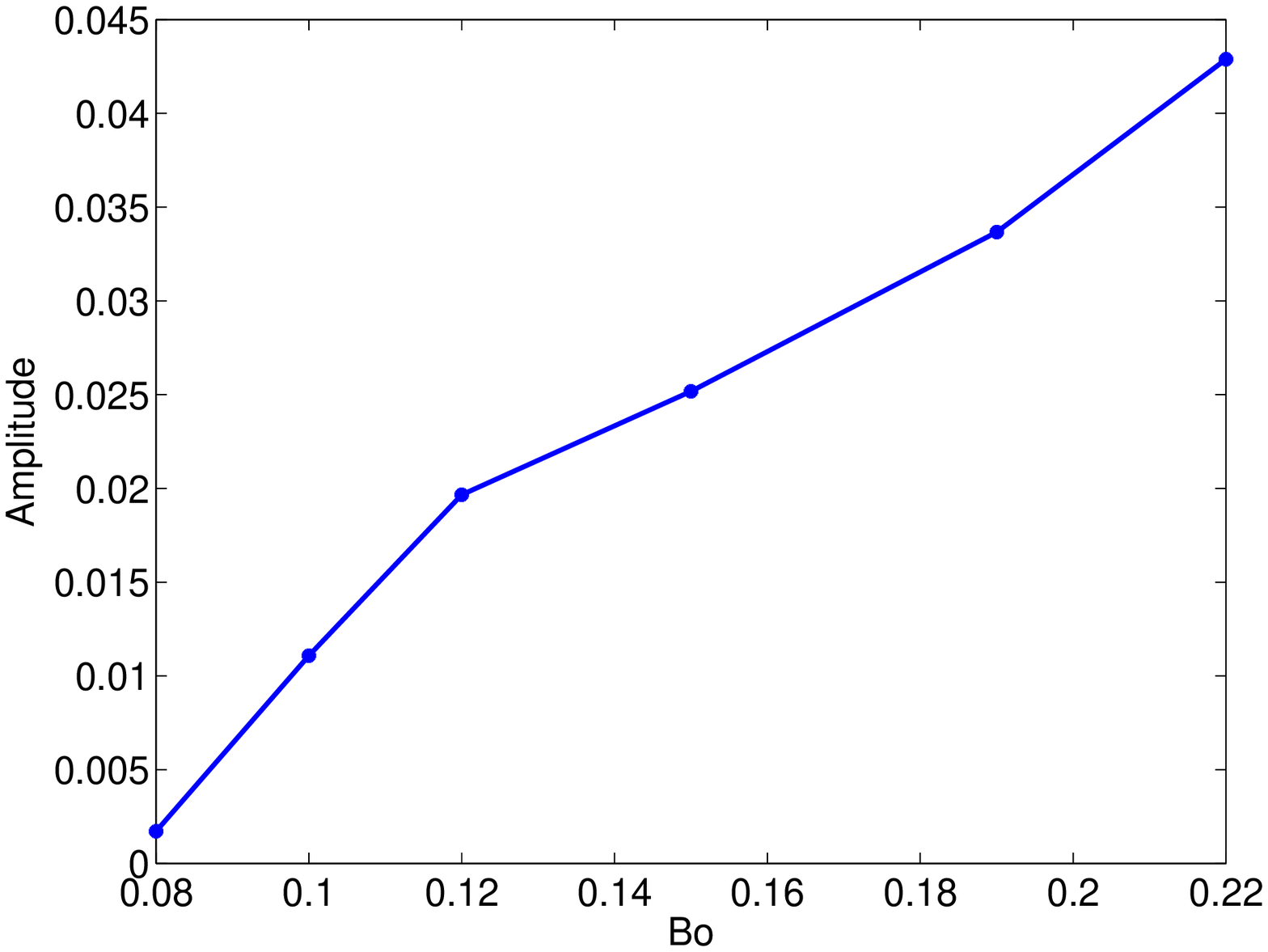}
  \caption{\small\em Amplitude of the fitting sinusoidal wave as function of the Bond number for the profiles of Figure~\ref{fig:fr1_15cont}.}
  \label{fig:ampBo}
\end{figure}

The dependence of the amplitude of the computed oscillatory tail on the Bond number can be observed in Figure~\ref{fig:ampBo}. This shows the amplitude of the fitting sinusoidal wave as function of the Bond number for the profiles of Figure~\ref{fig:fr1_15cont}.

\subsubsection*{Multi-hump generalised solitary waves}

\begin{figure}
  \centering
  \subfigure[]{
  \includegraphics[width=0.485\textwidth]{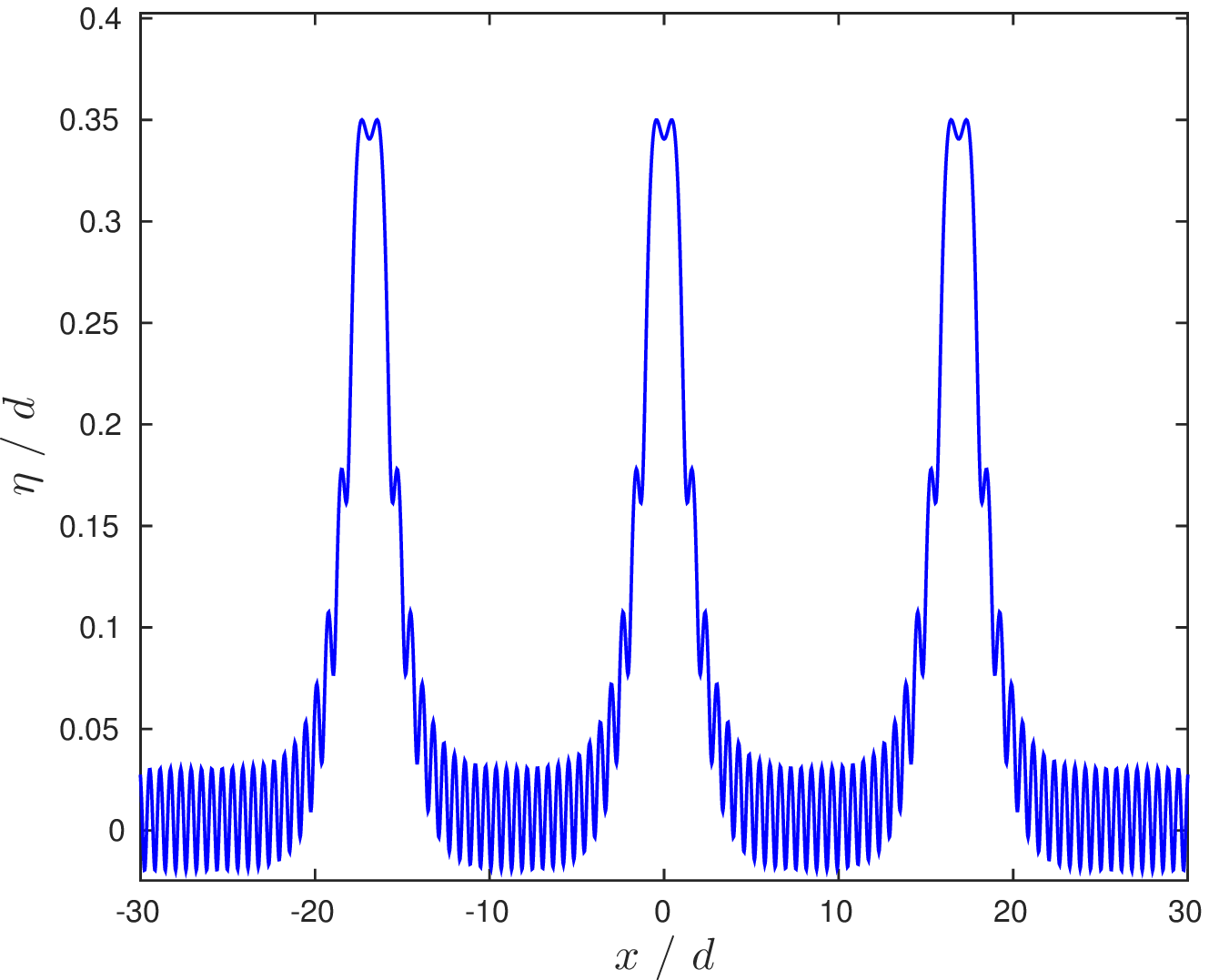}}
  \subfigure[]{
  \includegraphics[width=0.485\textwidth]{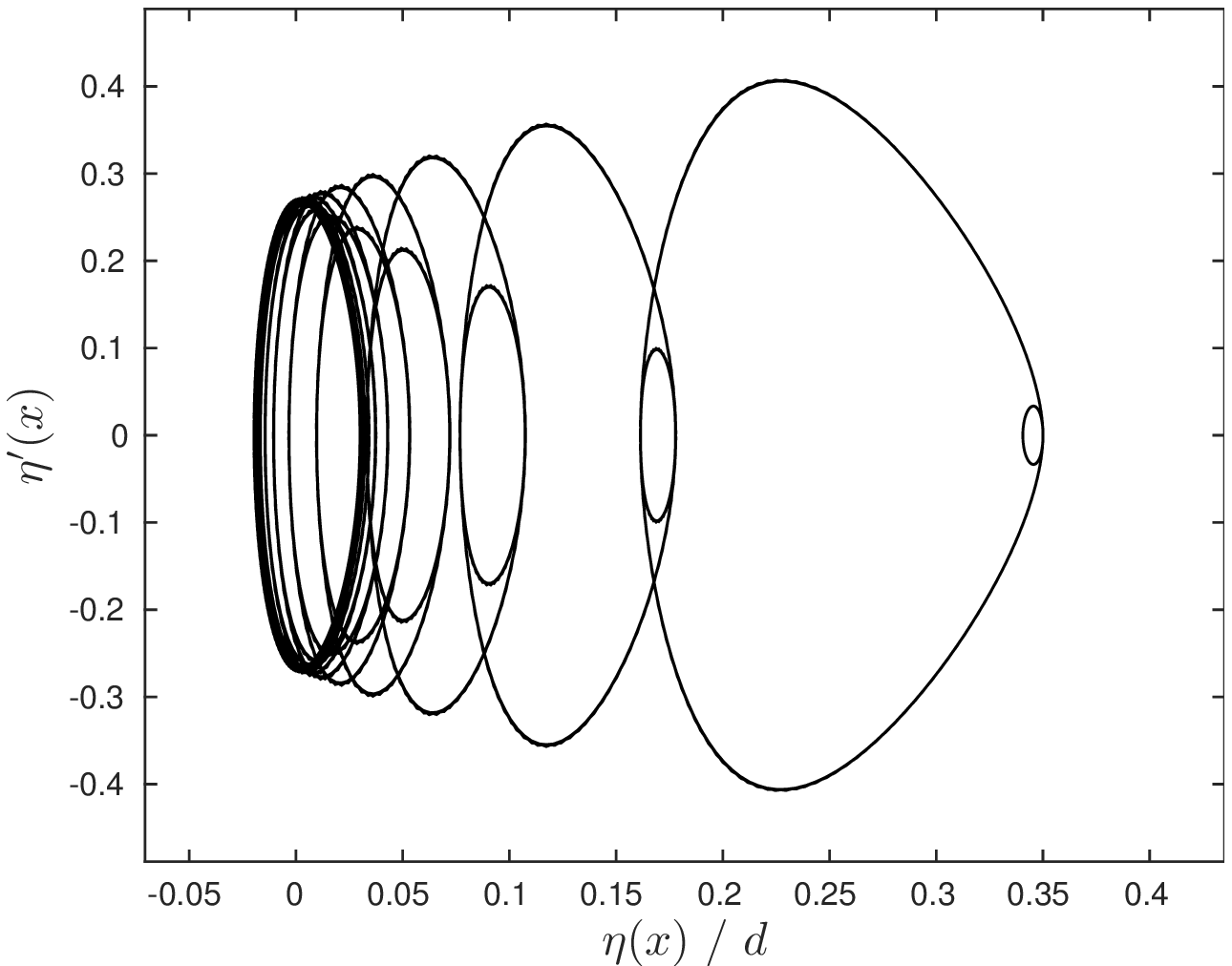}}
  \caption{\small\em Left: a three-pulsed generalised solitary wave solution for $\Fr = 1.17$ and $\Bo = 0.12$. Right: Phase space plot. The effective parameters are: $d_\infty = 1.0042$, $\Fr_\infty = 1.1995$, $\Bo_\infty = 0.118995$.}
  \label{fig:multipos2}
\end{figure}

As shown in \cite{Clamond2015a}, multi-hump generalised solitary waves (that is, homoclinic orbits with several loops near a resonance) can be computed using separated bumps as initial iteration. They are illustrated in Figure~\ref{fig:multipos2}. We refer to \cite{Clamond2015a} for other interesting examples.

\section{Conclusions and perspectives}\label{sec:concl}

This study is a continuation of a previous work \cite{Clamond2015a}. The goal here was to explain in details the numerical techniques used in \cite{Clamond2015a} and to provide some additional insights. 

The Babenko equation was discretised with a Fourier-type pseudo-spectral method \cite{Boyd2000}. The resulting discrete nonlinear and nonlocal system was solved using the \acf{lm} method. The reason to use this technology is to overcome the drawbacks of the direct application of Newton's (and quasi-Newton's) methods in this and related problems \cite{Boyd2007}. The rank-deficient character of the Jacobian, due to the translational invariance of the solitary waves, was the main reason for this choice. Using this formulation, we succeeded to compute, by natural continuation in the Bond number $\Bo$, the generalised solitary waves of elevation for $\Bo < 1/3$ \cite{Champneys2002}. Above the critical Bond number $\Bo = 1/3$, we found the classical localised solitary waves of depression which propagate with subcritical speeds $\Fr < 1$, in agreement with the predictions of the KdV5 model. Here, we completed the numerical results with additional information provided by the computations: ({\em i}) on the asymptotic decay of the classical solitary waves; ({\em ii}) on the oscillatory tails of the computed generalised solitary wave profiles and ({\em iii}) on the internal hydrodynamics structure. In the latter case, we unveil what happens with various physical fields (velocities, pressure, accelerations) under a generalised solitary wave. This was made using Cauchy-type integral representations, available when we take the full advantage of the conformal mapping technique, \cite{Dutykh2013b}.

The present work opens several research directions. First of all, the variability of solutions arising in several directions is a subject to be explored further. Then, the dynamics and stability of these solutions have to be studied. This requires the corresponding unsteady formulation along with an efficient discretisation procedure. The generalised solitary waves are homoclinic to Stokes waves. We conjecture that most of the solutions computed in this study might be unstable, since the periodic (gravity) Stokes wave is unstable \cite{McLean1982}, mainly because of the Bespalov--Talanov--Benjamin--Feir instability \cite{Benjamin1967a, Bespalov1966}. 

The role of the surface tension has to be studied properly. Can it suppress some instabilities? Finally, the corresponding Babenko-type formulation for periodic capillary-gravity waves has to be developed as well with the same questions about their dynamics and stability. It will be slightly different from the Babenko equation derived in \cite{Clamond2015a} due to some extra parameters, which characterise periodic waves. The numerical procedure for the computation should also change in some suitable way.

The code employed to obtain all the numerical results presented in this manuscript (but also the results of our previous work \cite{Clamond2015a}) is freely available to download at the following URL address 
\cite{Clamond2015b}:
\begin{itemize}
  \item \url{https://github.com/dutykh/BabenkoCG/}
\end{itemize}
We invite the readers to use this code to compute and to study generalised solitary waves.


\subsection*{Acknowledgments}
\addcontentsline{toc}{subsection}{Acknowledgments}

D.~\textsc{Clamond} \& D.~\textsc{Dutykh} would like to acknowledge the support of CNRS under the PEPS InPhyNiTi 2015 project FARA. A.~\textsc{Dur\'an} has been supported by project MTM2014-54710-P. D.~\textsc{Dutykh} \& A.~\textsc{Dur\'an} acknowledge the hospitality of Laboratoire J.A.~Dieudonn\'e UMR 7351 as well as of the University of Nice Sophia Antipolis where important parts of this work have been performed. All the authors would like to thank Professor Taras~\textsc{Lakoba} (University of Vermont, USA) for very stimulating discussions on the numerical methods for nonlinear travelling waves. 

\addcontentsline{toc}{section}{References}
\bibliographystyle{abbrv}
\bibliography{biblio}

\end{document}